 \documentclass[twocolumn,longbibliography,aps,prl,preprintnumbers,amsmath,amssymb]{revtex4-2}

\usepackage[dvipsnames]{xcolor} 
\usepackage{comment} 
\usepackage{amsmath}
\usepackage{xcolor}

\usepackage{graphicx}
\usepackage{dcolumn}
\usepackage{bm}

\newcommand{\cabb}{Cs$_2$AgBiBr$_6$}

\newcommand{\cubic}{$Fm\bar{3}m$}
\newcommand{\tetra}{$I4$/$m$}
\newcommand{\mono}{$P12_1$/$n1$}
\newcommand{\monotwo}{$I12/m1$}
\newcommand{\triclinic}{$I\bar{1}$}

\newcommand{\wav}{cm$^{-1}$}
\newcommand{\etal}{{\it et al.}}

\begin{document}


\title{Low-temperature structural instabilities of the halide double perovskite \cabb{} investigated via x-ray diffraction and infrared phonons}

\author{Collin Tower}

\author{Fereidoon S. Razavi}%
\author{Jeremy Dion}%

\author{Maureen Reedyk}%

\affiliation{%
 Brock University, Department of Physics, St. Catharines, ON, Canada
}%

\author{J\"urgen Nuss}

\affiliation{
 Max Planck Institute for Solid State Research, Stuttgart, Germany
}%
\author{Reinhard K. Kremer}

\affiliation{
 Max Planck Institute for Solid State Research, Stuttgart, Germany
}%

\date{\today}

\begin{abstract}
The halide double perovskite \cabb{} has been proposed as a potential replacement for organic halide perovskites for photovoltaic applications. Further investigation of its dielectric response, optical properties  and structural stability is thus warranted. \cabb{} exhibits a well-documented structural phase transition at 120~K but indications for an additional lower temperature ($\sim$~40~K) phase transition have also been reported. On the basis of measurements of the specific heat capacity, temperature dependent powder X-ray diffraction, low frequency capacitance, and infrared reflectivity we clarify the previously reported splitting of phonon modes in the Raman spectrum at $\sim$40~K as due to a subtle structural phase transition from the tetragonal \tetra{} structure to a monoclinic \mono{} crystal structure. The infrared active vibrational modes are experimentally investigated in the three structural regimes. In the cubic structure at room temperature the four IR active modes are observed at 135,$\sim$85, 55, and $\sim$25~\wav{}. As the symmetry reduces to tetragonal a minute splitting of these modes is found, however below 40~K additional mode splitting is observed indicating a further reduction in symmetry.
\end{abstract}

\maketitle

\section{\label{sec:Intro}Introduction}
Halide perovskite materials have come into increased scientific focus due to their potential 
for photovoltaic device applications such as in solar cell technology\cite{SolarBook}, or X-ray detection devices~\cite{X-ray_1}. Lead halide perovskites with the composition A$^+$Pb$^{2+}$X$_3^-$,  where A$^+$ can be an organic cation, typically methylammonium (MA), formamidinium or an alkali cation, and X$^-$ is a halide anion with Cl$^-$, Br$^-$, and I$^-$ are broadly investigated examples.
While MAPbI$_3$ has reached noteworthy photovoltaic efficiencies its integration into photovoltaic devices has been held back by issues of thermal instability~\cite{MAPbI3_ThermalStability}, and ecological concern over the incorporation of toxic lead~\cite{DoublePerovskite_Defect_Tolerance}. 
As the thermal stability of inorganic materials is typically higher than that of organic materials the former problem 
is addressed by replacing the organic cation with an inorganic alkali metal such as Cs. Inorganic halide perovskites however, still exhibit stability issues through structural transformations and oxidization~\cite{CsPbI3_Stability}. 

Substitution with less toxic elements 
with the same valence electron configuration as lead such as tin and germanium has been investigated. However, due to the higher energy of their outer s orbitals these materials readily oxidize to a non-photoactive state~\cite{Pb_Replace}. Bi$^{3+}$ offers a less toxic alternative to lead with the same orbital structure. However, to substitute Pb$^{+2}$ with trivalent Bi$^{3+}$ the composition must be modified by adding a monovalent cation to maintain charge neutrality. The resulting double perovskites have the composition A$_2$BB'X$_6$, where A and B are monovalent metal cations, typically Cs, and  Cu, Ag, or Au, respectively, B' is a trivalent pnictogen, such as  Sb or Bi, and X is a halide Cl$^-$, Br$^-$, or I$^-$ anion.~\cite{DoublePerovskite_combinations} The Goldschmidt tolerance and octahedral factors suggest that Ag$^+$ has an ideal ionic radius for creating stable octahedral coordination with Br and I~\cite{Perovskite_tolerance} and as such, a halide double perovskite that has garnered great interest is \cabb{}.
There is a well documented structural phase transition in \cabb{}  where upon lowering temperature the crystal structure changes from a highly symmetric cubic structure space group \cubic{} to a tetragonal crystal structure. 
This transition occurs at around $\sim$~120~K and affects many of the opto-electronic properties including the exciton energy, the band gap and the carrier lifetime.~\cite{Schade2019} From neutron powder diffraction data, Schade \textit{et al.} refined the crystal structure at 30~K assuming the tetragonal space group \tetra.~\cite{Schade2019}

The low temperature structural behavior of \cabb{} was further investigated by Keshavarz \etal{} who noted a peak in the thermal expansion coefficient at $\sim$40~K and proposed that it was due to an additional structural anomaly of unknown origin.~\cite{CenterFreq_Hardening} 
In contrast to the 120~K structural phase transition causing a sharp anomaly in the heat capacity ($C_{\rm P}$), Keshavarz \etal{} reported that the $\sim$~40~K transition could not be seen in heat capacity  measurements, while Cohen \etal{} detected it in the first temperature derivative of $C_{\rm P}$ and proposed that it is weakly first order.\cite{cohendiverging}
He \etal{} carried out extensive X-ray and neutron diffraction measurements to investigate this low temperature phase transition and suggested it is caused by a phonon instability of the tetragonal phase creating a complex ground state with a very large unit cell comprising several hundred atoms.~\cite{ComplexGroundState} 

The double perovskites have been predicted to have ultra low thermal conductivity attributed to strong anharmonicity in their vibrational properties.
For \cabb{} this strong anharmonicity has been suggested to be the cause of the cubic to tetragonal phase transition due to the occurrence
of a soft phonon branch.~\cite{ThermalCond_DFT1,ThermalCond_DFT2}
The Raman active phonons of \cabb{} as it passes through both phase transitions have been thoroughly studied experimentally.\cite{cohendiverging} Splitting of some and a number of additional Raman modes were observed suggesting a symmetry lowering of the crystal structure below 40~K, possibly to a monoclinic crystal structure.
The infrared phonon spectrum has been predicted via theoretical DFT calculations~\cite{DFT_CASC}. 

Herein we study experimentally the effect of the structural instabilities in \cabb{} on its infrared (IR) optical properties via the far infrared (FIR) vibrations. Knowledge of the IR active modes imparts further insight into the optical properties of \cabb{} and provides valuable information which can aid first principles calculations of the phonon dispersion as experimental values can be compared with the calculated values to evaluate the accuracy of the DFT calculations and refine the parameters used. As the crystals are centrosymmetric, the IR active modes determined experimentally in this work together with the complementary Raman active modes which have been measured extensively, for example by Cohen \etal{} \cite{cohendiverging}, give the nearly complete phonon frequencies at the gamma point. As phonons carry thermal energy, they play an important role in properties such as the heat capacity, thermal conductivity, expansion and transport. Confidence in DFT predictions of such properties will thus be aided by full knowledge of the experimentally determined gamma point frequencies.

We have measured the IR active modes and analyzed the results using the Kramers-Kronig analysis. The results agree reasonably well with the theoretical DFT predictions\cite{DFT_CASC,ThermalCond_DFT1,ThermalCond_DFT2,DFT_Cubic_tetra}. Additional phonon modes below the $\sim$~40~K structural transition are observed in the IR spectra. To gain further insight we performed extensive low-temperature X-ray powder and single crystal diffraction measurements. These confirm the structural phase transition from a cubic to a tetragonal crystal structure at $\sim$120~K. Subtle signatures are seen in various quantities (cell volume, lattice parameters) consistent with an additional structural phase transition  below $\sim$~40~K. Taking into account the additional Raman modes found below $\sim$40~K by  Cohen \textit{et al.}~\cite{cohendiverging} and the appearance of the additional IR modes we observe below 40~K, we are lead to conclude that the low temperature crystal structure is monoclinic and can be described in space group \mono{}.
 
\section{\label{experimental} Experimental}
\begin{figure}[]
\includegraphics[width=0.45\textwidth,keepaspectratio]{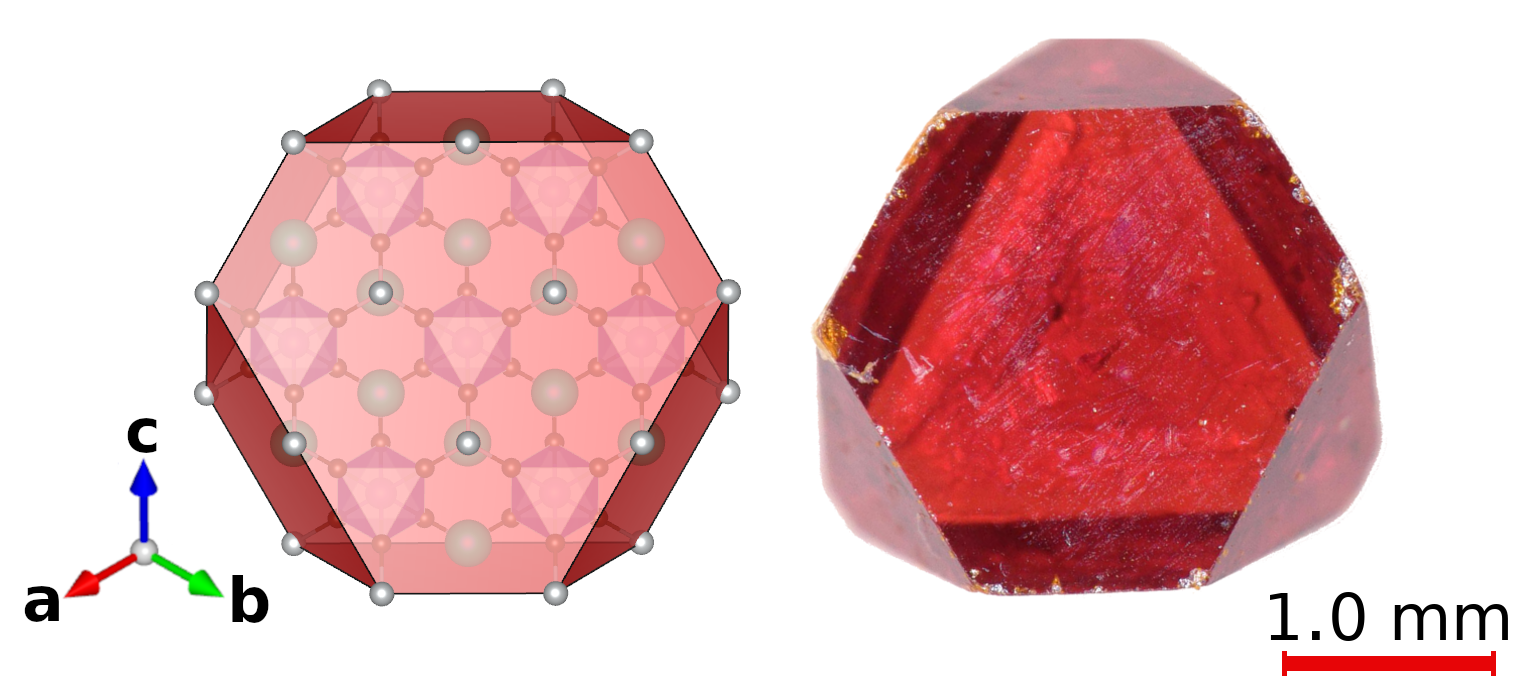}
\caption{\label{fig:CrystalPics} Left: An example crystal geometry of \cabb{} taken along the eight permutations of the (111) plane. Right: A crystal synthesized using the controlled cooling method.}
\end{figure}

\begin{figure}[]
\centering
\includegraphics[width=9cm]{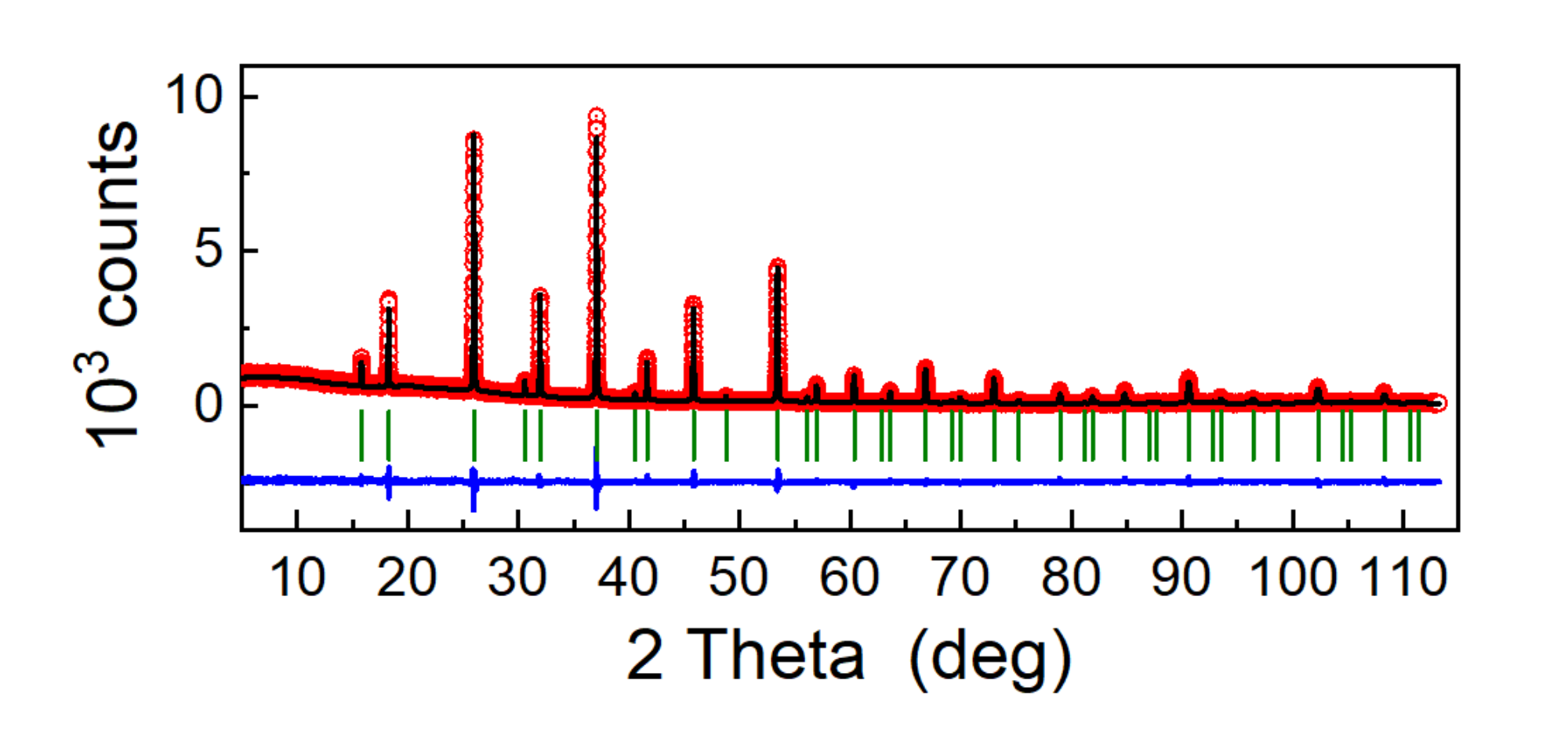}
\caption{\label{fig:RTCobalt} PXRD pattern of a powder specimen obtained by crushing crystals of \cabb, collected at 298~K using Co $K_{\alpha 1}$ radiation ($\lambda$~=~1.789010~\AA). The red circles display the measured, the black solid line represents the pattern calculated  using the positions of the Bragg angles indicated by the green vertical bars. The blue solid line highlights the difference between measured and calculated pattern.
Cs, Ag, Bi, and Br atoms occupy Wyckoff sites 8$c$, 4$b$, 4$a$, and 24$e$, respectively,  in space group no. 225 \cubic{} with $Z$~=~4 (See TABLE \ref{tab:Wyckoff}) .}
\end{figure}

Crystals of \cabb{} were synthesized following the controlled cooling method outlined by Yin \etal\cite{Crystal_Growth}. Details are included in the Supplemental Material.\cite{Supplement} 
A typical crystal is shown in FIG. \ref{fig:CrystalPics}.
In order to check for impurities a random selection of crystals was crushed and checked by Powder X-ray Diffraction (PXRD) at room temperature. FIG. \ref{fig:RTCobalt} shows a room-temperature powder X-ray diffraction pattern of \cabb{} in comparison with a Rietveld profile refinement assuming space group \cubic. The pattern  collected using Co$K_{\alpha 1}$ radiation  ($\lambda$~=~1.789010~\AA) proves absence of impurities or splitting of Bragg reflections at high Bragg angles assuring cubic symmetry at room temperature.

Heat capacity and capacitance measurements at 50 Hz were carried out with a physical property measurement system (PPMS, Quantum Design) from 4-300 K. The dielectric permittivity was determined from the capacitance. Further details can be found in the Supplemental Material.\cite{Supplement}

The lattice parameters of \cabb{} were determined as a function of temperature from profile refinements (Rietveld refinement~\cite{Fullprof}) of powder X-ray diffraction patterns collected of two powder samples with a Bruker D8 Discovery X-ray diffractometer system (Bragg-Brentano scattering geometry) using monochromatized CuK$\alpha_1$ radiation). 
One powder sample was prepared by mildly crushing small crystals in an agate mortar. The particle size of the sample was adjusted between 32~$\mu$m and 63~$\mu$m by straining the powder through sieves of the respective mesh size. The second sample was prepared from a powder formed in lieu of bulk crystals in one of our synthesis attempts. Additional information can be found in the Supplemental Material.\cite{Supplement}

Low temperature single-crystal X-ray diffraction measurements were performed at select temperatures using Mo$K\alpha_1$ radiation with  cooling via an Oxford Cryosystems device.\cite{Cakmak2009}
The reflection intensities were integrated with the SAINT subroutine available in the Bruker Suite software.\cite{Bruker2019} An absorption correction\cite{Sheldrick2016} was applied and the crystal structures were solved by direct methods and  full-matrix least-square fitting using standard software (SHELXTL, JANA2006)\cite{Sheldrick2008,Sheldrick2015,Petricek2014} For more details see  Ref.~\cite{Supplement}.

Temperature dependent IR reflectance was measured over a range of 20-11000 \wav\ using a Bruker-IFS-66v/s Fourier transform spectrometer with in house modifications for temperature control. Details of the source, detector and beamsplitter combinations used can be found in the Supplemental Material.\cite{Supplement}
The sample was annealed at 100 $^{\circ}C$ for several days before the IR measurements were performed to ensure it was free of water contamination, as prior to annealing the mid-infrared (MIR) range showed evidence thereof. See Supplemental Material for further details.\cite{Supplement}
The absolute reflectivity was determined by using a metallic film deposited on the sample surface {\it in situ} as reference\cite{Homes}.
Kramers-Kronig (KK) analysis was performed to transform from the experimentally determined reflectivity spectra to the complex dielectric function.

\section{Results and Discussion}
\subsection{Heat Capacity}

\begin{figure}[]
\centering
\includegraphics[width=0.45\textwidth,keepaspectratio]{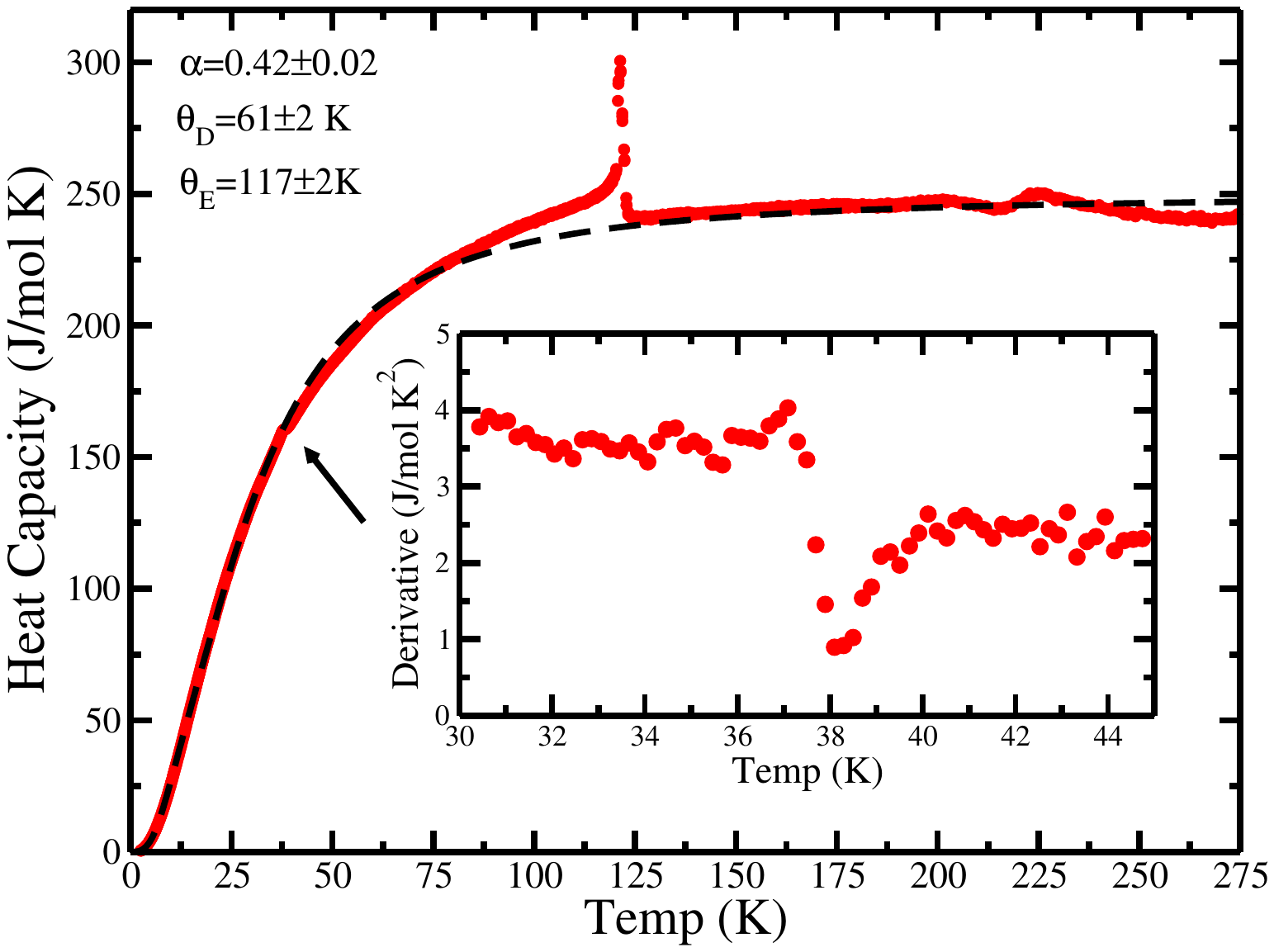}
\caption{\label{fig:HeatCap} The heat capacity of \cabb{} displaying 
a sharp anomaly at $\sim$120~K. The inset shows the numerical derivative with respect to temperature  which manifests as a slight inflection  in the heat capacity near $\sim$38~K.}
\end{figure}

The heat capacity of \cabb{} as a function of temperature is shown in FIG. \ref{fig:HeatCap}. It displays a pronounced $\lambda$ anomaly near 120~K due to the crystal structure transition from cubic to tetragonal (\cubic{} $\rightarrow$ \tetra{}) described previously by Schade \textit{et al.}.\cite{Schade2019} An  analog anomaly near  $\sim$40~K is not observed in the direct heat capacity data. However, the derivative with respect to the temperature reveals an inflection near $\sim$38~K (see inset in FIG. \ref{fig:HeatCap}) corresponding well to the temperature below which e.g. additional Raman peaks were detected.\cite{cohendiverging}

The heat capacity was fit to a weighted Einstein and Debye model given by Eqn.\ref{eq:HC}. The model includes three parameters, a weight factor between the Debye and Einstein model $\alpha$, and the Debye and Einstein temperatures $\theta_{\rm{D}}$ and $\theta_{\rm{E}}$. $T$ is the temperature, $R$ is the ideal gas constant and $N$ the number of atoms in the formula unit. 

\begin{equation}{\label{eq:HC}}
\begin{split}
C_p ={} 9\alpha NR \left( \frac{T}{\theta_D} \right)^3 \int_0^{\theta_D/T} \frac{x^4e^x}{(e^x-1)^2}dx + \\
 &\hspace*{-4.25cm} + 3 (1-\alpha) NR \left( \frac{\theta_E}{T} \right)^2 \frac{e^{\theta_E/T}}{(e^{\theta_E/T}-1)^2} 
\end{split}
\end{equation}

This fit, with parameters given in FIG. \ref{fig:HeatCap} agrees well at low temperatures deviating slightly at the $\sim$40~K phase transition, then deviating significantly at the 120~K structural phase transition before recovering at the high temperature Dulong-Petit law value of $\sim$250~J/molK just above the 120~K tetragonal to cubic phase transition. The features observed in the experimental curve near 220 K are attributed to anomalies in the heat capacity of the ApiezonN vacuum grease.\cite{N-Apiezon}
The fit converged to  a 42$\pm$2~\%  Debye contribution signifying that the acoustic phonons  have a slightly weaker contribution than the optical phonons to the heat capacity. 

The Einstein and Debye temperatures found above agree reasonably well with experimental and theoretical determinations of the phonon frequencies. The Debye frequency of 42$\pm$1 \wav{} is comparable to the maximum acoustic phonon frequency of $\sim$50 \wav\ determined by the DFT calculations of Zelewski \textit{et al.}\cite{DFT_Cubic_tetra}. 
The Einstein frequency of 81$\pm$1 \wav{} is in good agreement with the average of the experimental optical phonon frequencies of 91 \wav, as derived from the room-temperature Raman data reported by Cohen et al.\cite{cohendiverging} and the infrared modes identified in this study.

\subsection{Low-Temperature Crystal Structure}
\begin{figure}[]
\centering
\includegraphics[width=9.0cm]{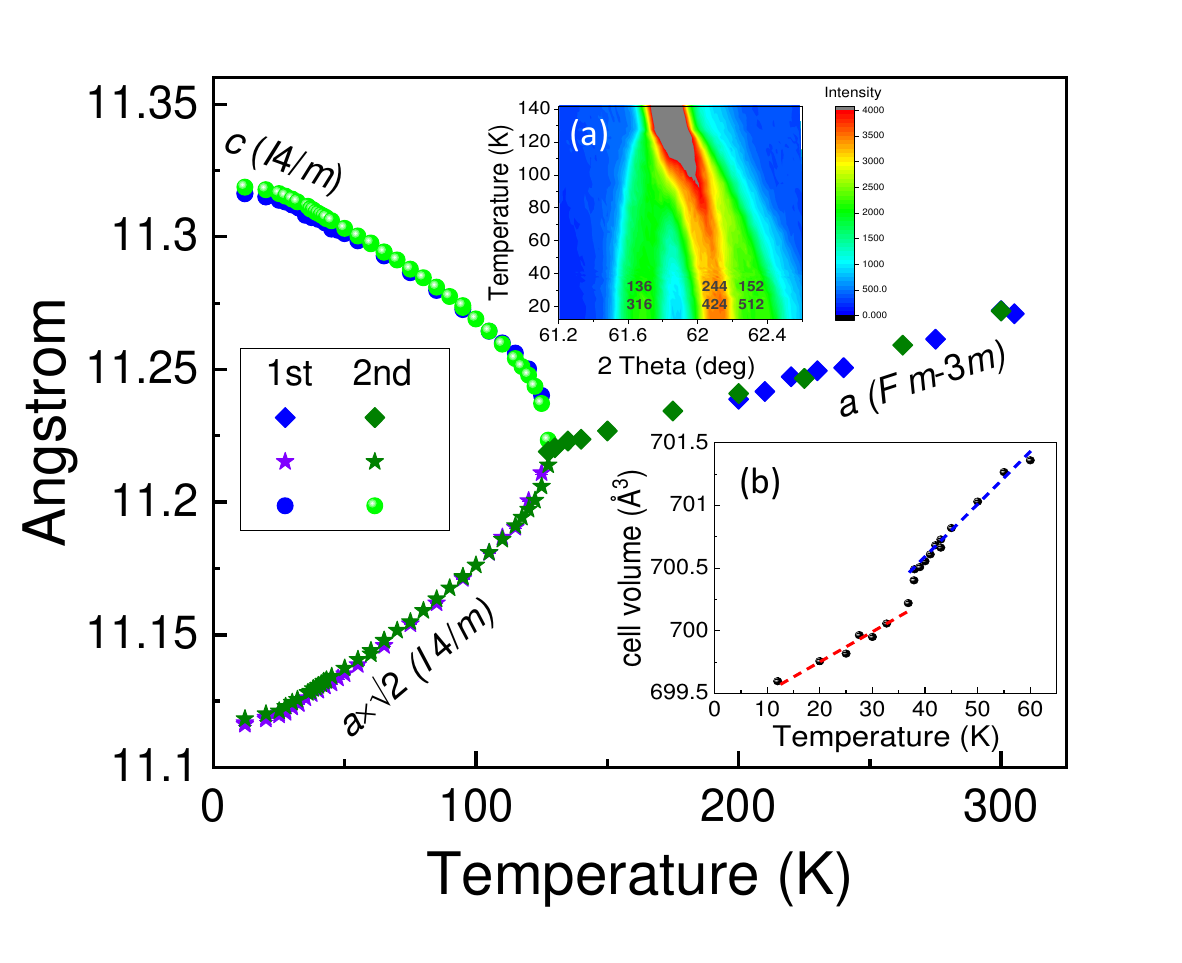}
\caption{\label{fig:TetLattPar} Lattice parameters as a function of temperature as obtained by Rietveld profile refinements of two sets of X-ray diffraction patterns collected using X-rays of wavelength of $\lambda$~=~1.540598~\AA. The inset (a) displays a contour plot highlighting the splitting of the (642)$_{\rm cubic}$ Bragg reflection below $\sim$120~K. 
The inscribed Bragg indices are with respect to the tetragonal space group \tetra. The lower inset (b) shows the temperature dependence of  the tetragonal cell volume below 60~K. The dashed lines are guides to the eye.}
\end{figure}

\begin{figure}[]
\centering
\includegraphics[width=9cm]{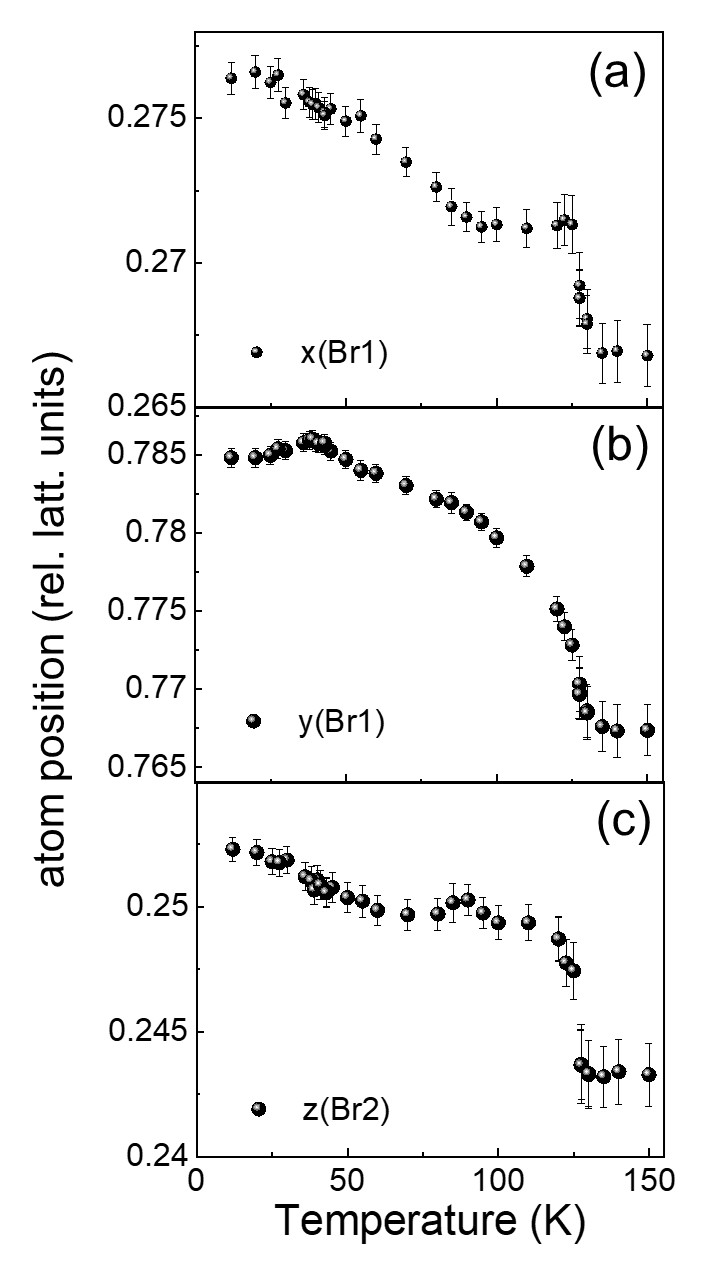}
\caption{\label{fig:AtomPosBr1rev} Atom positional parameters $x$, $y$ of the Br1, (a) and (b), 
and $z$ of the  Br2 atoms, (c), in \cabb{} as obtained from Rietveld profile refinements of the powder diffraction data collected using Cu $K_{\alpha 1}$ radiation assuming space group \tetra.}
\end{figure}

A first approach to explore the low-temperature crystal structure of \cabb{} was made based on PXRD patterns collected between room temperature and 12~K on two samples obtained by crushing a set of larger crystals and  a polycrystalline powder formed in lieu of bulk crystals in one synthesis attempt. Initially the powder diffraction patterns were analysed by  Rietveld profile refinements assuming the tetragonal space group \tetra{} for $T<$~120~K and \cubic{} for $T>$120~K. Within error bars the lattice parameters of the two samples (see FIG.~\ref{fig:TetLattPar}) are identical and correspond well to those reported before by Schade \textit{et al.}\cite{Schade2019} The splitting of the lattice parameters at $\sim$120~K reflects  the cubic-to-tetragonal structural phase transition. The Br atom positional parameters also respond to the cubic-to-tetragonal  structural phase transition. In FIG.~\ref{fig:AtomPosBr1rev} (a), (b), and (c) we summarize the $x$, $y$, and $z$
positional parameters of the bromine atoms labelled Br1 and Br2, respectively. All exhibit a clear increase below $\sim$125~K.

Below $\sim$40~K the profile refinements assuming the space group \tetra, requiring a splitting of the single Br cubic site into two sites (Br1 and Br2), also fits the diffraction patterns fairly well, indicating no dramatic symmetry lowering of the crystal structure, in agreement with the finding reported by Schade \textit{et al.}.\cite{Schade2019}
However, while changes in the \tetra{} refined lattice parameters at around 40~K are not immediately evident, the temperature dependence of the cell volume exhibits  a decrease and slight change of slope at this temperature (see FIG. ~\ref{fig:TetLattPar}(b)).
This finding is paralleled by  an anomaly of the atom positional parameter $y$ of the Br1 atoms (Wyckoff site 8h - See TABLE \ref{tab:Wyckoff} ) at the same temperature (see FIG. \ref{fig:AtomPosBr1rev}). Whereas $y$ of Br1 atom passes through a peak at 38~K, the $x$ positional parameter tends to saturation below this temperature. Similarly, the
$z$ positional parameter of the Br2 atoms (Wyckoff site 4e)
levels off, reaching a value of 0.2523(5) in the temperature region below 40~K.

The single crystal X-ray structure analysis (for more details see the Supplemental Material, Ref. \cite{Supplement}) confirms the tetragonal space group \tetra{} for the crystal structures at 60~K and 100~K. Also an intensity data set collected at 32~K can be readily refined assuming tetragonal symmetry and space group \tetra. A similarly acceptable refinement can, however, also be achieved in the triclinic space group \triclinic. This finding, along with the anomalies observed in the heat capacity, the cell volume and the bromine atom positional parameters, suggested attempting a refinement of the powder diffraction patterns collected below 40~K assuming lower than tetragonal symmetry. 

\begin{figure}[]
\centering
\includegraphics[width=9.0cm]{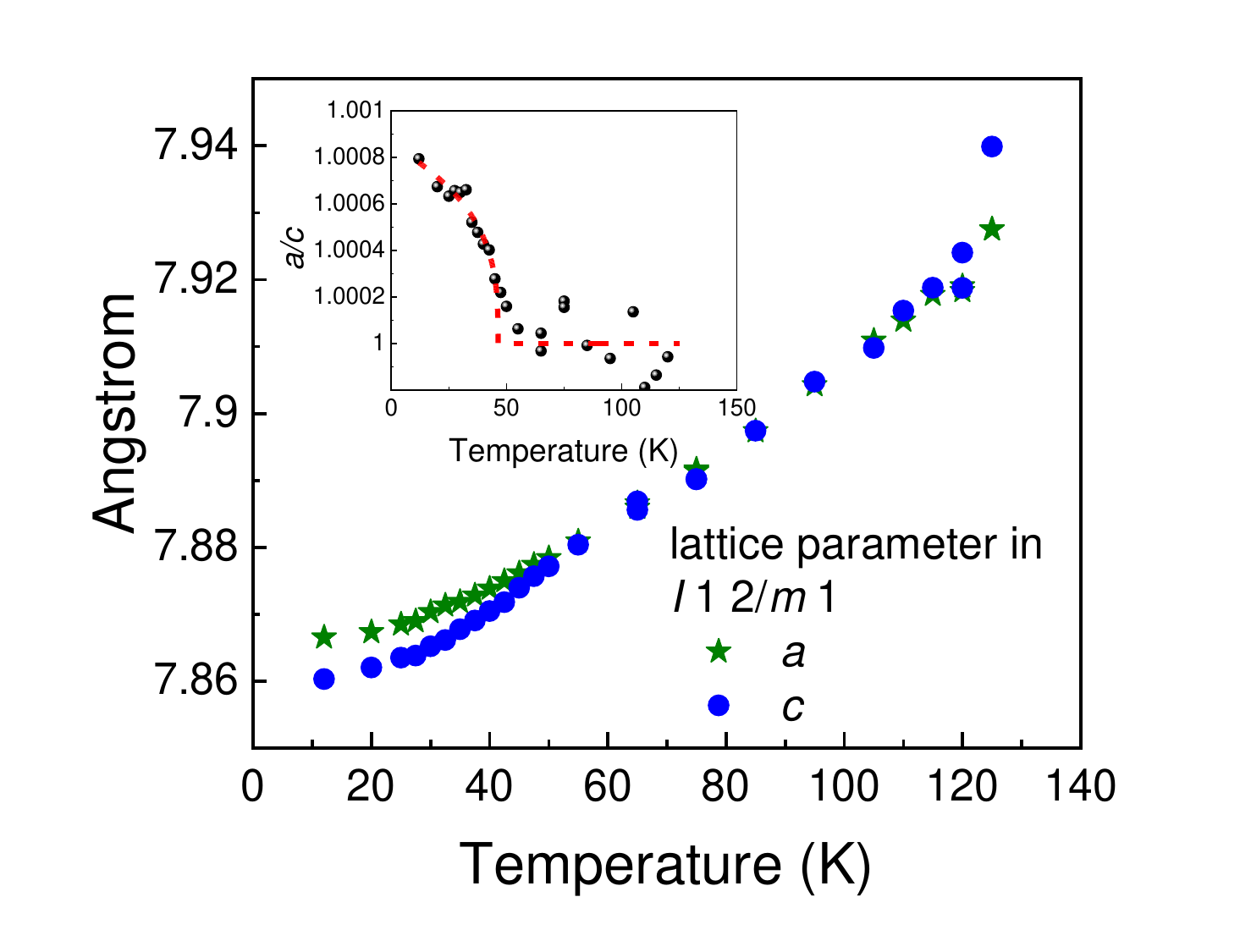}
\caption{\label{fig:AacTemp} Lattice parameters $a$ and $c$ obtained from Le Bail fits\cite{LeBail} of the x-ray powder diffraction patterns ( assuming the monoclinic space group \monotwo{} (space group no. 12). The inset shows the ratio $a$/$c$. The red dashed line is a fit to a critical power law indicating a critical temperature $T_{\rm{c,mono}}$~$\approx$~46~K.}
\end{figure}

Motivated by the commonly encountered double perovskite transformation from \tetra{} to \monotwo{} (see Supplemental Material for a discussion of phase transitions in double perovskites\cite{Supplement})
we first performed Le Bail fits of the powder diffraction patterns assuming monoclinic symmetry  (\monotwo{}, space group no. 12)  to trace minute changes of the lattice parameters as a function of temperature.\cite{LeBail}.
FIG.~\ref{fig:AacTemp} displays the $a$ and $c$ lattice parameter as a function of temperature obtained from the Le Bail fits of the powder diffraction patterns ($\lambda$~=~1.540598~\AA) assuming the special setting of monoclinic space group  \monotwo{} (no. 12). These correspond  to the lattice parameter $a$~=~$b$ in the tetragonal system. While the fitted lattice parameters $a$ and $c$ are identical between 120~K and $\sim$50~K a minute splitting develops at lower temperatures. The temperature dependence of the splitting ratio $a$/$c$ (see inset in FIG.~\ref{fig:AacTemp}) is similar to that of a critical power law with a critical temperature of $\approx$~46~K. For $T~\rightarrow$~0, the splitting amounts to $\approx$~0.1~\% and the monoclinic angle $\beta$ becomes 89.931(2)$^{\rm o}$ (see also TABLE~\ref{Table2}, 3rd col. from left).

Another frequently observed low-temperature monoclinic crystal structure \mono{} resulted in refinements with reliability parameters very close to those of the tetragonal structure with reliability parameters comparble to those of the refinement in  the monoclinic space group \monotwo.
This observation and the examples of the transition metal oxide double perovskites discussed in the Supplemental Material\cite{Supplement} prompted us to try a profile refinement of the powder diffraction patterns collected below $\sim$40~K  assuming the space group \triclinic{} which lead to a slight reduction of the reliability factors compared to the refinement using space group \monotwo{} and a minute cell distortion away from the (almost) orthorhombic cell found with the space group \monotwo{} or \mono.
For the sake of clarity (all angles remain close to 90~$^{\rm o}$ in  setting \triclinic) we carried out the refinements in the  \triclinic{} setting of the triclinic space group $P\bar{\rm 1}$.

\begin{figure}[]
\centering
\includegraphics[width=9cm]{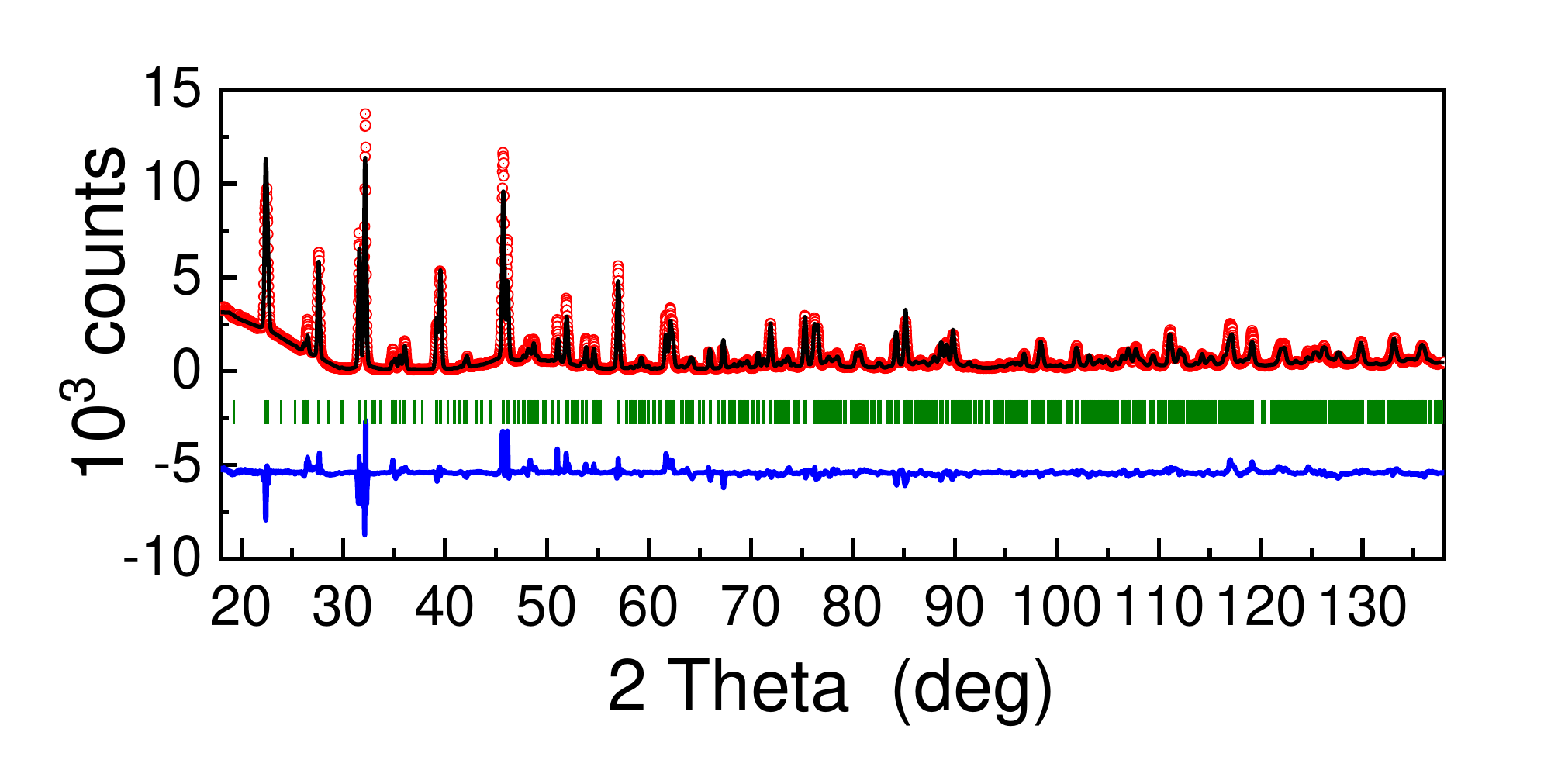}
\caption{\label{fig:T20Krefine} PXRD  pattern of a sample of crushed crystals of \cabb{} collected at 20~K using Cu $K_{\alpha 1}$ radiation ($\lambda$~=~1.540598~\AA). The red circles display the measured data, the black solid line represents the pattern calculated  using the positions of the Bragg angles indicated by the green vertical bars. The blue solid line below highlights the difference between measured and calculated pattern.
Cs, Ag, Bi, and Br atoms occupy Wyckoff sites 4$e$, 2$b$, 2$a$, and 4$e$, respectively,  in space group no. 14, setting $P\rm{1 2}_1/n 1$. (See TABLE \ref{tab:Wyckoff}).}
\end{figure}

\begin{figure}[]
\includegraphics[width=0.45\textwidth,keepaspectratio]{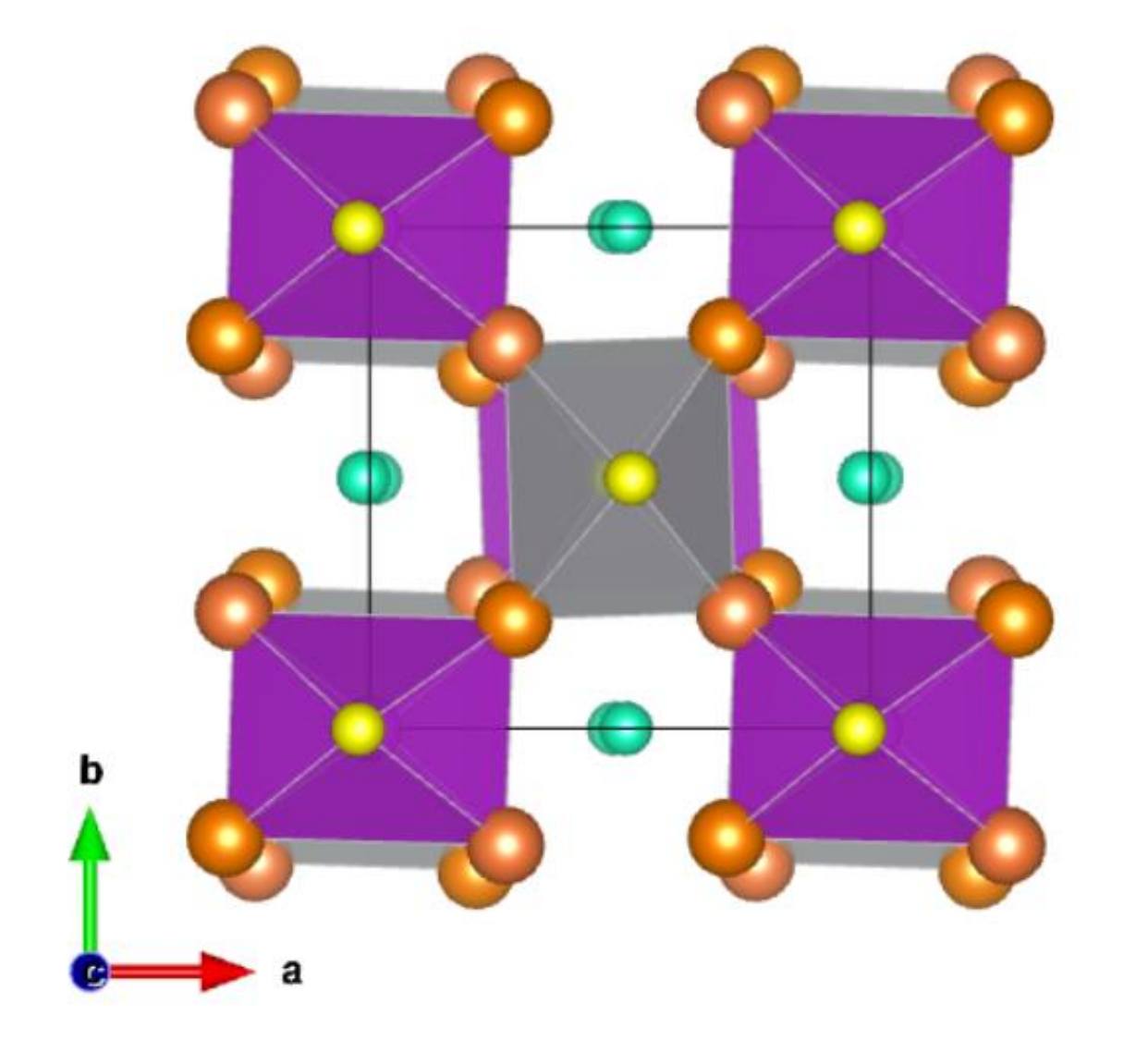}
\caption{\label{fig:Structure} Crystal structure of \cabb{} projected along [001] as refined from the PXRD pattern collected at 20~K. The refinement was done assuming monoclinic symmetry ((space group no. 14  \mono)  assuming isotropic displacement parameters for all atoms. The Br atoms are represented by the yellow (Br1) and the orange (Br2 and Br3) spheres. The Ag and Bi atoms center the Br octahedra  depicted in grey and violet respectively. Cs atoms are shown as cyan spheres.}
\end{figure}

FIG.\ref{fig:T20Krefine} displays the powder diffraction pattern collected at 20~K in comparison with the Rietveld profile refinement assuming the space group \mono.  FIG.\ref{fig:Structure} displays the corresponding crystal structure of \cabb{} (\mono{} setting of the monoclinic space group no. 14).

The differences between the tetragonal, monoclinic and triclinic structure solutions refined are minute. Checking the triclinic structure solution with the PLATON software identified a possible tetragonal super symmetry \tetra.\cite{Platon}
TABLE \ref{Table2} and TABLE \ref{Table3} summarize the structure parameters and the calculated inter atomic distances at 20~K in comparison with those obtained from the refinements assuming space group \mono.

\begin{table}[]
\caption{Structural parameters of \cabb{} as obtained from a Rietveld profile refinement of the PXRD pattern collected at 20~K (wavelength Cu$K_{{\alpha}1}$). As discussed in the main text, four different space groups commonly found for double perovskites were imposed. Error bars are quoted if isotropic displacement parameters $B$ were refined. In some cases isotropic displacement parameters of the Br atoms were constrained to be identical. Isotropic displacement parameters marked with an (*) result from an anisotropic refinement assuming $\beta_{12}$~=~$\beta_{13}$~=~$\beta_{23}$~=~0.}
\label{Table2}
\begin{ruledtabular}
\begin{tabular}{l  c c c c }
space grp.   &  \triclinic  &   \monotwo & \mono  & \tetra \\
no.   &  2 &  12 & 14 & 87\\
\hline
\\
$a$  ({\AA})  & 7.9374(3)  & 7.8702(2)  & 7.8603(3)  &  7.86409(7)\\
$b$  ({\AA})  &  7.9265(3) & 11.3182(2) &  7.8672(3)  & 7.86409(7)\\
$c$  ({\AA})   & 11.1266(4) &  7.8584(2) & 11.3179(2)  & 11.3181(2)\\

$\alpha$ ($^{\rm o}$)   & 
90.017(4)    &    90     & 90           &  90\\
$\beta$ ($^{\rm o}$)    &  
89.914(4)    & 89.931(2) &  90.106(2)   & 90\\
$\gamma$ ($^{\rm o}$)  &  
91.031(2)    &  90       & 90           & 90\\

$V_{\rm{cell}}$  ({\AA}$^3$) & 699.96(5)& 700.00(3) &  699.88(4) & 699.96(1) \\

Cs   & &  & &  \\
$x$  & 0.500(4)    &  0.5       & 0.4882(7)  & 0.5   \\
$y$  & 0.500(3)    &  0.2526(8) & 1.000(5)   &  0.0  \\ 
$z$  & 0.250(2)    &  0         & 0.2546(3)  & 0.25  \\
$B_{\rm{iso}}$ ({\AA}$^2$)    & 0.82(3)    &  0.65(4) & 1.26(5)  & 0.85(3)*\\ 
\\
Ag  &  & & &   \\
$x$ & 0        &   0        &   0     &   0\\
$y$   & 0.5      &   0.5      &   0     &     0\\ 
$z$  & 0        &   0        & 0.5     & 0.5\\
$B_{\rm{iso}}$ ({\AA}$^2$) &  
    0.23(12)  &  0.2       & 0.2  & 0.2\\ 
\\    
Bi   & &  & &  \\
$x$  & 0.5   &  0     &   0     & 0   \\
$y$  & 0    &  0     &   0     &  0  \\ 
$z$  & 0    &  0     &   0     & 0   \\
$B_{\rm{iso}}$ ({\AA}$^2$)   & 1.08(9)  &  0.69(4) & 1.78(5)  & 0.76(4)*\\ 
\\
Br1   &  & & &    \\
$x$  &  0.252(4)  &  0         & 0.978(1)   & 0.2834(6) \\
$y$  &  0.248(4)  &  0.2491(7) & -0.00(1)  & 0.7785(6)\\
$z$  &  1.000(6)  &  0         & 0.2508(7)  &  0 \\
$B_{\rm{iso}}$ ({\AA}$^2$)  & 
    0.95(5)*    &  0.84(5)   & 1.27(9)  & 1.10(3)\\ 
\\
Br2  &  & & &    \\
$x$   &  0.244(3)     & 0.281(1)  & 0.287(4)  & 0 \\
$y$  &  0.744(3)     &  0        & 0.218(4)  & 0\\
$z$   &  0.027(1)    &  0.216(1) & -0.001(4)  &  0.2495(7) \\
$B_{\rm{iso}}$ ({\AA}$^2$)   &
         0.95(5)*  &   0.84(5) & 2.65(6) & 1.10(3)\\ 
\\
Br3   &  & & &    \\
$x$  &  0.536(3)   & 0.225(1)   & 0.729(3)   & - \\
$y$  &  0.037(3)   &  0         & 0.233(4)   & - \\
$z$  &  0.246(2)   &  0.716(1)  & 1.001(4)   &  - \\
$B_{\rm{iso}}$ ({\AA}$^2$)    & 
     1.6(1)*    &  1.5(1)    & 2.65(6)    & -\\ 
\\
$R_{\rm P}$ (\%)  & 11.6  & 11.8  & 13.3 & 12.3\\
$R_{\rm F}$ (\%)  & 9.2   & 10.7  & 15.4 & 10.8\\
$\chi^2$          & 20.5  & 20.0  & 24.4 & 22.6\\
\end{tabular}
\end{ruledtabular}
\end{table}

To discern a likely crystal structure out of the candidate symmetry groups investigated via Reitveld refinement, optical measurements were taken into consideration. Based upon the Raman active modes identified by Cohen \etal\cite{cohendiverging} and the results of our infrared spectroscopy investigation presented below we conclude that \mono{} best explains the experimental vibrational spectra.

\begin{table}[]
\caption{Inter atomic distances for \cabb{} as obtained from a Rietveld profile refinement of the PXRD pattern collected at 20~K (see FIG.~\ref{fig:T20Krefine}, wavelength Cu$K_{{\alpha}1}$) assuming monoclinic (space group no. 14) \mono{} and triclinic symmetry (space group no. 2, \triclinic). Distances have been calculated using the atom fractional coordinates and lattice parameters listed in TABLE~\ref{Table2}.}
\label{Table3}
\begin{ruledtabular}
\begin{tabular}{c c  c  }
atoms  &  distances (\AA) (\mono )  &   distances (\AA) (\triclinic) \\
\hline
Cs~-~Br1  & 	3.85(1)  &    3.927(54)\\
Cs~-~Br1  & 	3.94(9)  &    3.930(54)\\ 
Cs~-~Br1  & 	3.95(9)  & 	3.934(54)\\ 
Cs~-~Br1  & 	4.01(1)  & 	3.937(54)\\ 
Cs~-~Br2  & 	3.79(4)  & 	3.756(33)\\ 
Cs~-~Br2  & 	3.71(4)  &    3.761(33)\\
Cs~-~Br3  &     3.82(4)  & 	3.677(34)\\ 
Cs~-~Br3  &     3.90(4)  & 	3.715(340)\\ 
\\
Ag~-~Br1  & 	2.826(8) (2$\times$)  &    2.852(40) (2$\times$)\\
Ag~-~Br2  & 	2.78(3) (2$\times$)  &    2.731(24) (2$\times$)\\ 
Ag~-~Br3  & 	2.81(3) (2$\times$) & 	2.846(16) (2$\times$)\\
\\
Bi~-~Br1  & 	2.84(1) (2$\times$)  &    2.807(40) (2$\times$)\\
Bi~-~Br2  & 	2.83(3) (2$\times$)  &    2.863(24) (2$\times$)\\ 
Bi~-~Br3  & 	2.81(3) (2$\times$) & 	2.773(16) (2$\times$)\\
\end{tabular}
\end{ruledtabular}
\end{table}

\subsection{Dielectric Permittivity}
The temperature dependence of the low frequency limit complex dielectric function: 
\begin{equation}{\label{eq:Dielectric}}
\Tilde{\epsilon} = \epsilon_1 + i \epsilon_2
\end{equation}
where $\epsilon_1$ and $\epsilon_2$ are its real and imaginary parts respectively,  was extracted from the dielectric capacitance and loss measured between 5~K and 300~K  at a frequency of 50 Hz using Eqn.\ref{eq:Cap} and Eqn.\ref{eq:Loss} :
\begin{equation}{\label{eq:Cap}}
\epsilon_1 = \frac{C}{\epsilon_o}\frac{t}{A}
\end{equation}
\begin{equation}{\label{eq:Loss}}
\epsilon_2 = \frac{1}{\epsilon_o \omega} \frac{1}{R} \frac{t}{A}.
\end{equation}
 The system was modeled as a capacitor in parallel with a resistor, where $\omega$ is the frequency, and $C$ and $R$ are respectively the measured frequency dependent capacitance and resistance. $A$ is the average area of the sample faces, $t$ is the distance between the faces and $\epsilon_o$ is the permittivity of free space.

Changes in the crystal structure of a material may affect its dielectric response. 
Neutron scattering measurements have shown that the cubic-tetragonal structural phase transition in \cabb{} occurs before the soft optic mode frequency becomes zero suggesting the phase transition is weakly first order.\cite{zhang2025phonon} The Lyddane, Sachs, Teller equation relates the dielectric properties to the phonon frequencies, predicting that a complete collapse of the soft mode to zero frequency would be accompanied by a divergence in the permittivity.\cite{scott1974soft} With the soft mode frequency remaining finite  an anomaly might be expected at the phase transition temperature.

\begin{figure}[]
\includegraphics[width=0.5\textwidth,keepaspectratio]{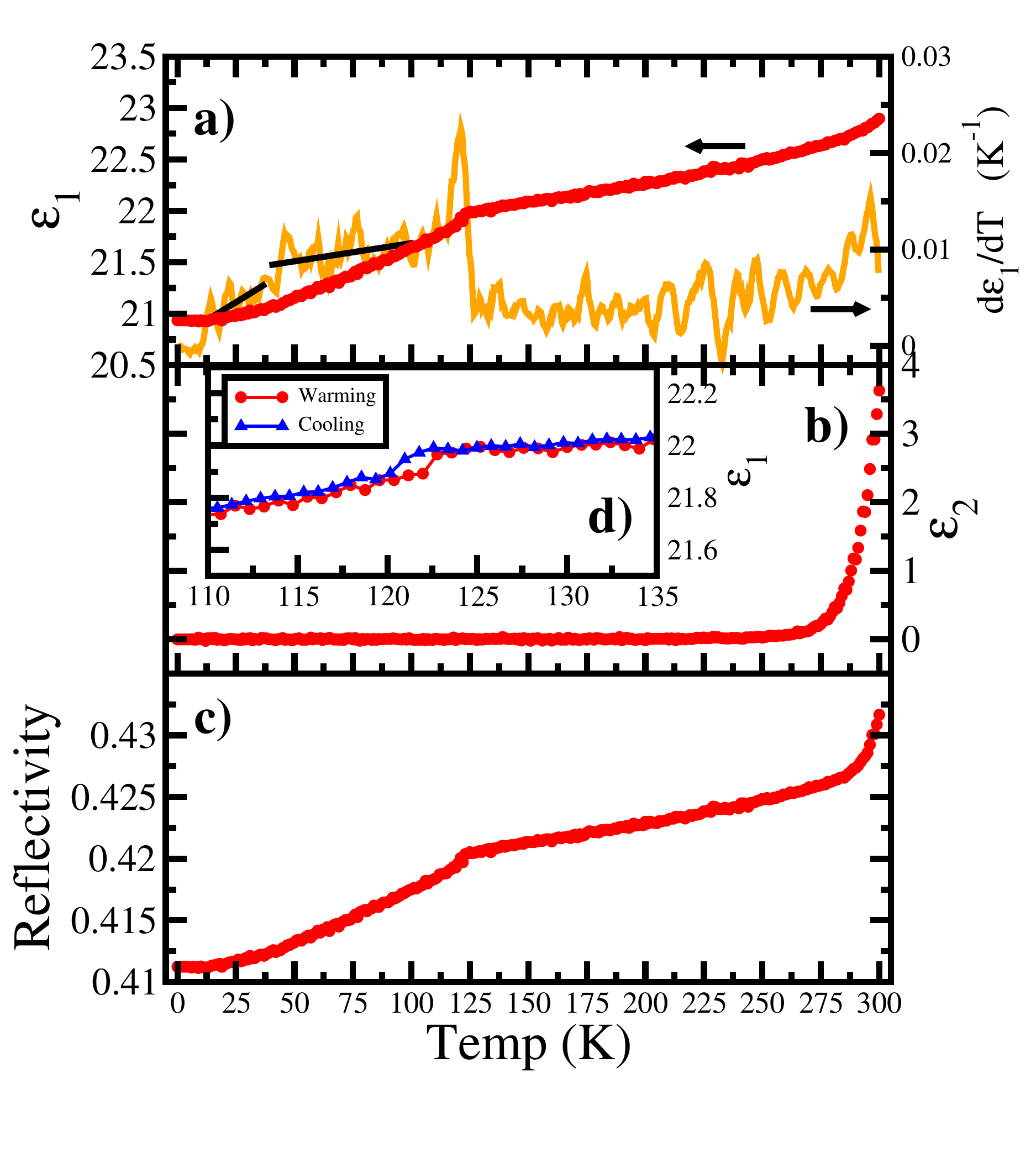}
\caption{\label{fig:50HzCap}(a) Temperature dependence of the real part of the dielectric function measured at 50 Hz and its derivative with arrows pointing to their respective axes. The black straight lines are fits to the derivative data above and below the low-$T$ phase transition. (b) Temperature dependence of the imaginary part of the dielectric function measured at 50~Hz. The dielectric function was subsequently used to determine the near DC (50 Hz) reflectivity of \cabb{} (c) to input in the Kramers-Kronig analysis. (d) Detail of  $\epsilon_1$ measured at 50 Hz near the 120 K transition exhibiting a slight hysteresis between cooling and warming given as blue triangles and red circles respectively.}
\end{figure}

As shown in FIG.\ref{fig:50HzCap} the 120~K structural phase transition is present as a clear change in the slope in the real part of the 50 Hz dielectric permittivity and as a peak in its derivative shown overlaid as the orange curve. The slight hysteresis between warming and cooling (see FIG.\ref{fig:50HzCap}(d)) agrees with the assignment of a weak first order phase transition. Whereas anomalous behavior near 40~K  is not readily visible in the real part of the dielectric response, a change of the slope is clearly observed in the derivative with respect to temperature. 

As displayed in  FIG.\ref{fig:50HzCap}(c), the 50 Hz
dielectric function was subsequently used to determine the normal incidence reflectivity {\bf{R}} of \cabb{} via Eqn.\ref{eq:ComplexReflectance} by taking the magnitude squared of both sides. The reflectivity {\bf{R}} was then used to guide the low frequency extrapolation of the Kramers-Kronig (KK) analysis on \cabb{} which will be described in detail below.

\begin{equation}{\label{eq:ComplexReflectance}}
\sqrt{{\bf{R}}}e^{i\theta}=\frac{\sqrt{\Tilde{\epsilon}}-1}{\sqrt{\Tilde{\epsilon}}+1}
\end{equation}

\subsection{Infrared Reflectivity}

The temperature dependence of the IR reflectivity of an annealed single crystal of \cabb{} was measured with the goal of  analyzing the IR active modes in \cabb. The reflectivity was measured using near normal incidence conditions with the light propagating perpendicular to the (111) crystal face in the \cubic{} cubic phase (see FIG.\ref{fig:CrystalPics}), which transforms to (101) in the tetragonal \tetra{} phase and (100) in the proposed monoclinic \mono{} phase. This should allow the observation of all IR active phonons, as shown in TABLE \ref{Table:SymmetryModes}, however, while a mode may be group theoretically predicted, actual oscillator strengths may be too weak to experimentally resolve.

\begin{figure}[]
\includegraphics[width=0.5\textwidth,keepaspectratio]{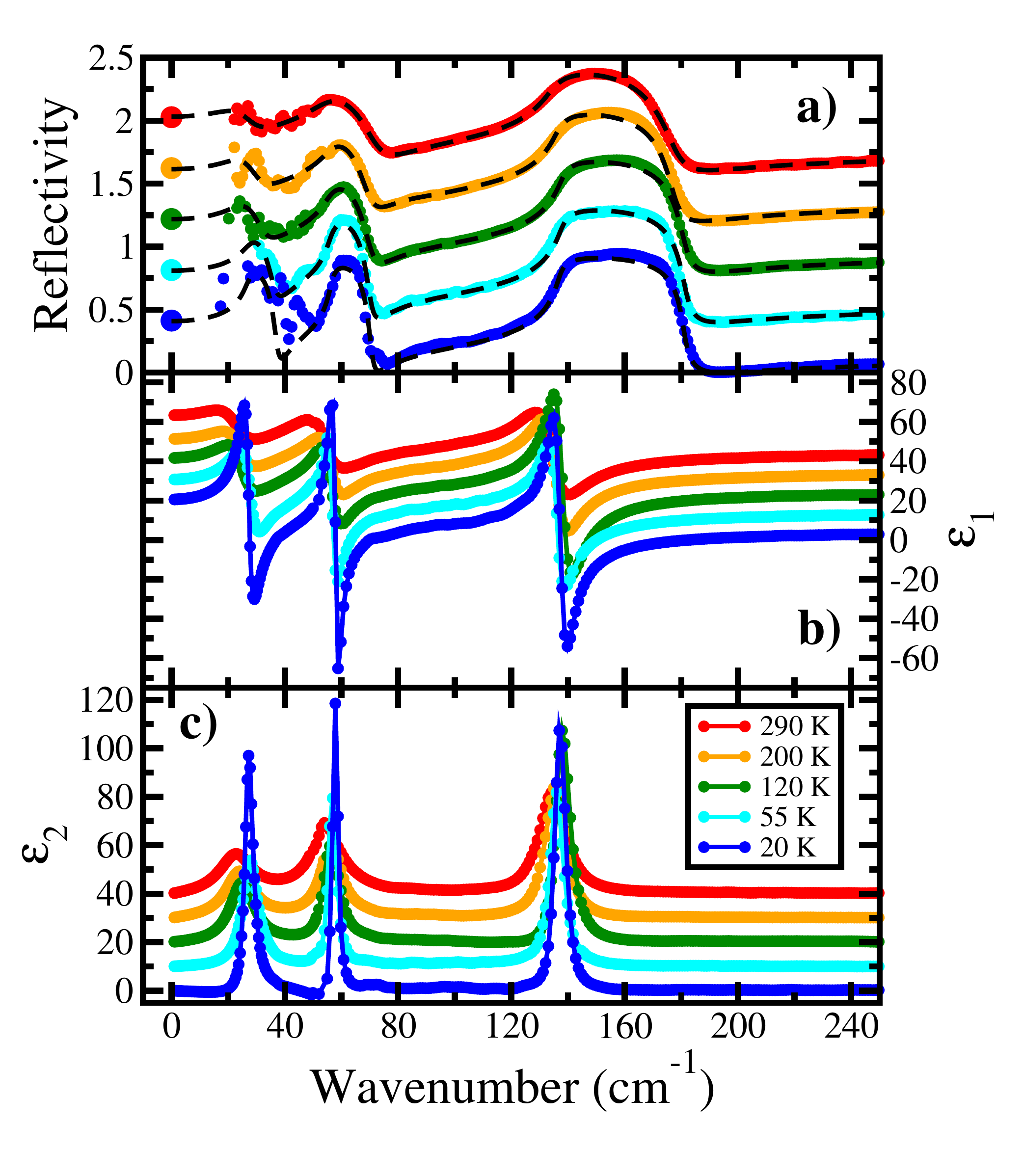}
\caption{\label{fig:KK_example} (a) Reflectivity spectra of \cabb{} at select temperatures. Each subsequent curve is shifted by 0.4 for clarity.
The dashed lines overlapping the experimental reflectivity curves are fits to the dielectric model Eqn.~(\ref{eq:Lorentz}). The fit in the region 0-50~\wav{} aided by the near DC (50 Hz)  reflectivity obtained from the capacitance-loss measurements and shown as large solid circles was used to extrapolate the spectra for KK analyses. (b) and (c) Corresponding real and imaginary parts, respectively, of the KK calculated dielectric function, similarly consecutively shifted by 10 for clarity.}
\end{figure}

Temperature dependent spectra were obtained in a range from 20~\wav{} to 11000~\wav. FIG.\ref{fig:KK_example}(a) shows typical reflectivity spectra in the far infrared region of interest. As the low-intensity vibrational modes appear only very weakly in the reflectivity spectra, we performed a KK transformation to obtain clearly identifiable resonance features in the dielectric function.
The magnitude of the reflectivity ${\bf{R}}(\omega)$ and the phase $\theta(\omega)$ are related via Eqn.~(\ref{eq:Kramers}). 

\begin{equation}{\label{eq:Kramers}}
\theta(\omega) = -\frac{\omega}{\pi}\int_0^\infty d\Omega \frac{ \ln({\bf{R}}(\Omega) / {\bf{R}}(\omega)) }{\Omega^2-\omega^2}
\end{equation}
From the reflectivity and phase the complex dielectric function (Eqn. \ref{eq:ComplexReflectance}) was determined. It may be noted that the KK integration necessitates a suitable choice of high and low frequency extrapolations. 

 Reflectivity data obtained from ellipsometry data measured by J{\"o}bsis \etal.~\cite{UV_extrap} expanded the reflectivity to a range from 20~\wav{} to 33000~\wav. For more details see  Ref.~\cite{Supplement}. To further extend the range to X-rays a method described by Tanner\cite{Xray_extrap} was employed. Tanner's method assumes an atomic number density weighted contribution using the elemental scattering factors to the dielectric function. The complex part of the scattering factors for the elements were obtained from Henke \etal~\cite{XrayFactors} and the KK relations were applied to obtain the full scattering factor from which the X-ray reflectivity from ~80 000~\wav{} to $\omega_{max} =$ 250 000 000~\wav{} was obtained. The gap between UV to X-ray energies was modeled by using a cubic spline. FIG. S2 of the  Supplemental Material.\cite{Supplement} provides an example of these extrapolations.

To extrapolate the high frequency limit to infinity the $R \propto \omega^{-4}$ relation was used. This corrects the phase at $\omega$ due to contributions from $\omega_{max}$ to $\infty$, details of this approach can be found in standard optical texts such as Appendix G of Wooten's Optical Properties of Solids\cite{WootenOptical}.

The low frequency extrapolation was guided by the number of expected IR active modes. In the cubic phase four such modes are expected. Although we observe these modes, the mode at $\sim$85~\wav{} is very weak, and the lowest frequency mode occurs at $\sim$25~\wav{} in a region  (20-50~\wav) of very low signal-to-noise ratio for the spectrometer used. The reflectivity spectra were fitted to the Lorentz dielectric model: 
\begin{equation}{\label{eq:Lorentz}}
\Tilde{\epsilon} = \epsilon_{\infty} + \sum_j^N \frac{f_{j}}{\omega_{oj}^2-\omega^2-i \gamma_{j} \omega}
\end{equation}
where $\epsilon_{\infty}$ accounts for high frequency absorption processes beyond the spectral range, and the $\omega_{oj}$ are the frequencies of the oscillator contributions with corresponding scattering rates $\gamma_{j}$ and oscillator strengths $f_{j}$. The resulting fits to the reflectance are given for select temperatures in FIG. \ref{fig:KK_example}(a).  To keep the fit more physical in the low frequency region without measured data the model was forced to fit the "zero" frequency reflectivity, shown as large solid circles in FIG. \ref{fig:KK_example}(a), obtained via capacitance-loss measurements (50 Hz $\approx 2 \times 10^{-9}$~\wav). The fit in the range of 0-50~\wav{} was then used as the low frequency extension for KK analysis. Choosing a suitable frequency range, the dielectric function was then extracted using KK analysis. The real and imaginary parts of $\Tilde{\epsilon}$ are shown for select temperatures in FIG. \ref{fig:KK_example}(b) and (c) respectively.

\subsubsection{IR Vibrational Modes}
In the cubic phase above 120~K with space group \cubic{} group theory predicts four IR active vibrational modes all of $T_{1u}$ symmetry.
Experimentally, three IR active phonon modes are clearly visible in our spectra with frequencies of 25, 55, and $\sim$135~\wav{} (see FIG. \ref{fig:KK_example}) at room temperature. A fourth phonon mode (see below) occurs at $\sim$85~\wav{} at room temperature, however, has a very low intensity and is just barely resolved in the reflectivity measurements, but is better observed in the dielectric function (see FIG. \ref{Fig.pdf2Split}(a)). The 25~\wav{} mode occurs in a region of low signal-to-noise ratio at low wavenumbers. Its high frequency tail is just observed. 

Zelewski \textit{et al.}\cite{DFT_Cubic_tetra} determined the phonon dispersion relations of \cabb{} in the cubic and the tetragonal phase by density functional theory (DFT) calculations. Through analyzing the DFT calculated band diagrams at the $\Gamma$-point, the frequencies of each mode were matched to the known Raman frequencies assuring the correct degeneracy leaving only the silent modes and IR active modes unassigned. The frequencies of the IR active modes were determined to be 30, 57, 99, and 149 \wav{} agreeing quite well with our experimentally determined values.  
Rodrigues \etal{} used DFT calculations to determine the infrared reflectance and absorbance spectra of the the isotypic compound Cs$_2$AgSbBr$_6$~\cite{DFT_CASC}. Clearly visible IR modes are found at 25~cm$^{-1}$, 80~cm$^{-1}$, and 212~cm$^{-1}$. Like in \cabb{}, the fourth mode is very weak in intensity and located at $\sim$133~cm$^{-1}$. As expected for vibrational modes the replacement of the heavier Bi by the lighter element Sb shifts the energies of the three highest IR modes to higher wavenumber whereas the 25~cm$^{-1}$ mode remains at the same energy. This is likely because the low wavenumber IR phonon is a collective mode involving the entire basis. 
Although all modes in the cubic structure possess T$_{1u}$ symmetry and may therefore involve the full basis, as the translational irreducible representation of each atomic site contributes, atomic displacement calculations based on a short-range force-field model for
 the double perovskite Ba$_2$MgWO$_6$ show that the low-wavenumber mode has a larger contribution from the A-site cation (Cs in the case of \cabb) relative to the higher-wavenumber modes.\cite{short-range}.

All of the IR active modes in the (m$\overline3$m) cubic structure are T$_{1u}$ triply degenerate modes which are expected to split into single and doubly degenerate A$_u$ and E$_u$ modes, respectively, in the (4/m) tetragonal phase (see TABLE \ref{Table:SymmetryModes}). Experimentally,  below 120~K, subtle evidence for splitting of the modes appears in the reflectivity data. FIG.\ref{fig:70cm-1_optical}(a) shows a detail of the temperature dependence of the reflectance in the vicinity of the 55 \wav{} phonon. A small feature starts to develop on the high frequency edge  just above 70 \wav\ at temperatures below 120 K, as shown for 80 K in FIG.\ref{fig:70cm-1_optical}(b). The KK extracted dielectric function 
exhibits greater sensitivity to the vibrational modes, specifically the complex part, which displays more local behaviour of the modes, compared to the more global effects of the reflectivity. FIG.\ref{fig:70cm-1_optical}(c) and (d) show $\epsilon_1$ and $\epsilon_2$ in the vicinity of this mode. FIG.\ref{fig:70cm-1_optical}(e) shows $\epsilon_2$ in the spectral  region corresponding to FIG.\ref{fig:70cm-1_optical}(b), confirming that this feature is more prominent in the imaginary dielectric function.  FIG.\ref{fig:70cm-1_optical}(d) shows that a distinct peak is formed at 69 \wav{} at the lowest temperatures.

Simultaneous fitting of the Kramers–Kronig-derived dielectric function to the real and imaginary components of Eqn. \ref{eq:Lorentz}, followed by analysis of the peak positions, reveals clear splitting of the three highest-frequency experimentally resolved T$_{1u}$ modes at 120 K. See
FIG.\ref{Fig.pdf2Split}(a),(c),(d) and (e).
Note that, as shown in FIG.\ref{Fig.pdf2Split}(b), for the 135 \wav{} mode the splitting manifests as an asymmetry to the lineshape below 120 K. In the cubic phase a single Lorentzian is able to capture the lineshape. Further splittings were observed at the 40 K phase transition. 

\subsubsection{40 K Phase Transition}

Predicated upon our PXRD results a symmetry lowering below $\sim$40~K is clearly supported. 
The monoclinic and triclinic space groups considered in our discussion of the PXRD results below 40 K would all result in an increase in the number of IR active modes. For the monoclinic space group \mono{} group theory predicts 17$A_u$ and 16$B_u$ modes whereas for the monoclinic space group $I\rm{1 2}/m\rm{1}$ 15 IR active modes (5$A_u$ + 10$B_u$) are expected.
For the triclinic space group $I\bar{\rm 1}$ group theory also predicts 15 modes of $A_u$ symmetry \cite{Bilbao-Generic-1,Bilbao-Generic-2,Bilbao-Generic-3,Bilbao-Generic-SAM}.

Strong support for a monoclinic \mono{} symmetry of the low temperature crystal structure of \cabb{} can be derived from the distinct increase of the number of Raman modes observed by Cohen \textit{et al.} whose measurements suggest a total of 24 Raman active modes\cite{cohendiverging}.
Group theory predicts for the monoclinic I2/m structure only 12 Raman active modes (6A$_g$+6B$_g$). Similarly the triclinic \triclinic{} structure only gives 12 Raman active modes (12$A_g$) 
whereas for the space group \mono{} 24 Raman modes are expected (12$A_g$ + 12$B_g$), in quantitative agreement with Cohen \textit{et al's.} findings.

Note that the number of modes given above for the \monotwo{}, \triclinic, and \mono{} space groups correspond to the unit cells of TABLE \ref{Table2} which agree with the number of modes of known oxide and halide double perovskites which form in these space groups such as Sr$_2$ScSbO$_6$ with an \monotwo{} to \mono{} transition and Rb$_2$KScF$_6$ with an \tetra{} to \mono{} transition\cite{Faik2012,Additional_Structural_Transition}. A 40 atom (20 atom in the primitive cell) I$\bar{1}$ or I2/m cell could also give 24 Raman active modes with possible structures for these groups given by the Wyckoff sites in TABLE \ref{tab:Wyckoff}. This would, however, require a substantial increase in the number of free parameters compared to the \monotwo{} and \triclinic{} refinements given in TABLE \ref{Table2} and therefore, although we cannot explicitly eliminate them, we are inclined to suggest they are not as favored as \mono.

\begin{figure*}
\includegraphics[keepaspectratio,width=1\textwidth]{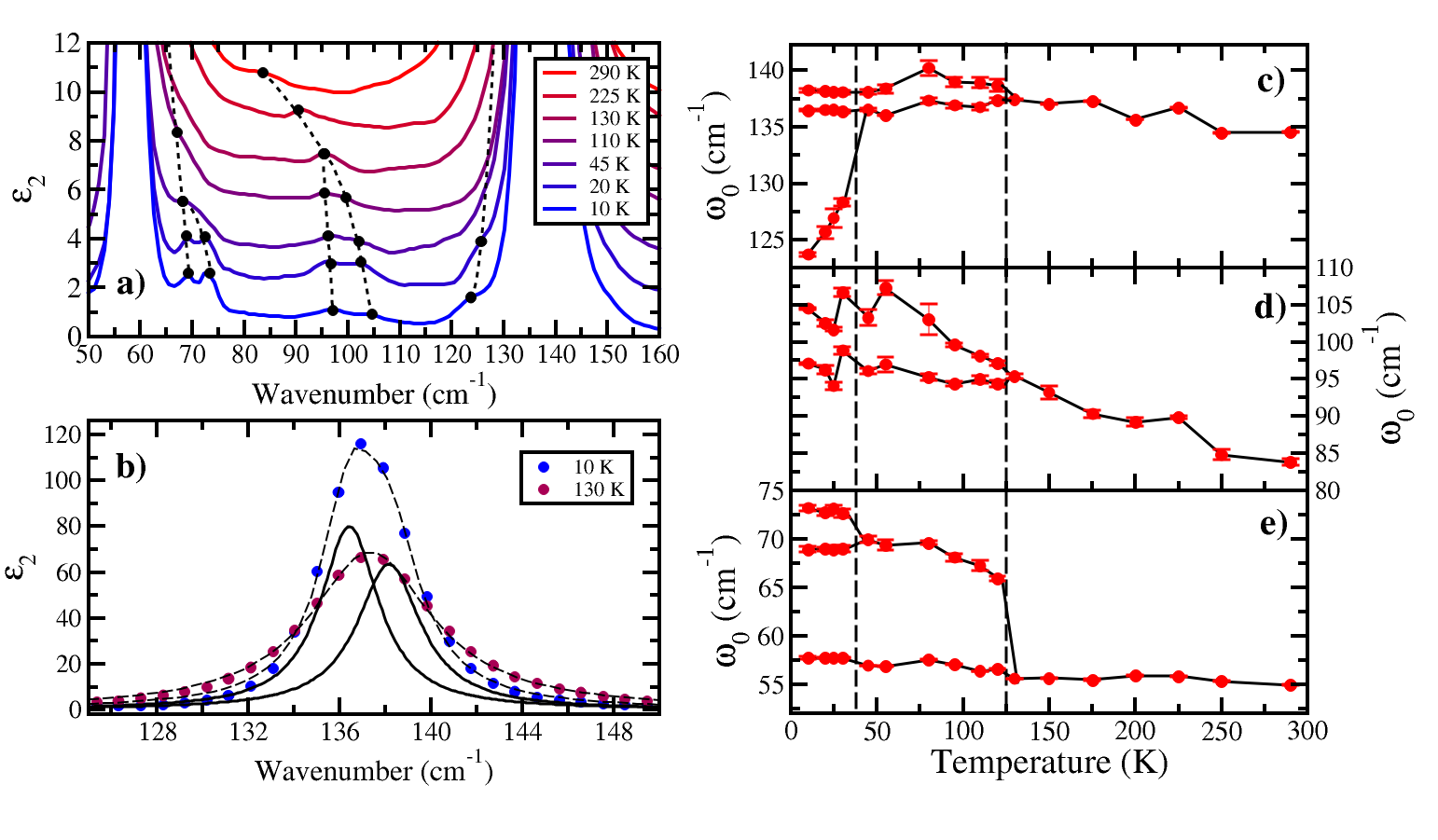}
\caption{\label{Fig.pdf2Split}}
a) The KK extracted complex part of the dielectric function at select temperatures above and below the 38 K and 120 K phase transitions. Black dots give the peak locations of fitting the phonon splitting of the three resolved T$_{1u}$ to Eqn.\ref{eq:Lorentz}. Black dotted lines are to guide the eye to phonon splitting. b) Example of Eqn.\ref{eq:Lorentz} Lorentz fit of the ~135 \wav{} phonon at 130 K in the cubic structure and 10 K in the monoclinic structure. Dashed lines indicate the fit, solid lines show the fit components of the 10 K spectra. The splitting on the low-wavenumber edge is not evident on this scale due to low intensity seen in a). c),d),e) Temperature evolution of the center frequency of the 135, 85, and 55 \wav{} phonons and splittings, obtained from Eqn.\ref{eq:Lorentz} Lorentz fit. Vertical dashed lines indicate the 38 K and 120 K phase transitions.
\end{figure*}

\begin{figure}[]
\includegraphics[width=0.5\textwidth,keepaspectratio]{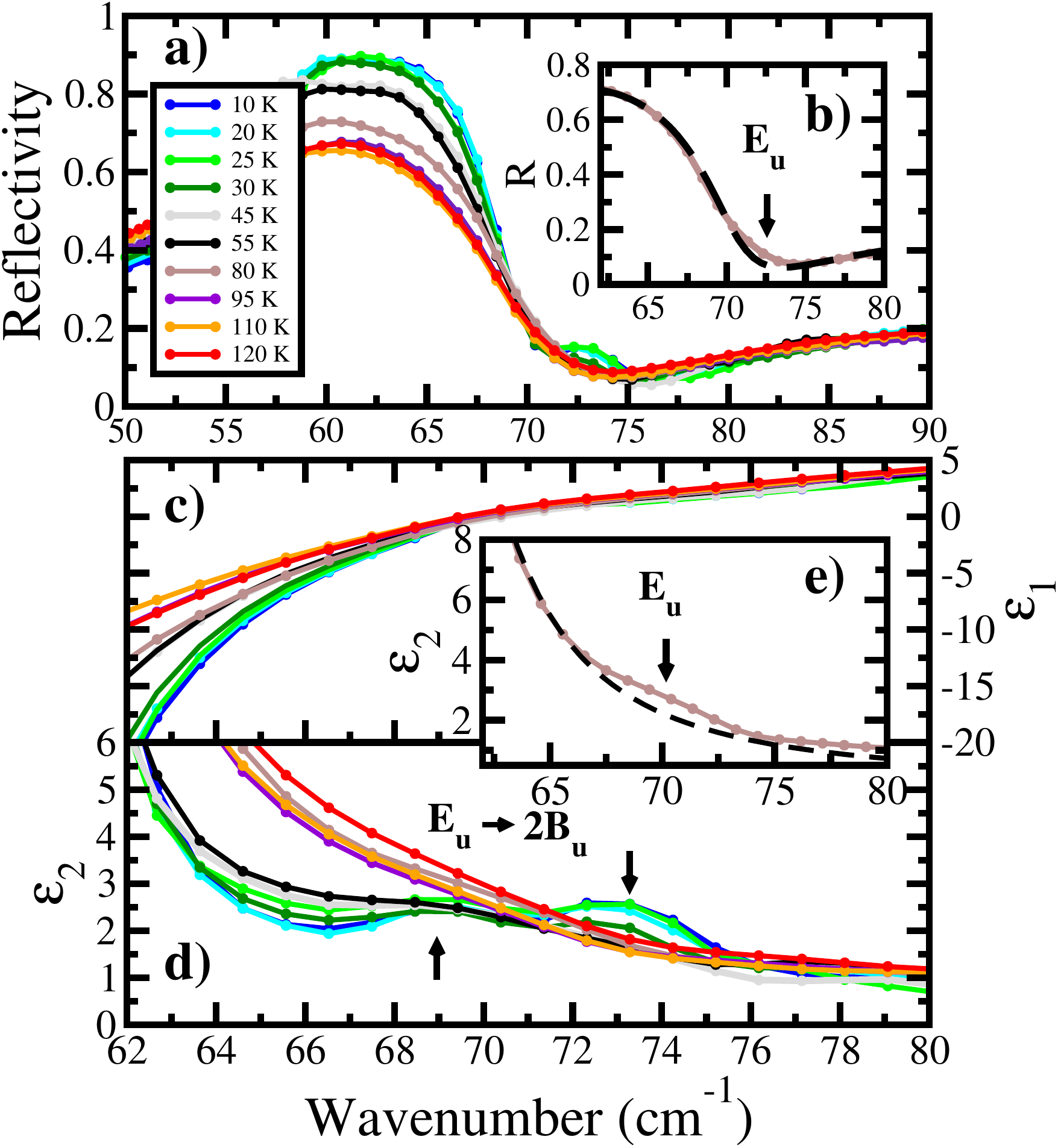}
\caption{\label{fig:70cm-1_optical} Details of the reflectivity, (a), and the KK extracted dielectric function, (c) and (d), below 120~K around the 69~\wav{} $E_u$ phonon which splits from the $T_{1u}$ mode ($T_{1u}$ $\rightarrow$ $A_u$ +$E_u$). The Insets show, as an example for 80~K, the 69 \wav{} $E_u$ mode detected in the tetragonal phase as a deviation from the Lorentz dielectric model fit to the four phonon modes of the cubic phase (dashed black lines) for both the reflectivity (b) and complex part of the dielectric function (e). It is just barely discernible from the large tail of the 55~\wav{} phonon. Below the 40~K transition another mode appears near 73~\wav{}, believed to be the $E_u$ mode splitting into two $B_u$ modes under monoclinic symmetry.
}
\end{figure}

To identify the IR modes expected at low temperature, we again consider TABLE \ref{Table:SymmetryModes}, where the (2/m) columns correspond to \mono. Reading across the row starting with the IR active  T$_{1u}$ in the cubic structure (m$\overline3$m) we see that each T$_{1u}$ mode splits into an A$_{u}$ mode and an E$_{u}$ mode in the tetragonal structure (4/m). The doubly degenerate E$_{u}$ mode further splits in the monoclinic structure (2/m) such that at the lowest temperatures  one A$_{u}$ mode and two B$_{u}$ modes are expected. The resolution of a typical low temperature far infrared spectroscopy measurement limits the ability to observe small splittings and weaker modes. In FIG. \ref{fig:KK_example}(a) in the 20 K reflectivity spectrum there is, however, experimental evidence of an additional mode at $\approx 73$ \wav{} for temperatures below 40 K which can be seen more clearly in FIG: \ref{fig:70cm-1_optical}(a) at temperatures below 40 K. This mode is assigned to the further splitting of the $\approx 69$ \wav{} E$_u$ mode which split from the 55 \wav{} T$_{1u}$ mode at 120 K. This splitting is more distinguishable in the KK extracted dielectric function shown in FIG.\ref{Fig.pdf2Split} which shows E$_u$ $\rightarrow$   2B$_u$ splitting of the 55 \wav{} and 135 \wav{} modes below 40 K. Further splitting of the weak 85 \wav{} mode below 40 K could not be resolved.

\begin{table}[b]
\caption{\label{tab:Wyckoff} Wyckoff sites of \cabb{} in \cubic{} (cubic), \tetra{} (tetragonal) and \mono{} (monoclinic) space groups. Possible Wyckoff positions for an \monotwo{} and \triclinic{} structure which would give 24 Raman active modes are also listed. The Wyckoff positions in braces are the Wyckoff symbols of P$2_1$/n in the standard setting of group 14 (P$2_1$/c)}.
\begin{ruledtabular}
\begin{tabular}{cccccc}
 & \cubic & \tetra & P2$_1$/n \{P2$_1$/c\} & I$2/m$ & \triclinic\\
\hline
Cs & 8c & 4d  & 4e \{4e\} & 8j & 4i (2x)\\
\hline
Ag & 4b & 2b & 2b \{2d\} & 2b,2d  & 2d,2g\\
\hline
Bi & 4a & 2a & 2a \{2a\} & 2a,2c  & 2b,2e \\
\hline
Br & 24e  & 8h,4e & 4e \{4e\} (3x) & 4i (4x),4g,4h   & 4i (6x) \\
\end{tabular}
\end{ruledtabular}
\end{table}

\begin{table*}
\caption{ Vibrational mode splitting in the three phases of \cabb{} Fm$\bar{3}$m I4/m and P2$_1$/c given by grouped rows. Raman active, IR active, acoustic, and silent modes are italic text, bold text, marked with an asterisk, and marked with a dagger respectively. Blank rows for point groups m$\bar{3}$m and 4/m indicate additional modes in the 2/m symmetry from a primitive cell doubling from I4/m to P2$_1$/c. Listed beside each mode is the corresponding linear or quadratic basis function if applicable, and the Wyckoff sites that may contribute to the mode.}
\label{Table:SymmetryModes}
\begin{ruledtabular}
\begin{tabular}{ccc|ccc|ccc}
\multicolumn{3}{c|}{m$\bar{3}$m}&\multicolumn{3}{c|}{4/m} &\multicolumn{3}{c}{2/m} \\
 Mode Irr Rep & Basis Fnc &Wyckoff &&&&&&\\ \hline
$A_{1g}$ & x$^2$+y$^2$+z$^2$ & 24e &
$A_{g}$ & x$^2$+y$^2$,z$^2$ & 8h(2x),4e  & 
$A_{g}$ & x$^2$,y$^2$,z$^2$, xy & 4e(3x) 
\\ 
\hline
$E_{g}$ & (2z$^2$-x$^2$-y$^2$,x$^2$-y$^2$)  & 24e &
$A_{g}$ & x$^2$+y$^2$,z$^2$ & 8h(2x),4e & 
$A_{g}$ & x$^2$,y$^2$,z$^2$, xy & 4e(3x)  
\\
&&&$B_{g}$ & xy,x$^2$-y$^2$ & 8h(2x),4d &
$A_{g}$ & x$^2$,y$^2$,z$^2$, xy & 4e(3x)
\\
\hline
2$T_{2g}$ & (xy,xz,yz) & 8c,24e & 
2$B_{g}$ & xy,x$^2$-y$^2$ & 8h(2x),4d &
2$A_{g}$ & x$^2$,y$^2$,z$^2$,xy & 4e(3x) 
\\
&&&2$E_{g}$ & (xz,yz) & 8h,4e,4d &
4$B_{g}$ & xz, yz & 4e(3x)
\\
\hline
4\textbf{T$_{1u}$} & (x,y,z) & 8c,4b,4a,24e(2x) &
4\textbf{A$_u$} & z & 8h,4e,4d,2b,2a &
4\textbf{A$_u$} & z & 4e(3x),2d(3x),2a(3x)
\\
&&&4\textbf{E$_u$} & (x,y) & 8h(2x),4e,4d,2b,2a &
8\textbf{B$_u$} & x,y & 4e(3x),2d(3x), 2a(3x)
\\
\hline
$T_{1u}*$ & (x,y,z) & 8c,4b,4a,24e(2x) &
A$_u*$ & z & 8h,4e,4d,2b,2a &
A$_u*$ & z & 4e(3x),2d(3x),2a(3x)
\\
&&&E$_u*$ & (x,y) & 8h(2x),4e,4d,2b,2a &
2B$_u*$ & x, y & 4e(3x),2d(3x),2a(3x)
\\
\hline
$T_{1g}$ $\dagger$ & & 24e &
$A_{g}$ & x$^2$+y$^2$,z$^2$ & 8h(2x),4e  & 
$A_{g}$ & x$^2$, y$^2$,z$^2$,xy & 4e(3x)  
\\
&&&$E_{g}$ & (xz,yz) & 8h,4e,4d &
2$B_{g}$ & xz,yz & 4e(3x)
\\
\hline
$T_{2u}$ $\dagger$ &  & 24e &
B$_u$ $\dagger$ &  & 8h &
\textbf{A$_u$} & z & 4e(3x), 2d(3x), 2a(3x)
\\
&&&\textbf{E$_u$} & (x,y) & 8h(2x),4e,4d,2b,2a &
2\textbf{B$_u$} & x,y & 4e(3x),2d(3x),2a(3x)
\\
\hline
&&&&&&6$A_{g}$ & x$^2$, y$^2$, z$^2$, xy & 4e(3x) 
\\
\hline
&&&&&&6$B_{g}$ & xz,yz & 4e(3x)
\\
\hline
&&&&&&12\textbf{A$_u$} & z & 4e(3x),2d(3x),2a(3x)
\\
\hline
&&&&&&6\textbf{B$_u$} & x,y & 4e(3x),2d(3x),2a(3x)
\\
\end{tabular}
\end{ruledtabular}
\end{table*}

\section{\label{sec:Conclusion}Conclusion}
In summary, we have investigated thermal, structural, dielectric and infrared optical properties of the bromide double perovskite \cabb{}. 
Our low-temperature x-ray investigations and the IR reflectivity spectra provide compelling evidence that 
apart from the thoroughly investigated
cubic to tetragonal structural phase transition at $\sim$120~K in \cabb{},  another low-temperature structural phase transition takes place at $\sim$40~K which leads to a symmetry lowering of the crystal structure. Our x-ray powder diffraction data
point to a monoclinic or triclinic crystal structure of the low temperature phase with a slight preference - judging from the somewhat reduced reliability factors - to a triclinic $I\bar{\rm 1}$ description.
Our findings regarding the temperature dependence of the cell volume substantiate the report by Keshavarz \etal~\cite{CenterFreq_Hardening} who observed a  thermal expansion anomaly  at $\sim$~40~K.
Our results concur in part with the findings of He \textit{et al.} in that the transition involves a more complex network of the Bi-Br octahedrals as the structure is altered below the 40 K transition, however we suggest based on our infrared and literature Raman spectra that this structural transition is the symmetry lowering from I4/m to \mono{} rather than to a several hundred atom unit cell as predicted by their DFT modeling.

The appearance of additional IR active modes is consistent with a lower symmetry for the low-temperature crystal structure. All monoclinic or triclinic descriptions induce a growth of the number of IR active modes. The monoclinic and the triclinic description refinements to the PXRD patterns of the low-temperature phase are almost equally good. The somewhat reduced reliability factors favor the triclinic description. The symmetry of perovskite structures is often over or underestimated\cite{PerovskitesGroupTheory}, however, judging from the significantly increased number of Raman modes found by Cohen \textit{et al.}\cite{cohendiverging} 
which is in quantitative agreement with that expected for the monoclinic space group \mono{} we suggest that \mono{} provides the more consistent description of the crystal structure of \cabb{} below $\sim40~K$.

\section{\label{sec:Acknowledgments}Acknowledgements}

\begin{acknowledgments}
We acknowledge the support of Brock University and the Natural Sciences and Engineering Research Council of Canada (NSERC), [funding reference number RGPIN-2022-05214].
 Scholarship funding was provided to
C.T. from the Natural Sciences and Engineering Research Council of Canada (NSERC CGS-D) and J.D. from the Government of Ontario (OGS).
\end{acknowledgments}

\clearpage


\begin{thebibliography}{58}%
\makeatletter
\providecommand \@ifxundefined [1]{%
 \@ifx{#1\undefined}
}%
\providecommand \@ifnum [1]{%
 \ifnum #1\expandafter \@firstoftwo
 \else \expandafter \@secondoftwo
 \fi
}%
\providecommand \@ifx [1]{%
 \ifx #1\expandafter \@firstoftwo
 \else \expandafter \@secondoftwo
 \fi
}%
\providecommand \natexlab [1]{#1}%
\providecommand \enquote  [1]{``#1''}%
\providecommand \bibnamefont  [1]{#1}%
\providecommand \bibfnamefont [1]{#1}%
\providecommand \citenamefont [1]{#1}%
\providecommand \href@noop [0]{\@secondoftwo}%
\providecommand \href [0]{\begingroup \@sanitize@url \@href}%
\providecommand \@href[1]{\@@startlink{#1}\@@href}%
\providecommand \@@href[1]{\endgroup#1\@@endlink}%
\providecommand \@sanitize@url [0]{\catcode `\\12\catcode `\$12\catcode `\&12\catcode `\#12\catcode `\^12\catcode `\_12\catcode `\%12\relax}%
\providecommand \@@startlink[1]{}%
\providecommand \@@endlink[0]{}%
\providecommand \url  [0]{\begingroup\@sanitize@url \@url }%
\providecommand \@url [1]{\endgroup\@href {#1}{\urlprefix }}%
\providecommand \urlprefix  [0]{URL }%
\providecommand \Eprint [0]{\href }%
\providecommand \doibase [0]{https://doi.org/}%
\providecommand \selectlanguage [0]{\@gobble}%
\providecommand \bibinfo  [0]{\@secondoftwo}%
\providecommand \bibfield  [0]{\@secondoftwo}%
\providecommand \translation [1]{[#1]}%
\providecommand \BibitemOpen [0]{}%
\providecommand \bibitemStop [0]{}%
\providecommand \bibitemNoStop [0]{.\EOS\space}%
\providecommand \EOS [0]{\spacefactor3000\relax}%
\providecommand \BibitemShut  [1]{\csname bibitem#1\endcsname}%
\let\auto@bib@innerbib\@empty
\bibitem [{\citenamefont {Fu}\ \emph {et~al.}(2019)\citenamefont {Fu}, \citenamefont {Ho-Baillie}, \citenamefont {Mulmudi},\ and\ \citenamefont {Trang}}]{SolarBook}%
  \BibitemOpen
  \bibfield  {author} {\bibinfo {author} {\bibfnamefont {K.}~\bibnamefont {Fu}}, \bibinfo {author} {\bibfnamefont {A.}~\bibnamefont {Ho-Baillie}}, \bibinfo {author} {\bibfnamefont {H.~K.}\ \bibnamefont {Mulmudi}},\ and\ \bibinfo {author} {\bibfnamefont {P.~T.~T.}\ \bibnamefont {Trang}},\ }\href@noop {} {\emph {\bibinfo {title} {Perovskite Solar Cells}}}\ (\bibinfo  {publisher} {Apple Academic Press},\ \bibinfo {year} {2019})\BibitemShut {NoStop}%
\bibitem [{\citenamefont {Hu}\ \emph {et~al.}(2018)\citenamefont {Hu}, \citenamefont {Deng}, \citenamefont {Hu}, \citenamefont {Zhao}, \citenamefont {Zhang}, \citenamefont {Tan}, \citenamefont {Niu}, \citenamefont {Wu},\ and\ \citenamefont {Tang}}]{X-ray_1}%
  \BibitemOpen
  \bibfield  {author} {\bibinfo {author} {\bibfnamefont {Q.}~\bibnamefont {Hu}}, \bibinfo {author} {\bibfnamefont {Z.}~\bibnamefont {Deng}}, \bibinfo {author} {\bibfnamefont {M.}~\bibnamefont {Hu}}, \bibinfo {author} {\bibfnamefont {A.}~\bibnamefont {Zhao}}, \bibinfo {author} {\bibfnamefont {Y.}~\bibnamefont {Zhang}}, \bibinfo {author} {\bibfnamefont {Z.}~\bibnamefont {Tan}}, \bibinfo {author} {\bibfnamefont {G.}~\bibnamefont {Niu}}, \bibinfo {author} {\bibfnamefont {H.}~\bibnamefont {Wu}},\ and\ \bibinfo {author} {\bibfnamefont {J.}~\bibnamefont {Tang}},\ }\bibfield  {title} {\bibinfo {title} {X-ray scintillation in lead-free double perovskite crystals},\ }\href@noop {} {\bibfield  {journal} {\bibinfo  {journal} {Science China. Chemistry}\ }\textbf {\bibinfo {volume} {61.12}},\ \bibinfo {pages} {1581–} (\bibinfo {year} {2018})}\BibitemShut {NoStop}%
\bibitem [{\citenamefont {Ava}\ \emph {et~al.}(2019)\citenamefont {Ava}, \citenamefont {Mamun}, \citenamefont {Marsillac},\ and\ \citenamefont {Namkoong}}]{MAPbI3_ThermalStability}%
  \BibitemOpen
  \bibfield  {author} {\bibinfo {author} {\bibfnamefont {T.}~\bibnamefont {Ava}}, \bibinfo {author} {\bibfnamefont {A.}~\bibnamefont {Mamun}}, \bibinfo {author} {\bibfnamefont {S.}~\bibnamefont {Marsillac}},\ and\ \bibinfo {author} {\bibfnamefont {G.}~\bibnamefont {Namkoong}},\ }\bibfield  {title} {\bibinfo {title} {A review: Thermal stability of methylammonium lead halide based perovskite solar cells},\ }\href@noop {} {\bibfield  {journal} {\bibinfo  {journal} {Applied Sciences}\ }\textbf {\bibinfo {volume} {9.1}},\ \bibinfo {pages} {188} (\bibinfo {year} {2019})}\BibitemShut {NoStop}%
\bibitem [{\citenamefont {Slavney}\ \emph {et~al.}(2016)\citenamefont {Slavney}, \citenamefont {Hu}, \citenamefont {Lindenberg},\ and\ \citenamefont {Karunadasa}}]{DoublePerovskite_Defect_Tolerance}%
  \BibitemOpen
  \bibfield  {author} {\bibinfo {author} {\bibfnamefont {A.}~\bibnamefont {Slavney}}, \bibinfo {author} {\bibfnamefont {T.}~\bibnamefont {Hu}}, \bibinfo {author} {\bibfnamefont {A.}~\bibnamefont {Lindenberg}},\ and\ \bibinfo {author} {\bibfnamefont {H.}~\bibnamefont {Karunadasa}},\ }\bibfield  {title} {\bibinfo {title} {A bismuth-halide double perovskite with long carrier recombination lifetime for photovoltaic applications},\ }\href@noop {} {\bibfield  {journal} {\bibinfo  {journal} {Journal of the American Chemical Society}\ }\textbf {\bibinfo {volume} {138.7}},\ \bibinfo {pages} {2138} (\bibinfo {year} {2016})}\BibitemShut {NoStop}%
\bibitem [{\citenamefont {Wang}\ \emph {et~al.}(2018)\citenamefont {Wang}, \citenamefont {Zhang}, \citenamefont {Kan}, \citenamefont {Wang},\ and\ \citenamefont {Zhao}}]{CsPbI3_Stability}%
  \BibitemOpen
  \bibfield  {author} {\bibinfo {author} {\bibfnamefont {Y.}~\bibnamefont {Wang}}, \bibinfo {author} {\bibfnamefont {T.}~\bibnamefont {Zhang}}, \bibinfo {author} {\bibfnamefont {M.}~\bibnamefont {Kan}}, \bibinfo {author} {\bibfnamefont {Y.~L.~T.}\ \bibnamefont {Wang}},\ and\ \bibinfo {author} {\bibfnamefont {Y.}~\bibnamefont {Zhao}},\ }\bibfield  {title} {\bibinfo {title} {Efficient $\mathrm{\alpha}$-$\mathrm{ CsPbI_{3}}$ photovoltaics with surface terminated organic cations},\ }\href@noop {} {\bibfield  {journal} {\bibinfo  {journal} {Joule}\ }\textbf {\bibinfo {volume} {2.10}},\ \bibinfo {pages} {2065} (\bibinfo {year} {2018})}\BibitemShut {NoStop}%
\bibitem [{\citenamefont {Sun}\ \emph {et~al.}(2016)\citenamefont {Sun}, \citenamefont {Li}, \citenamefont {Yang},\ and\ \citenamefont {Li}}]{Pb_Replace}%
  \BibitemOpen
  \bibfield  {author} {\bibinfo {author} {\bibfnamefont {P.}~\bibnamefont {Sun}}, \bibinfo {author} {\bibfnamefont {Q.}~\bibnamefont {Li}}, \bibinfo {author} {\bibfnamefont {L.}~\bibnamefont {Yang}},\ and\ \bibinfo {author} {\bibfnamefont {Z.}~\bibnamefont {Li}},\ }\bibfield  {title} {\bibinfo {title} {Theoretical insights into a potential lead-free hybrid perovskite: Substituting $\mathrm{Pb^{2+}}$ with $\mathrm{Ge^{2+}}$},\ }\href@noop {} {\bibfield  {journal} {\bibinfo  {journal} {Nanoscale}\ }\textbf {\bibinfo {volume} {8.3}},\ \bibinfo {pages} {1503} (\bibinfo {year} {2016})}\BibitemShut {NoStop}%
\bibitem [{\citenamefont {McClure}\ \emph {et~al.}(2016)\citenamefont {McClure}, \citenamefont {Ball}, \citenamefont {Windl},\ and\ \citenamefont {Woodward}}]{DoublePerovskite_combinations}%
  \BibitemOpen
  \bibfield  {author} {\bibinfo {author} {\bibfnamefont {E.}~\bibnamefont {McClure}}, \bibinfo {author} {\bibfnamefont {M.}~\bibnamefont {Ball}}, \bibinfo {author} {\bibfnamefont {W.}~\bibnamefont {Windl}},\ and\ \bibinfo {author} {\bibfnamefont {P.}~\bibnamefont {Woodward}},\ }\bibfield  {title} {\bibinfo {title} {$\mathrm{Cs_{2}AgBiBr_{6}}$ $\mathrm{(X = Br, Cl)}$: New visible light absorbing, lead-free halide perovskite semiconductors},\ }\href@noop {} {\bibfield  {journal} {\bibinfo  {journal} {Chemistry of Materials}\ }\textbf {\bibinfo {volume} {28.5}},\ \bibinfo {pages} {1348} (\bibinfo {year} {2016})}\BibitemShut {NoStop}%
\bibitem [{\citenamefont {Igbari}\ \emph {et~al.}(2019)\citenamefont {Igbari}, \citenamefont {Wang},\ and\ \citenamefont {Liao}}]{Perovskite_tolerance}%
  \BibitemOpen
  \bibfield  {author} {\bibinfo {author} {\bibfnamefont {F.}~\bibnamefont {Igbari}}, \bibinfo {author} {\bibfnamefont {Z.}~\bibnamefont {Wang}},\ and\ \bibinfo {author} {\bibfnamefont {L.}~\bibnamefont {Liao}},\ }\bibfield  {title} {\bibinfo {title} {Progress of lead‐free halide double perovskites},\ }\href@noop {} {\bibfield  {journal} {\bibinfo  {journal} {Advanced Energy Materials}\ }\textbf {\bibinfo {volume} {9.12}},\ \bibinfo {pages} {1803150} (\bibinfo {year} {2019})}\BibitemShut {NoStop}%
\bibitem [{\citenamefont {Schade}\ \emph {et~al.}(2019)\citenamefont {Schade}, \citenamefont {Wright}, \citenamefont {Johnson}, \citenamefont {Dollmann}, \citenamefont {Wenger}, \citenamefont {Nayak}, \citenamefont {Prabhakaran}, \citenamefont {Herz}, \citenamefont {Nicholas}, \citenamefont {Snaith},\ and\ \citenamefont {Radaelli}}]{Schade2019}%
  \BibitemOpen
  \bibfield  {author} {\bibinfo {author} {\bibfnamefont {L.}~\bibnamefont {Schade}}, \bibinfo {author} {\bibfnamefont {A.}~\bibnamefont {Wright}}, \bibinfo {author} {\bibfnamefont {R.}~\bibnamefont {Johnson}}, \bibinfo {author} {\bibfnamefont {M.}~\bibnamefont {Dollmann}}, \bibinfo {author} {\bibfnamefont {B.}~\bibnamefont {Wenger}}, \bibinfo {author} {\bibfnamefont {P.}~\bibnamefont {Nayak}}, \bibinfo {author} {\bibfnamefont {D.}~\bibnamefont {Prabhakaran}}, \bibinfo {author} {\bibfnamefont {L.}~\bibnamefont {Herz}}, \bibinfo {author} {\bibfnamefont {R.}~\bibnamefont {Nicholas}}, \bibinfo {author} {\bibfnamefont {H.}~\bibnamefont {Snaith}},\ and\ \bibinfo {author} {\bibfnamefont {P.}~\bibnamefont {Radaelli}},\ }\bibfield  {title} {\bibinfo {title} {Structural and optical properties of $\mathrm{Cs_{2}AgBiBr_{6}}$ double perovskite},\ }\href@noop {} {\bibfield  {journal} {\bibinfo  {journal} {ACS Energy Letters}\ }\textbf {\bibinfo {volume} {4.1}},\ \bibinfo {pages} {299} (\bibinfo {year} {2019})}\BibitemShut
  {NoStop}%
\bibitem [{\citenamefont {Keshavarz}\ \emph {et~al.}(2020)\citenamefont {Keshavarz}, \citenamefont {Debroye}, \citenamefont {Ottesen}, \citenamefont {Martin}, \citenamefont {Zhang}, \citenamefont {Fron}, \citenamefont {Kuechler}, \citenamefont {Steele}, \citenamefont {Bremholm}, \citenamefont {Van~de Vondel}, \citenamefont {Wang}, \citenamefont {Bonn}, \citenamefont {Roeffaers}, \citenamefont {Wiedmann},\ and\ \citenamefont {Hofkens}}]{CenterFreq_Hardening}%
  \BibitemOpen
  \bibfield  {author} {\bibinfo {author} {\bibfnamefont {M.}~\bibnamefont {Keshavarz}}, \bibinfo {author} {\bibfnamefont {E.}~\bibnamefont {Debroye}}, \bibinfo {author} {\bibfnamefont {M.}~\bibnamefont {Ottesen}}, \bibinfo {author} {\bibfnamefont {C.}~\bibnamefont {Martin}}, \bibinfo {author} {\bibfnamefont {H.}~\bibnamefont {Zhang}}, \bibinfo {author} {\bibfnamefont {E.}~\bibnamefont {Fron}}, \bibinfo {author} {\bibfnamefont {R.}~\bibnamefont {Kuechler}}, \bibinfo {author} {\bibfnamefont {J.~A.}\ \bibnamefont {Steele}}, \bibinfo {author} {\bibfnamefont {M.}~\bibnamefont {Bremholm}}, \bibinfo {author} {\bibfnamefont {J.}~\bibnamefont {Van~de Vondel}}, \bibinfo {author} {\bibfnamefont {H.}~\bibnamefont {Wang}, \bibfnamefont {I}}, \bibinfo {author} {\bibfnamefont {M.}~\bibnamefont {Bonn}}, \bibinfo {author} {\bibfnamefont {M.~B.~J.}\ \bibnamefont {Roeffaers}}, \bibinfo {author} {\bibfnamefont {S.}~\bibnamefont {Wiedmann}},\ and\ \bibinfo {author} {\bibfnamefont {J.}~\bibnamefont {Hofkens}},\ }\bibfield  {title}
  {\bibinfo {title} {Tuning the structural and optoelectronic properties of $\mathrm{{C}s_2AgBiBr_6}$ double-perovskite single crystals through alkali-metal substitution},\ }\bibfield  {journal} {\bibinfo  {journal} {Advanced Materials}\ }\textbf {\bibinfo {volume} {32}},\ \href {https://doi.org/10.1002/adma.202001878} {10.1002/adma.202001878} (\bibinfo {year} {2020})\BibitemShut {NoStop}%
\bibitem [{\citenamefont {Cohen}\ \emph {et~al.}(2022)\citenamefont {Cohen}, \citenamefont {Brenner}, \citenamefont {Klarbring}, \citenamefont {Sharma}, \citenamefont {Fabini}, \citenamefont {Korobko}, \citenamefont {Nayak}, \citenamefont {Hellman},\ and\ \citenamefont {Yaffe}}]{cohendiverging}%
  \BibitemOpen
  \bibfield  {author} {\bibinfo {author} {\bibfnamefont {A.}~\bibnamefont {Cohen}}, \bibinfo {author} {\bibfnamefont {T.~M.}\ \bibnamefont {Brenner}}, \bibinfo {author} {\bibfnamefont {J.}~\bibnamefont {Klarbring}}, \bibinfo {author} {\bibfnamefont {R.}~\bibnamefont {Sharma}}, \bibinfo {author} {\bibfnamefont {D.~H.}\ \bibnamefont {Fabini}}, \bibinfo {author} {\bibfnamefont {R.}~\bibnamefont {Korobko}}, \bibinfo {author} {\bibfnamefont {P.~K.}\ \bibnamefont {Nayak}}, \bibinfo {author} {\bibfnamefont {O.}~\bibnamefont {Hellman}},\ and\ \bibinfo {author} {\bibfnamefont {O.}~\bibnamefont {Yaffe}},\ }\bibfield  {title} {\bibinfo {title} {Diverging expressions of anharmonicity in halide perovskites},\ }\href@noop {} {\bibfield  {journal} {\bibinfo  {journal} {Advanced Materials}\ ,\ \bibinfo {pages} {2107932}} (\bibinfo {year} {2022})}\BibitemShut {NoStop}%
\bibitem [{\citenamefont {He}\ \emph {et~al.}(2024)\citenamefont {He}, \citenamefont {Krogstad}, \citenamefont {Gupta}, \citenamefont {Lanigan-Atkins}, \citenamefont {Mao}, \citenamefont {Ye}, \citenamefont {Liu}, \citenamefont {Hong}, \citenamefont {Chi}, \citenamefont {Wei}, \citenamefont {Huang}, \citenamefont {Rosenkranz}, \citenamefont {Osborn},\ and\ \citenamefont {Delaire}}]{ComplexGroundState}%
  \BibitemOpen
  \bibfield  {author} {\bibinfo {author} {\bibfnamefont {X.}~\bibnamefont {He}}, \bibinfo {author} {\bibfnamefont {M.}~\bibnamefont {Krogstad}}, \bibinfo {author} {\bibfnamefont {M.~K.}\ \bibnamefont {Gupta}}, \bibinfo {author} {\bibfnamefont {T.}~\bibnamefont {Lanigan-Atkins}}, \bibinfo {author} {\bibfnamefont {C.}~\bibnamefont {Mao}}, \bibinfo {author} {\bibfnamefont {F.}~\bibnamefont {Ye}}, \bibinfo {author} {\bibfnamefont {Y.}~\bibnamefont {Liu}}, \bibinfo {author} {\bibfnamefont {T.}~\bibnamefont {Hong}}, \bibinfo {author} {\bibfnamefont {S.}~\bibnamefont {Chi}}, \bibinfo {author} {\bibfnamefont {H.}~\bibnamefont {Wei}}, \bibinfo {author} {\bibfnamefont {J.}~\bibnamefont {Huang}}, \bibinfo {author} {\bibfnamefont {S.}~\bibnamefont {Rosenkranz}}, \bibinfo {author} {\bibfnamefont {R.}~\bibnamefont {Osborn}},\ and\ \bibinfo {author} {\bibfnamefont {O.}~\bibnamefont {Delaire}},\ }\bibfield  {title} {\bibinfo {title} {Multiple lattice instabilities and complex ground state in $\mathrm{Cs_{2}AgBiBr_{6}}$},\
  }\href {https://doi.org/10.1103/PRXEnergy.3.013014} {\bibfield  {journal} {\bibinfo  {journal} {PRX Energy}\ }\textbf {\bibinfo {volume} {3}},\ \bibinfo {pages} {013014} (\bibinfo {year} {2024})}\BibitemShut {NoStop}%
\bibitem [{\citenamefont {Zheng}\ \emph {et~al.}(2024)\citenamefont {Zheng}, \citenamefont {Lin}, \citenamefont {Lin}, \citenamefont {Hautier}, \citenamefont {Guo},\ and\ \citenamefont {Huang}}]{ThermalCond_DFT1}%
  \BibitemOpen
  \bibfield  {author} {\bibinfo {author} {\bibfnamefont {J.}~\bibnamefont {Zheng}}, \bibinfo {author} {\bibfnamefont {C.}~\bibnamefont {Lin}}, \bibinfo {author} {\bibfnamefont {C.}~\bibnamefont {Lin}}, \bibinfo {author} {\bibfnamefont {G.}~\bibnamefont {Hautier}}, \bibinfo {author} {\bibfnamefont {R.}~\bibnamefont {Guo}},\ and\ \bibinfo {author} {\bibfnamefont {B.}~\bibnamefont {Huang}},\ }\bibfield  {title} {\bibinfo {title} {Unravelling ultralow thermal conductivity in perovskite $\mathrm{Cs_{2}AgBiBr_{6}}$: dominant wave-like phonon tunnelling and strong anharmonicity},\ }\href@noop {} {\bibfield  {journal} {\bibinfo  {journal} {Npj Computational Materials}\ }\textbf {\bibinfo {volume} {10}},\ \bibinfo {pages} {30} (\bibinfo {year} {2024})}\BibitemShut {NoStop}%
\bibitem [{\citenamefont {Klarbring}\ \emph {et~al.}(2020)\citenamefont {Klarbring}, \citenamefont {Hellman}, \citenamefont {Abrikosov},\ and\ \citenamefont {Simak}}]{ThermalCond_DFT2}%
  \BibitemOpen
  \bibfield  {author} {\bibinfo {author} {\bibfnamefont {J.}~\bibnamefont {Klarbring}}, \bibinfo {author} {\bibfnamefont {O.}~\bibnamefont {Hellman}}, \bibinfo {author} {\bibfnamefont {I.}~\bibnamefont {Abrikosov}},\ and\ \bibinfo {author} {\bibfnamefont {S.}~\bibnamefont {Simak}},\ }\bibfield  {title} {\bibinfo {title} {Anharmonicity and ultralow thermal conductivity in lead-free halide double perovskites},\ }\href@noop {} {\bibfield  {journal} {\bibinfo  {journal} {Physical Review Letters}\ }\textbf {\bibinfo {volume} {125}},\ \bibinfo {pages} {045701} (\bibinfo {year} {2020})}\BibitemShut {NoStop}%
\bibitem [{\citenamefont {Rodrigues}\ \emph {et~al.}(2021)\citenamefont {Rodrigues}, \citenamefont {Escanhoela}, \citenamefont {B.~Fragoso}, \citenamefont {Ferrer}, \citenamefont {\'{A}nlvarez Galv\'{a}n}, \citenamefont {Fern\'{a}nndez-D\'{i}az}, \citenamefont {Souza}, \citenamefont {Ferreira}, \citenamefont {Pecharrom\'{a}n},\ and\ \citenamefont {Alonso}}]{DFT_CASC}%
  \BibitemOpen
  \bibfield  {author} {\bibinfo {author} {\bibfnamefont {J.}~\bibnamefont {Rodrigues}}, \bibinfo {author} {\bibfnamefont {C.}~\bibnamefont {Escanhoela}}, \bibinfo {author} {\bibfnamefont {G.~S.}\ \bibnamefont {B.~Fragoso}}, \bibinfo {author} {\bibfnamefont {M.}~\bibnamefont {Ferrer}}, \bibinfo {author} {\bibfnamefont {C.}~\bibnamefont {\'{A}nlvarez Galv\'{a}n}}, \bibinfo {author} {\bibfnamefont {M.}~\bibnamefont {Fern\'{a}nndez-D\'{i}az}}, \bibinfo {author} {\bibfnamefont {J.}~\bibnamefont {Souza}}, \bibinfo {author} {\bibfnamefont {F.}~\bibnamefont {Ferreira}}, \bibinfo {author} {\bibfnamefont {C.}~\bibnamefont {Pecharrom\'{a}n}},\ and\ \bibinfo {author} {\bibfnamefont {J.~A.}\ \bibnamefont {Alonso}},\ }\bibfield  {title} {\bibinfo {title} {Experimental and theoretical investigations on the structural, electronic, and vibrational properties of $\mathrm{Cs_{2}AgSbCl_{6}}$ double perovskite},\ }\href@noop {} {\bibfield  {journal} {\bibinfo  {journal} {Industrial and Engineering Chemistry Research}\ }\textbf
  {\bibinfo {volume} {60}},\ \bibinfo {pages} {18918–} (\bibinfo {year} {2021})}\BibitemShut {NoStop}%
\bibitem [{\citenamefont {Zelewski}\ \emph {et~al.}(2019)\citenamefont {Zelewski}, \citenamefont {Urban}, \citenamefont {Surrente}, \citenamefont {Maude}, \citenamefont {Kuc}, \citenamefont {Schade}, \citenamefont {Johnson}, \citenamefont {Dollmann}, \citenamefont {Nayak}, \citenamefont {Snaith}, \citenamefont {Radaelli}, \citenamefont {Kudrawiec}, \citenamefont {Nicholas}, \citenamefont {Plochocka},\ and\ \citenamefont {Baranowski}}]{DFT_Cubic_tetra}%
  \BibitemOpen
  \bibfield  {author} {\bibinfo {author} {\bibfnamefont {S.~J.}\ \bibnamefont {Zelewski}}, \bibinfo {author} {\bibfnamefont {J.~M.}\ \bibnamefont {Urban}}, \bibinfo {author} {\bibfnamefont {A.}~\bibnamefont {Surrente}}, \bibinfo {author} {\bibfnamefont {D.~K.}\ \bibnamefont {Maude}}, \bibinfo {author} {\bibfnamefont {A.}~\bibnamefont {Kuc}}, \bibinfo {author} {\bibfnamefont {L.}~\bibnamefont {Schade}}, \bibinfo {author} {\bibfnamefont {R.~D.}\ \bibnamefont {Johnson}}, \bibinfo {author} {\bibfnamefont {M.}~\bibnamefont {Dollmann}}, \bibinfo {author} {\bibfnamefont {P.~K.}\ \bibnamefont {Nayak}}, \bibinfo {author} {\bibfnamefont {H.~J.}\ \bibnamefont {Snaith}}, \bibinfo {author} {\bibfnamefont {P.}~\bibnamefont {Radaelli}}, \bibinfo {author} {\bibfnamefont {R.}~\bibnamefont {Kudrawiec}}, \bibinfo {author} {\bibfnamefont {R.~J.}\ \bibnamefont {Nicholas}}, \bibinfo {author} {\bibfnamefont {P.}~\bibnamefont {Plochocka}},\ and\ \bibinfo {author} {\bibfnamefont {M.}~\bibnamefont {Baranowski}},\ }\bibfield  {title}
  {\bibinfo {title} {Revealing the nature of photoluminescence emission in the metal-halide double perovskite $\mathrm{Cs_{2}AgBiBr_{6}}$},\ }\href@noop {} {\bibfield  {journal} {\bibinfo  {journal} {Journal of Materials Chemistry}\ }\textbf {\bibinfo {volume} {7}},\ \bibinfo {pages} {8350–} (\bibinfo {year} {2019})}\BibitemShut {NoStop}%
\bibitem [{\citenamefont {Yin}\ \emph {et~al.}(2019)\citenamefont {Yin}, \citenamefont {Wu}, \citenamefont {Pan}, \citenamefont {Yang}, \citenamefont {Li}, \citenamefont {J.~Luo},\ and\ \citenamefont {Tang}}]{Crystal_Growth}%
  \BibitemOpen
  \bibfield  {author} {\bibinfo {author} {\bibfnamefont {L.}~\bibnamefont {Yin}}, \bibinfo {author} {\bibfnamefont {H.}~\bibnamefont {Wu}}, \bibinfo {author} {\bibfnamefont {W.}~\bibnamefont {Pan}}, \bibinfo {author} {\bibfnamefont {B.}~\bibnamefont {Yang}}, \bibinfo {author} {\bibfnamefont {P.}~\bibnamefont {Li}}, \bibinfo {author} {\bibfnamefont {G.~N.}\ \bibnamefont {J.~Luo}},\ and\ \bibinfo {author} {\bibfnamefont {J.}~\bibnamefont {Tang}},\ }\bibfield  {title} {\bibinfo {title} {Controlled cooling for synthesis of $\mathrm{Cs_{2}AgBiBr_{6}}$ single crystals and its application for x‐ray detection},\ }\href@noop {} {\bibfield  {journal} {\bibinfo  {journal} {Advanced Optical Materials}\ }\textbf {\bibinfo {volume} {7.19}},\ \bibinfo {pages} {1900491} (\bibinfo {year} {2019})}\BibitemShut {NoStop}%
\bibitem [{Sup()}]{Supplement}%
  \BibitemOpen
  \href@noop {} {}\bibinfo {note} {See Supplemental Material [url] for information regarding synthesis, experimental techniques used, a discussion of water contamination of the MIR reflectance spectrum, phase transitions in double perovskites, the results of the low temperature single crystal X-ray diffraction measurements, and powder transmission measurements, which includes Refs.\cite{weng2019lead,wang2021thickness,bartalucci2023probing,liu2023situ,jockel2024solar,Vasala2015,Fu2010,Battle1981,Battle1983,Zhou2009,Ji2020,Sikarwar2023,Dunitz1988,Stratton,poynting-cross}}\BibitemShut {NoStop}%
\bibitem [{\citenamefont {Rodr\'{\i}guez-Carvajal}(1993)}]{Fullprof}%
  \BibitemOpen
  \bibfield  {author} {\bibinfo {author} {\bibfnamefont {J.}~\bibnamefont {Rodr\'{\i}guez-Carvajal}},\ }\bibfield  {title} {\bibinfo {title} {Recent advances in magnetic structure determination by neutron powder diffraction},\ }\href {https://doi.org/10.1016/0921-4526(93)90108-I} {\bibfield  {journal} {\bibinfo  {journal} {Physica B}\ }\textbf {\bibinfo {volume} {192}},\ \bibinfo {pages} {55} (\bibinfo {year} {1993})}\BibitemShut {NoStop}%
\bibitem [{\citenamefont {Cakmak}\ \emph {et~al.}(2009)\citenamefont {Cakmak}, \citenamefont {Nuss},\ and\ \citenamefont {Jansen}}]{Cakmak2009}%
  \BibitemOpen
  \bibfield  {author} {\bibinfo {author} {\bibfnamefont {G.}~\bibnamefont {Cakmak}}, \bibinfo {author} {\bibfnamefont {J.}~\bibnamefont {Nuss}},\ and\ \bibinfo {author} {\bibfnamefont {M.}~\bibnamefont {Jansen}},\ }\bibfield  {title} {\bibinfo {title} {$\mathrm{LiB_{6}O_{9}F}$, the first lithium fluorooxoborate – crystal structure and ionic conductivity},\ }\href {https://doi.org/https://doi.org/10.1002/zaac.200900056} {\bibfield  {journal} {\bibinfo  {journal} {Zeitschrift für anorganische und allgemeine Chemie}\ }\textbf {\bibinfo {volume} {635}},\ \bibinfo {pages} {631} (\bibinfo {year} {2009})},\ \Eprint {https://arxiv.org/abs/https://onlinelibrary.wiley.com/doi/pdf/10.1002/zaac.200900056} {https://onlinelibrary.wiley.com/doi/pdf/10.1002/zaac.200900056} \BibitemShut {NoStop}%
\bibitem [{Bru(2019)}]{Bruker2019}%
  \BibitemOpen
  \bibfield  {title} {\bibinfo {title} {Bruker $\mathrm{S}$uite, version 2019/1. $\mathrm{B}$ruker $\mathrm{AXS}$ $\mathrm{I}$nc.},\ }\href@noop {} {\bibfield  {journal} {\bibinfo  {journal} {Bruker, Madison, WI}\ } (\bibinfo {year} {2019})}\BibitemShut {NoStop}%
\bibitem [{\citenamefont {Sheldrick}(2016)}]{Sheldrick2016}%
  \BibitemOpen
  \bibfield  {author} {\bibinfo {author} {\bibfnamefont {G.~M.}\ \bibnamefont {Sheldrick}},\ }\bibfield  {title} {\bibinfo {title} {Sadabs bruker-axs area detector scaling and absorption, version 2016/2},\ }\href@noop {} {\bibfield  {journal} {\bibinfo  {journal} {University of G\"ottingen}\ } (\bibinfo {year} {2016})}\BibitemShut {NoStop}%
\bibitem [{\citenamefont {Sheldrick}(2008)}]{Sheldrick2008}%
  \BibitemOpen
  \bibfield  {author} {\bibinfo {author} {\bibfnamefont {G.~M.}\ \bibnamefont {Sheldrick}},\ }\bibfield  {title} {\bibinfo {title} {{A short history of {\it SHELX}}},\ }\href {https://doi.org/10.1107/S0108767307043930} {\bibfield  {journal} {\bibinfo  {journal} {Acta Crystallographica Section A}\ }\textbf {\bibinfo {volume} {64}},\ \bibinfo {pages} {112} (\bibinfo {year} {2008})}\BibitemShut {NoStop}%
\bibitem [{\citenamefont {Sheldrick}(2015)}]{Sheldrick2015}%
  \BibitemOpen
  \bibfield  {author} {\bibinfo {author} {\bibfnamefont {G.~M.}\ \bibnamefont {Sheldrick}},\ }\bibfield  {title} {\bibinfo {title} {{Crystal structure refinement with {\it SHELXL}}},\ }\href {https://doi.org/10.1107/S2053229614024218} {\bibfield  {journal} {\bibinfo  {journal} {Acta Crystallographica Section C}\ }\textbf {\bibinfo {volume} {71}},\ \bibinfo {pages} {3} (\bibinfo {year} {2015})}\BibitemShut {NoStop}%
\bibitem [{\citenamefont {Petříček}\ \emph {et~al.}(2014)\citenamefont {Petříček}, \citenamefont {Dušek},\ and\ \citenamefont {Palatinus}}]{Petricek2014}%
  \BibitemOpen
  \bibfield  {author} {\bibinfo {author} {\bibfnamefont {V.}~\bibnamefont {Petříček}}, \bibinfo {author} {\bibfnamefont {M.}~\bibnamefont {Dušek}},\ and\ \bibinfo {author} {\bibfnamefont {L.}~\bibnamefont {Palatinus}},\ }\bibfield  {title} {\bibinfo {title} {Crystallographic computing system jana2006: General features},\ }\href {https://doi.org/doi:10.1515/zkri-2014-1737} {\bibfield  {journal} {\bibinfo  {journal} {Zeitschrift für Kristallographie - Crystalline Materials}\ }\textbf {\bibinfo {volume} {229}},\ \bibinfo {pages} {345} (\bibinfo {year} {2014})}\BibitemShut {NoStop}%
\bibitem [{\citenamefont {Homes}\ \emph {et~al.}(2003)\citenamefont {Homes}, \citenamefont {Reedyk}, \citenamefont {Crandles},\ and\ \citenamefont {Timusk}}]{Homes}%
  \BibitemOpen
  \bibfield  {author} {\bibinfo {author} {\bibfnamefont {C.}~\bibnamefont {Homes}}, \bibinfo {author} {\bibfnamefont {M.}~\bibnamefont {Reedyk}}, \bibinfo {author} {\bibfnamefont {D.}~\bibnamefont {Crandles}},\ and\ \bibinfo {author} {\bibfnamefont {T.}~\bibnamefont {Timusk}},\ }\bibfield  {title} {\bibinfo {title} {Technique for measuring the reflectance of irregular, submillimeter-sized samples},\ }\href@noop {} {\bibfield  {journal} {\bibinfo  {journal} {Applied Optics}\ }\textbf {\bibinfo {volume} {32}},\ \bibinfo {pages} {2976} (\bibinfo {year} {2003})}\BibitemShut {NoStop}%
\bibitem [{\citenamefont {Schnelle}\ \emph {et~al.}(1999)\citenamefont {Schnelle}, \citenamefont {Engelhardt},\ and\ \citenamefont {Gmelin}}]{N-Apiezon}%
  \BibitemOpen
  \bibfield  {author} {\bibinfo {author} {\bibfnamefont {W.}~\bibnamefont {Schnelle}}, \bibinfo {author} {\bibfnamefont {J.}~\bibnamefont {Engelhardt}},\ and\ \bibinfo {author} {\bibfnamefont {E.}~\bibnamefont {Gmelin}},\ }\bibfield  {title} {\bibinfo {title} {Specific heat capacity of apiezon n high vacuum grease and of duran borosilicate glass},\ }\href@noop {} {\bibfield  {journal} {\bibinfo  {journal} {Cryogenics (Guildford)}\ }\textbf {\bibinfo {volume} {39.3}},\ \bibinfo {pages} {271–275} (\bibinfo {year} {1999})}\BibitemShut {NoStop}%
\bibitem [{\citenamefont {Le~Bail}(2005)}]{LeBail}%
  \BibitemOpen
  \bibfield  {author} {\bibinfo {author} {\bibfnamefont {A.}~\bibnamefont {Le~Bail}},\ }\bibfield  {title} {\bibinfo {title} {Whole powder pattern decomposition methods and applications: A retrospection},\ }\href {https://doi.org/10.1154/1.2135315} {\bibfield  {journal} {\bibinfo  {journal} {Powder Diffraction}\ }\textbf {\bibinfo {volume} {20}},\ \bibinfo {pages} {316–} (\bibinfo {year} {2005})}\BibitemShut {NoStop}%
\bibitem [{\citenamefont {Spek}(2001)}]{Platon}%
  \BibitemOpen
  \bibfield  {author} {\bibinfo {author} {\bibfnamefont {A.~L.}\ \bibnamefont {Spek}},\ }\bibfield  {title} {\bibinfo {title} {Platon, a multipurpose crystallographic tool},\ }\href {http://www.platonsoft.nl/platon/pl000000.html} {\bibfield  {journal} {\bibinfo  {journal} {Utrecht University, Utrecht, The Netherlands}\ } (\bibinfo {year} {2001})}\BibitemShut {NoStop}%
\bibitem [{\citenamefont {Zhang}\ \emph {et~al.}(2025)\citenamefont {Zhang}, \citenamefont {Gehring}, \citenamefont {Slavney}, \citenamefont {Vigil}, \citenamefont {Hong}, \citenamefont {Rodriguez-Rivera}, \citenamefont {Klarbring}, \citenamefont {Karunadasa}, \citenamefont {Toney},\ and\ \citenamefont {Weadock}}]{zhang2025phonon}%
  \BibitemOpen
  \bibfield  {author} {\bibinfo {author} {\bibfnamefont {Z.}~\bibnamefont {Zhang}}, \bibinfo {author} {\bibfnamefont {P.~M.}\ \bibnamefont {Gehring}}, \bibinfo {author} {\bibfnamefont {A.~H.}\ \bibnamefont {Slavney}}, \bibinfo {author} {\bibfnamefont {J.~A.}\ \bibnamefont {Vigil}}, \bibinfo {author} {\bibfnamefont {T.}~\bibnamefont {Hong}}, \bibinfo {author} {\bibfnamefont {J.~A.}\ \bibnamefont {Rodriguez-Rivera}}, \bibinfo {author} {\bibfnamefont {J.}~\bibnamefont {Klarbring}}, \bibinfo {author} {\bibfnamefont {H.~I.}\ \bibnamefont {Karunadasa}}, \bibinfo {author} {\bibfnamefont {M.~F.}\ \bibnamefont {Toney}},\ and\ \bibinfo {author} {\bibfnamefont {N.~J.}\ \bibnamefont {Weadock}},\ }\bibfield  {title} {\bibinfo {title} {Phonon lifetimes and mode softening in cubic $\mathrm{Cs_{2}AgBiBr_{6}}$},\ }\href@noop {} {\bibfield  {journal} {\bibinfo  {journal} {Newton}\ }\textbf {\bibinfo {volume} {1}} (\bibinfo {year} {2025})}\BibitemShut {NoStop}%
\bibitem [{\citenamefont {Scott}(1974)}]{scott1974soft}%
  \BibitemOpen
  \bibfield  {author} {\bibinfo {author} {\bibfnamefont {J.}~\bibnamefont {Scott}},\ }\bibfield  {title} {\bibinfo {title} {Soft-mode spectroscopy: Experimental studies of structural phase transitions},\ }\href@noop {} {\bibfield  {journal} {\bibinfo  {journal} {Reviews of Modern Physics}\ }\textbf {\bibinfo {volume} {46}},\ \bibinfo {pages} {83} (\bibinfo {year} {1974})}\BibitemShut {NoStop}%
\bibitem [{\citenamefont {J{\"o}bsis}\ \emph {et~al.}(2021)\citenamefont {J{\"o}bsis}, \citenamefont {Caselli}, \citenamefont {Askes}, \citenamefont {Garnett}, \citenamefont {Savenije}, \citenamefont {Rabouw},\ and\ \citenamefont {Hutter}}]{UV_extrap}%
  \BibitemOpen
  \bibfield  {author} {\bibinfo {author} {\bibfnamefont {H.}~\bibnamefont {J{\"o}bsis}}, \bibinfo {author} {\bibfnamefont {V.}~\bibnamefont {Caselli}}, \bibinfo {author} {\bibfnamefont {S.}~\bibnamefont {Askes}}, \bibinfo {author} {\bibfnamefont {E.}~\bibnamefont {Garnett}}, \bibinfo {author} {\bibfnamefont {T.}~\bibnamefont {Savenije}}, \bibinfo {author} {\bibfnamefont {F.}~\bibnamefont {Rabouw}},\ and\ \bibinfo {author} {\bibfnamefont {E.}~\bibnamefont {Hutter}},\ }\bibfield  {title} {\bibinfo {title} {Recombination and localization: unfolding the pathways behind conductivity losses in $\mathrm{Cs_{2}AgBiBr_{6}}$ thin films},\ }\href@noop {} {\bibfield  {journal} {\bibinfo  {journal} {Applied Physics Letters}\ }\textbf {\bibinfo {volume} {119}},\ \bibinfo {pages} {13190} (\bibinfo {year} {2021})}\BibitemShut {NoStop}%
\bibitem [{\citenamefont {Tanner}(2015)}]{Xray_extrap}%
  \BibitemOpen
  \bibfield  {author} {\bibinfo {author} {\bibfnamefont {D.}~\bibnamefont {Tanner}},\ }\bibfield  {title} {\bibinfo {title} {Use of x-ray scattering functions in kramers-kronig analysis of reflectance},\ }\href@noop {} {\bibfield  {journal} {\bibinfo  {journal} {Physical Review. B}\ }\textbf {\bibinfo {volume} {91}},\ \bibinfo {pages} {035123} (\bibinfo {year} {2015})}\BibitemShut {NoStop}%
\bibitem [{\citenamefont {Henke}\ \emph {et~al.}(1993)\citenamefont {Henke}, \citenamefont {Gullikson},\ and\ \citenamefont {Davis}}]{XrayFactors}%
  \BibitemOpen
  \bibfield  {author} {\bibinfo {author} {\bibfnamefont {B.}~\bibnamefont {Henke}}, \bibinfo {author} {\bibfnamefont {E.}~\bibnamefont {Gullikson}},\ and\ \bibinfo {author} {\bibfnamefont {J.}~\bibnamefont {Davis}},\ }\bibfield  {title} {\bibinfo {title} {X-ray interactions: Photoabsorption, scattering, transmission, and reflection at e = 50-30,000 ev, z = 1-92},\ }\href@noop {} {\bibfield  {journal} {\bibinfo  {journal} {Atomic Data and Nuclear Data Tables}\ }\textbf {\bibinfo {volume} {54}},\ \bibinfo {pages} {181–342} (\bibinfo {year} {1993})}\BibitemShut {NoStop}%
\bibitem [{\citenamefont {Wooten}(1972)}]{WootenOptical}%
  \BibitemOpen
  \bibfield  {author} {\bibinfo {author} {\bibfnamefont {F.}~\bibnamefont {Wooten}},\ }\href@noop {} {\emph {\bibinfo {title} {Optical Properties of Solids}}}\ (\bibinfo  {publisher} {Academic Press},\ \bibinfo {year} {1972})\BibitemShut {NoStop}%
\bibitem [{\citenamefont {Rodrigues}\ \emph {et~al.}(2018)\citenamefont {Rodrigues}, \citenamefont {Bezerra},\ and\ \citenamefont {Hernandes}}]{short-range}%
  \BibitemOpen
  \bibfield  {author} {\bibinfo {author} {\bibfnamefont {J.}~\bibnamefont {Rodrigues}}, \bibinfo {author} {\bibfnamefont {D.}~\bibnamefont {Bezerra}},\ and\ \bibinfo {author} {\bibfnamefont {A.}~\bibnamefont {Hernandes}},\ }\bibfield  {title} {\bibinfo {title} {Calculation of the optical phonons in ordered ba2mgwo6 perovskite using short‐range force field model},\ }\href@noop {} {\bibfield  {journal} {\bibinfo  {journal} {Journal of Raman Spectroscopy}\ }\textbf {\bibinfo {volume} {49}},\ \bibinfo {pages} {1822–1829} (\bibinfo {year} {2018})}\BibitemShut {NoStop}%
\bibitem [{\citenamefont {Aroyo}\ \emph {et~al.}(2011)\citenamefont {Aroyo}, \citenamefont {Perez-Mato}, \citenamefont {Orobengoa}, \citenamefont {Tasci}, \citenamefont {de~la Flor},\ and\ \citenamefont {Kirov}}]{Bilbao-Generic-1}%
  \BibitemOpen
  \bibfield  {author} {\bibinfo {author} {\bibfnamefont {M.~I.}\ \bibnamefont {Aroyo}}, \bibinfo {author} {\bibfnamefont {J.~M.}\ \bibnamefont {Perez-Mato}}, \bibinfo {author} {\bibfnamefont {D.}~\bibnamefont {Orobengoa}}, \bibinfo {author} {\bibfnamefont {E.}~\bibnamefont {Tasci}}, \bibinfo {author} {\bibfnamefont {G.}~\bibnamefont {de~la Flor}},\ and\ \bibinfo {author} {\bibfnamefont {A.}~\bibnamefont {Kirov}},\ }\bibfield  {title} {\bibinfo {title} {Crystallography online: Bilbao crystallographic server},\ }\href@noop {} {\bibfield  {journal} {\bibinfo  {journal} {Bulgarian Chemical Communications}\ }\textbf {\bibinfo {volume} {43}},\ \bibinfo {pages} {183} (\bibinfo {year} {2011})}\BibitemShut {NoStop}%
\bibitem [{\citenamefont {Aroyo}\ \emph {et~al.}(2006{\natexlab{a}})\citenamefont {Aroyo}, \citenamefont {Perez-Mato}, \citenamefont {Capillas}, \citenamefont {Kroumova}, \citenamefont {Ivantchev}, \citenamefont {Madariaga}, \citenamefont {Kirov},\ and\ \citenamefont {Wondratschek}}]{Bilbao-Generic-2}%
  \BibitemOpen
  \bibfield  {author} {\bibinfo {author} {\bibfnamefont {M.~I.}\ \bibnamefont {Aroyo}}, \bibinfo {author} {\bibfnamefont {J.~M.}\ \bibnamefont {Perez-Mato}}, \bibinfo {author} {\bibfnamefont {C.}~\bibnamefont {Capillas}}, \bibinfo {author} {\bibfnamefont {E.}~\bibnamefont {Kroumova}}, \bibinfo {author} {\bibfnamefont {S.}~\bibnamefont {Ivantchev}}, \bibinfo {author} {\bibfnamefont {G.}~\bibnamefont {Madariaga}}, \bibinfo {author} {\bibfnamefont {A.}~\bibnamefont {Kirov}},\ and\ \bibinfo {author} {\bibfnamefont {H.}~\bibnamefont {Wondratschek}},\ }\bibfield  {title} {\bibinfo {title} {Bilbao crystallographic server i: Databases and crystallographic computing programs},\ }\href@noop {} {\bibfield  {journal} {\bibinfo  {journal} {Zeitschrift für Kristallographie}\ }\textbf {\bibinfo {volume} {221}},\ \bibinfo {pages} {15} (\bibinfo {year} {2006}{\natexlab{a}})}\BibitemShut {NoStop}%
\bibitem [{\citenamefont {Aroyo}\ \emph {et~al.}(2006{\natexlab{b}})\citenamefont {Aroyo}, \citenamefont {Kirov}, \citenamefont {Capillas}, \citenamefont {Perez-Mato},\ and\ \citenamefont {Wondratschek}}]{Bilbao-Generic-3}%
  \BibitemOpen
  \bibfield  {author} {\bibinfo {author} {\bibfnamefont {M.~I.}\ \bibnamefont {Aroyo}}, \bibinfo {author} {\bibfnamefont {A.}~\bibnamefont {Kirov}}, \bibinfo {author} {\bibfnamefont {C.}~\bibnamefont {Capillas}}, \bibinfo {author} {\bibfnamefont {J.~M.}\ \bibnamefont {Perez-Mato}},\ and\ \bibinfo {author} {\bibfnamefont {H.}~\bibnamefont {Wondratschek}},\ }\bibfield  {title} {\bibinfo {title} {Bilbao crystallographic server ii: Representations of crystallographic point groups and space groups},\ }\href@noop {} {\bibfield  {journal} {\bibinfo  {journal} {Acta Crystallographica Section A}\ }\textbf {\bibinfo {volume} {62}},\ \bibinfo {pages} {115} (\bibinfo {year} {2006}{\natexlab{b}})}\BibitemShut {NoStop}%
\bibitem [{\citenamefont {Kroumova}\ \emph {et~al.}(2003)\citenamefont {Kroumova}, \citenamefont {Aroyo}, \citenamefont {Mato}, \citenamefont {Kirov}, \citenamefont {Capillas}, \citenamefont {Ivantchev},\ and\ \citenamefont {Wondratschek}}]{Bilbao-Generic-SAM}%
  \BibitemOpen
  \bibfield  {author} {\bibinfo {author} {\bibfnamefont {E.}~\bibnamefont {Kroumova}}, \bibinfo {author} {\bibfnamefont {M.~I.}\ \bibnamefont {Aroyo}}, \bibinfo {author} {\bibfnamefont {J.~M.~P.}\ \bibnamefont {Mato}}, \bibinfo {author} {\bibfnamefont {A.}~\bibnamefont {Kirov}}, \bibinfo {author} {\bibfnamefont {C.}~\bibnamefont {Capillas}}, \bibinfo {author} {\bibfnamefont {S.}~\bibnamefont {Ivantchev}},\ and\ \bibinfo {author} {\bibfnamefont {H.}~\bibnamefont {Wondratschek}},\ }\bibfield  {title} {\bibinfo {title} {Bilbao crystallographic server: useful databases and tools for phase transitions studies},\ }\href@noop {} {\bibfield  {journal} {\bibinfo  {journal} {Phase Transitions}\ }\textbf {\bibinfo {volume} {76}},\ \bibinfo {pages} {155} (\bibinfo {year} {2003})}\BibitemShut {NoStop}%
\bibitem [{\citenamefont {Faik}\ \emph {et~al.}(2012)\citenamefont {Faik}, \citenamefont {Orobengoa}, \citenamefont {Iturbe-Zabalo},\ and\ \citenamefont {Igartua}}]{Faik2012}%
  \BibitemOpen
  \bibfield  {author} {\bibinfo {author} {\bibfnamefont {A.}~\bibnamefont {Faik}}, \bibinfo {author} {\bibfnamefont {D.}~\bibnamefont {Orobengoa}}, \bibinfo {author} {\bibfnamefont {E.}~\bibnamefont {Iturbe-Zabalo}},\ and\ \bibinfo {author} {\bibfnamefont {J.}~\bibnamefont {Igartua}},\ }\bibfield  {title} {\bibinfo {title} {A study of the crystal structures and the phase transitions of the ordered double perovskites $\mathrm{Sr_{2}ScSbO_{6}}$ and $\mathrm{Ca_{2}ScSbO_{6}}$},\ }\href {https://doi.org/https://doi.org/10.1016/j.jssc.2012.04.019} {\bibfield  {journal} {\bibinfo  {journal} {Journal of Solid State Chemistry}\ }\textbf {\bibinfo {volume} {192}},\ \bibinfo {pages} {273} (\bibinfo {year} {2012})}\BibitemShut {NoStop}%
\bibitem [{\citenamefont {Flerov}\ \emph {et~al.}(2003)\citenamefont {Flerov}, \citenamefont {Burriel}, \citenamefont {Gorev}, \citenamefont {Isla},\ and\ \citenamefont {Voronov}}]{Additional_Structural_Transition}%
  \BibitemOpen
  \bibfield  {author} {\bibinfo {author} {\bibfnamefont {I.}~\bibnamefont {Flerov}}, \bibinfo {author} {\bibfnamefont {R.}~\bibnamefont {Burriel}}, \bibinfo {author} {\bibfnamefont {M.}~\bibnamefont {Gorev}}, \bibinfo {author} {\bibfnamefont {P.}~\bibnamefont {Isla}},\ and\ \bibinfo {author} {\bibfnamefont {V.}~\bibnamefont {Voronov}},\ }\bibfield  {title} {\bibinfo {title} {Low-temperature specific heat of the $\mathrm{Rb_{2}KScF_{6}}$ elpasolite},\ }\href@noop {} {\bibfield  {journal} {\bibinfo  {journal} {Physics of the Solid State}\ }\textbf {\bibinfo {volume} {45.1}},\ \bibinfo {pages} {167} (\bibinfo {year} {2003})}\BibitemShut {NoStop}%
\bibitem [{\citenamefont {Howard}\ \emph {et~al.}(2003)\citenamefont {Howard}, \citenamefont {Kennedy},\ and\ \citenamefont {Woodward}}]{PerovskitesGroupTheory}%
  \BibitemOpen
  \bibfield  {author} {\bibinfo {author} {\bibfnamefont {C.~J.}\ \bibnamefont {Howard}}, \bibinfo {author} {\bibfnamefont {B.~J.}\ \bibnamefont {Kennedy}},\ and\ \bibinfo {author} {\bibfnamefont {P.~M.}\ \bibnamefont {Woodward}},\ }\bibfield  {title} {\bibinfo {title} {Ordered double perovskites; a group-theoretical analysis},\ }\href@noop {} {\bibfield  {journal} {\bibinfo  {journal} {Acta crystallographica. Section B, Structural science}\ }\textbf {\bibinfo {volume} {59}},\ \bibinfo {pages} {463–471} (\bibinfo {year} {2003})}\BibitemShut {NoStop}%
\bibitem [{\citenamefont {Weng}\ \emph {et~al.}(2019)\citenamefont {Weng}, \citenamefont {Qin}, \citenamefont {Umar}, \citenamefont {Wang}, \citenamefont {Zhang}, \citenamefont {Wang}, \citenamefont {Cui}, \citenamefont {Li}, \citenamefont {Zheng},\ and\ \citenamefont {Zhan}}]{weng2019lead}%
  \BibitemOpen
  \bibfield  {author} {\bibinfo {author} {\bibfnamefont {Z.}~\bibnamefont {Weng}}, \bibinfo {author} {\bibfnamefont {J.}~\bibnamefont {Qin}}, \bibinfo {author} {\bibfnamefont {A.~A.}\ \bibnamefont {Umar}}, \bibinfo {author} {\bibfnamefont {J.}~\bibnamefont {Wang}}, \bibinfo {author} {\bibfnamefont {X.}~\bibnamefont {Zhang}}, \bibinfo {author} {\bibfnamefont {H.}~\bibnamefont {Wang}}, \bibinfo {author} {\bibfnamefont {X.}~\bibnamefont {Cui}}, \bibinfo {author} {\bibfnamefont {X.}~\bibnamefont {Li}}, \bibinfo {author} {\bibfnamefont {L.}~\bibnamefont {Zheng}},\ and\ \bibinfo {author} {\bibfnamefont {Y.}~\bibnamefont {Zhan}},\ }\bibfield  {title} {\bibinfo {title} {Lead-free $\mathrm{Cs_{2}AgBiBr_{6}}$ double perovskite-based humidity sensor with superfast recovery time},\ }\href@noop {} {\bibfield  {journal} {\bibinfo  {journal} {Advanced Functional Materials}\ }\textbf {\bibinfo {volume} {29}},\ \bibinfo {pages} {1902234} (\bibinfo {year} {2019})}\BibitemShut {NoStop}%
\bibitem [{\citenamefont {Wang}\ \emph {et~al.}(2021)\citenamefont {Wang}, \citenamefont {Gao}, \citenamefont {Zhang}, \citenamefont {Chen}, \citenamefont {Junkang}, \citenamefont {Shen}, \citenamefont {Au}, \citenamefont {Li}, \citenamefont {Cai},\ and\ \citenamefont {Yin}}]{wang2021thickness}%
  \BibitemOpen
  \bibfield  {author} {\bibinfo {author} {\bibfnamefont {B.-H.}\ \bibnamefont {Wang}}, \bibinfo {author} {\bibfnamefont {B.}~\bibnamefont {Gao}}, \bibinfo {author} {\bibfnamefont {J.-R.}\ \bibnamefont {Zhang}}, \bibinfo {author} {\bibfnamefont {L.}~\bibnamefont {Chen}}, \bibinfo {author} {\bibfnamefont {G.}~\bibnamefont {Junkang}}, \bibinfo {author} {\bibfnamefont {S.}~\bibnamefont {Shen}}, \bibinfo {author} {\bibfnamefont {C.-T.}\ \bibnamefont {Au}}, \bibinfo {author} {\bibfnamefont {K.}~\bibnamefont {Li}}, \bibinfo {author} {\bibfnamefont {M.-Q.}\ \bibnamefont {Cai}},\ and\ \bibinfo {author} {\bibfnamefont {S.-F.}\ \bibnamefont {Yin}},\ }\bibfield  {title} {\bibinfo {title} {Thickness-induced band-gap engineering in lead-free double perovskite $\mathrm{Cs_{2}AgBiBr_{6}}$ for highly efficient photocatalysis},\ }\href@noop {} {\bibfield  {journal} {\bibinfo  {journal} {Physical Chemistry Chemical Physics}\ }\textbf {\bibinfo {volume} {23}},\ \bibinfo {pages} {12439} (\bibinfo {year} {2021})}\BibitemShut
  {NoStop}%
\bibitem [{\citenamefont {Bartalucci}\ \emph {et~al.}(2023)\citenamefont {Bartalucci}, \citenamefont {Mal{\"a}r}, \citenamefont {Mehnert}, \citenamefont {Kleine~B{\"u}ning}, \citenamefont {G{\"u}nzel}, \citenamefont {Icker}, \citenamefont {B{\"o}rner}, \citenamefont {Wiebeler}, \citenamefont {Meier}, \citenamefont {Grimme} \emph {et~al.}}]{bartalucci2023probing}%
  \BibitemOpen
  \bibfield  {author} {\bibinfo {author} {\bibfnamefont {E.}~\bibnamefont {Bartalucci}}, \bibinfo {author} {\bibfnamefont {A.~A.}\ \bibnamefont {Mal{\"a}r}}, \bibinfo {author} {\bibfnamefont {A.}~\bibnamefont {Mehnert}}, \bibinfo {author} {\bibfnamefont {J.~B.}\ \bibnamefont {Kleine~B{\"u}ning}}, \bibinfo {author} {\bibfnamefont {L.}~\bibnamefont {G{\"u}nzel}}, \bibinfo {author} {\bibfnamefont {M.}~\bibnamefont {Icker}}, \bibinfo {author} {\bibfnamefont {M.}~\bibnamefont {B{\"o}rner}}, \bibinfo {author} {\bibfnamefont {C.}~\bibnamefont {Wiebeler}}, \bibinfo {author} {\bibfnamefont {B.~H.}\ \bibnamefont {Meier}}, \bibinfo {author} {\bibfnamefont {S.}~\bibnamefont {Grimme}}, \emph {et~al.},\ }\bibfield  {title} {\bibinfo {title} {Probing a hydrogen-$\pi$ interaction involving a trapped water molecule in the solid state},\ }\href@noop {} {\bibfield  {journal} {\bibinfo  {journal} {Angewandte Chemie International Edition}\ }\textbf {\bibinfo {volume} {62}},\ \bibinfo {pages} {e202217725} (\bibinfo {year}
  {2023})}\BibitemShut {NoStop}%
\bibitem [{\citenamefont {Liu}\ \emph {et~al.}(2023)\citenamefont {Liu}, \citenamefont {Wu}, \citenamefont {Zhang}, \citenamefont {Zhao}, \citenamefont {Li}, \citenamefont {Li}, \citenamefont {Wen},\ and\ \citenamefont {Wang}}]{liu2023situ}%
  \BibitemOpen
  \bibfield  {author} {\bibinfo {author} {\bibfnamefont {J.}~\bibnamefont {Liu}}, \bibinfo {author} {\bibfnamefont {Z.}~\bibnamefont {Wu}}, \bibinfo {author} {\bibfnamefont {F.}~\bibnamefont {Zhang}}, \bibinfo {author} {\bibfnamefont {M.}~\bibnamefont {Zhao}}, \bibinfo {author} {\bibfnamefont {C.}~\bibnamefont {Li}}, \bibinfo {author} {\bibfnamefont {J.}~\bibnamefont {Li}}, \bibinfo {author} {\bibfnamefont {B.}~\bibnamefont {Wen}},\ and\ \bibinfo {author} {\bibfnamefont {F.}~\bibnamefont {Wang}},\ }\bibfield  {title} {\bibinfo {title} {In situ growth of lead-free halide perovskites into $\mathrm{SiO_2}$ sub-microcapsules toward water-stable photocatalytic $\mathrm{CO_2}$ reduction},\ }\href@noop {} {\bibfield  {journal} {\bibinfo  {journal} {Nanoscale}\ }\textbf {\bibinfo {volume} {15}},\ \bibinfo {pages} {7023} (\bibinfo {year} {2023})}\BibitemShut {NoStop}%
\bibitem [{\citenamefont {J{\"o}ckel}\ \emph {et~al.}(2024)\citenamefont {J{\"o}ckel}, \citenamefont {Yoon}, \citenamefont {Frebel}, \citenamefont {Neguse}, \citenamefont {Rossa}, \citenamefont {Bett}, \citenamefont {Schubert}, \citenamefont {Widenmeyer}, \citenamefont {Balke-Gr{\"u}newald},\ and\ \citenamefont {Weidenkaff}}]{jockel2024solar}%
  \BibitemOpen
  \bibfield  {author} {\bibinfo {author} {\bibfnamefont {D.~M.}\ \bibnamefont {J{\"o}ckel}}, \bibinfo {author} {\bibfnamefont {S.}~\bibnamefont {Yoon}}, \bibinfo {author} {\bibfnamefont {A.}~\bibnamefont {Frebel}}, \bibinfo {author} {\bibfnamefont {S.~M.}\ \bibnamefont {Neguse}}, \bibinfo {author} {\bibfnamefont {J.~D.}\ \bibnamefont {Rossa}}, \bibinfo {author} {\bibfnamefont {A.~J.}\ \bibnamefont {Bett}}, \bibinfo {author} {\bibfnamefont {M.}~\bibnamefont {Schubert}}, \bibinfo {author} {\bibfnamefont {M.}~\bibnamefont {Widenmeyer}}, \bibinfo {author} {\bibfnamefont {B.}~\bibnamefont {Balke-Gr{\"u}newald}},\ and\ \bibinfo {author} {\bibfnamefont {A.}~\bibnamefont {Weidenkaff}},\ }\bibfield  {title} {\bibinfo {title} {Solar degradation and stability of lead-free light absorber $\mathrm{Cs_{2}AgBiBr_{6}}$ in ambient conditions},\ }\href@noop {} {\bibfield  {journal} {\bibinfo  {journal} {Advanced Photonics Research}\ }\textbf {\bibinfo {volume} {5}},\ \bibinfo {pages} {2300269} (\bibinfo {year}
  {2024})}\BibitemShut {NoStop}%
\bibitem [{\citenamefont {Vasala}\ and\ \citenamefont {Karppinen}(2015)}]{Vasala2015}%
  \BibitemOpen
  \bibfield  {author} {\bibinfo {author} {\bibfnamefont {S.}~\bibnamefont {Vasala}}\ and\ \bibinfo {author} {\bibfnamefont {M.}~\bibnamefont {Karppinen}},\ }\bibfield  {title} {\bibinfo {title} {$\mathrm{A_{2}B'B''O_{6}}$ perovskites: A review},\ }\href {https://doi.org/https://doi.org/10.1016/j.progsolidstchem.2014.08.001} {\bibfield  {journal} {\bibinfo  {journal} {Progress in Solid State Chemistry}\ }\textbf {\bibinfo {volume} {43}},\ \bibinfo {pages} {1} (\bibinfo {year} {2015})}\BibitemShut {NoStop}%
\bibitem [{\citenamefont {Fu}\ \emph {et~al.}(2010)\citenamefont {Fu}, \citenamefont {Götz},\ and\ \citenamefont {Ijdo}}]{Fu2010}%
  \BibitemOpen
  \bibfield  {author} {\bibinfo {author} {\bibfnamefont {W.}~\bibnamefont {Fu}}, \bibinfo {author} {\bibfnamefont {R.}~\bibnamefont {Götz}},\ and\ \bibinfo {author} {\bibfnamefont {D.}~\bibnamefont {Ijdo}},\ }\bibfield  {title} {\bibinfo {title} {On the symmetry and crystal structures of $\mathrm{Ba_{2}LaIrO_{6}}$},\ }\href {https://doi.org/https://doi.org/10.1016/j.jssc.2009.12.006} {\bibfield  {journal} {\bibinfo  {journal} {Journal of Solid State Chemistry}\ }\textbf {\bibinfo {volume} {183}},\ \bibinfo {pages} {419} (\bibinfo {year} {2010})}\BibitemShut {NoStop}%
\bibitem [{\citenamefont {Battle}(1981)}]{Battle1981}%
  \BibitemOpen
  \bibfield  {author} {\bibinfo {author} {\bibfnamefont {P.}~\bibnamefont {Battle}},\ }\bibfield  {title} {\bibinfo {title} {The crystal structures of $\mathrm{Ba_{2}LaRuO_{6}}$ and $\mathrm{Ca_{2}LaRuO_{6}}$},\ }\href {https://doi.org/https://doi.org/10.1016/0025-5408(81)90007-6} {\bibfield  {journal} {\bibinfo  {journal} {Materials Research Bulletin}\ }\textbf {\bibinfo {volume} {16}},\ \bibinfo {pages} {397} (\bibinfo {year} {1981})}\BibitemShut {NoStop}%
\bibitem [{\citenamefont {Battle}\ \emph {et~al.}(1983)\citenamefont {Battle}, \citenamefont {Goodenough},\ and\ \citenamefont {Price}}]{Battle1983}%
  \BibitemOpen
  \bibfield  {author} {\bibinfo {author} {\bibfnamefont {P.}~\bibnamefont {Battle}}, \bibinfo {author} {\bibfnamefont {J.}~\bibnamefont {Goodenough}},\ and\ \bibinfo {author} {\bibfnamefont {R.}~\bibnamefont {Price}},\ }\bibfield  {title} {\bibinfo {title} {The crystal structures and magnetic properties of $\mathrm{Ba_{2}LaRuO_{6}}$ and $\mathrm{Ca_{2}LaRuO_{6}}$},\ }\href {https://doi.org/https://doi.org/10.1016/0022-4596(83)90147-0} {\bibfield  {journal} {\bibinfo  {journal} {Journal of Solid State Chemistry}\ }\textbf {\bibinfo {volume} {46}},\ \bibinfo {pages} {234} (\bibinfo {year} {1983})}\BibitemShut {NoStop}%
\bibitem [{\citenamefont {Zhou}\ \emph {et~al.}(2009)\citenamefont {Zhou}, \citenamefont {Kennedy}, \citenamefont {Avdeev}, \citenamefont {Giachini},\ and\ \citenamefont {Kimpton}}]{Zhou2009}%
  \BibitemOpen
  \bibfield  {author} {\bibinfo {author} {\bibfnamefont {Q.}~\bibnamefont {Zhou}}, \bibinfo {author} {\bibfnamefont {B.~J.}\ \bibnamefont {Kennedy}}, \bibinfo {author} {\bibfnamefont {M.}~\bibnamefont {Avdeev}}, \bibinfo {author} {\bibfnamefont {L.}~\bibnamefont {Giachini}},\ and\ \bibinfo {author} {\bibfnamefont {J.~A.}\ \bibnamefont {Kimpton}},\ }\bibfield  {title} {\bibinfo {title} {Structural studies of the phases in $\mathrm{Ba_{2}LaIrO_{6}}$—new light on an old problem},\ }\href {https://doi.org/https://doi.org/10.1016/j.jssc.2009.08.026} {\bibfield  {journal} {\bibinfo  {journal} {Journal of Solid State Chemistry}\ }\textbf {\bibinfo {volume} {182}},\ \bibinfo {pages} {3195} (\bibinfo {year} {2009})}\BibitemShut {NoStop}%
\bibitem [{\citenamefont {Ji}\ \emph {et~al.}(2020)\citenamefont {Ji}, \citenamefont {Klarbring}, \citenamefont {Wang}, \citenamefont {Ning}, \citenamefont {Wang}, \citenamefont {Yin}, \citenamefont {Figueroa}, \citenamefont {Christensen}, \citenamefont {Etter}, \citenamefont {Ederth}, \citenamefont {Sun}, \citenamefont {Simak}, \citenamefont {Abrikosov},\ and\ \citenamefont {Gao}}]{Ji2020}%
  \BibitemOpen
  \bibfield  {author} {\bibinfo {author} {\bibfnamefont {F.}~\bibnamefont {Ji}}, \bibinfo {author} {\bibfnamefont {J.}~\bibnamefont {Klarbring}}, \bibinfo {author} {\bibfnamefont {F.}~\bibnamefont {Wang}}, \bibinfo {author} {\bibfnamefont {W.}~\bibnamefont {Ning}}, \bibinfo {author} {\bibfnamefont {L.}~\bibnamefont {Wang}}, \bibinfo {author} {\bibfnamefont {C.}~\bibnamefont {Yin}}, \bibinfo {author} {\bibfnamefont {J.~S.~M.}\ \bibnamefont {Figueroa}}, \bibinfo {author} {\bibfnamefont {C.~K.}\ \bibnamefont {Christensen}}, \bibinfo {author} {\bibfnamefont {M.}~\bibnamefont {Etter}}, \bibinfo {author} {\bibfnamefont {T.}~\bibnamefont {Ederth}}, \bibinfo {author} {\bibfnamefont {L.}~\bibnamefont {Sun}}, \bibinfo {author} {\bibfnamefont {S.~I.}\ \bibnamefont {Simak}}, \bibinfo {author} {\bibfnamefont {I.~A.}\ \bibnamefont {Abrikosov}},\ and\ \bibinfo {author} {\bibfnamefont {F.}~\bibnamefont {Gao}},\ }\bibfield  {title} {\bibinfo {title} {Lead-free halide double perovskite cs2agbibr6 with decreased band gap},\
  }\href {https://doi.org/https://doi.org/10.1002/anie.202005568} {\bibfield  {journal} {\bibinfo  {journal} {Angewandte Chemie International Edition}\ }\textbf {\bibinfo {volume} {59}},\ \bibinfo {pages} {15191} (\bibinfo {year} {2020})},\ \Eprint {https://arxiv.org/abs/https://onlinelibrary.wiley.com/doi/pdf/10.1002/anie.202005568} {https://onlinelibrary.wiley.com/doi/pdf/10.1002/anie.202005568} \BibitemShut {NoStop}%
\bibitem [{\citenamefont {Sikarwar}\ \emph {et~al.}(2023)\citenamefont {Sikarwar}, \citenamefont {Siwach}, \citenamefont {Phani~Chandra}, \citenamefont {Antharjanam},\ and\ \citenamefont {Chandiran}}]{Sikarwar2023}%
  \BibitemOpen
  \bibfield  {author} {\bibinfo {author} {\bibfnamefont {P.}~\bibnamefont {Sikarwar}}, \bibinfo {author} {\bibfnamefont {P.}~\bibnamefont {Siwach}}, \bibinfo {author} {\bibfnamefont {N.~V.}\ \bibnamefont {Phani~Chandra}}, \bibinfo {author} {\bibfnamefont {S.}~\bibnamefont {Antharjanam}},\ and\ \bibinfo {author} {\bibfnamefont {A.~K.}\ \bibnamefont {Chandiran}},\ }\bibfield  {title} {\bibinfo {title} {Dimensional reduction of $\mathrm{Cs_{2}AgBiBr_{6}}$ using alkyl ammonium cations $\mathrm{CH_{3}(CH_{2})_{n}NH_{3}^{+}}$ (n = 1, 2, 3, and 6) of varying chain lengths and their role in structural and optoelectronic properties},\ }\href {https://doi.org/10.1021/acs.inorgchem.3c00571} {\bibfield  {journal} {\bibinfo  {journal} {Inorganic Chemistry}\ }\textbf {\bibinfo {volume} {62}},\ \bibinfo {pages} {5836} (\bibinfo {year} {2023})},\ \bibinfo {note} {pMID: 36995096},\ \Eprint {https://arxiv.org/abs/https://doi.org/10.1021/acs.inorgchem.3c00571} {https://doi.org/10.1021/acs.inorgchem.3c00571} \BibitemShut
  {NoStop}%
\bibitem [{\citenamefont {Dunitz}\ \emph {et~al.}(1988)\citenamefont {Dunitz}, \citenamefont {Schomaker},\ and\ \citenamefont {Trueblood}}]{Dunitz1988}%
  \BibitemOpen
  \bibfield  {author} {\bibinfo {author} {\bibfnamefont {J.~D.}\ \bibnamefont {Dunitz}}, \bibinfo {author} {\bibfnamefont {V.}~\bibnamefont {Schomaker}},\ and\ \bibinfo {author} {\bibfnamefont {K.~N.}\ \bibnamefont {Trueblood}},\ }\bibfield  {title} {\bibinfo {title} {Interpretation of atomic displacement parameters from diffraction studies of crystals},\ }\href {https://doi.org/10.1021/j100315a002} {\bibfield  {journal} {\bibinfo  {journal} {The Journal of Physical Chemistry}\ }\textbf {\bibinfo {volume} {92}},\ \bibinfo {pages} {856} (\bibinfo {year} {1988})},\ \Eprint {https://arxiv.org/abs/https://doi.org/10.1021/j100315a002} {https://doi.org/10.1021/j100315a002} \BibitemShut {NoStop}%
\bibitem [{\citenamefont {Stratton}\ \emph {et~al.}(2015)\citenamefont {Stratton}, \citenamefont {Antennas},\ and\ \citenamefont {Society}}]{Stratton}%
  \BibitemOpen
  \bibfield  {author} {\bibinfo {author} {\bibfnamefont {J.~A.}\ \bibnamefont {Stratton}}, \bibinfo {author} {\bibfnamefont {I.}~\bibnamefont {Antennas}},\ and\ \bibinfo {author} {\bibfnamefont {P.}~\bibnamefont {Society}},\ }\href@noop {} {\emph {\bibinfo {title} {Electromagnetic theory}}}\ (\bibinfo  {publisher} {Hoboken, New Jersey: John Wiley \& Sons, Inc},\ \bibinfo {year} {2015})\BibitemShut {NoStop}%
\bibitem [{\citenamefont {Weber}(2014)}]{poynting-cross}%
  \BibitemOpen
  \bibfield  {author} {\bibinfo {author} {\bibfnamefont {H.}~\bibnamefont {Weber}},\ }\bibfield  {title} {\bibinfo {title} {The fresnel equations for lossy dielectrics and conservation of energy},\ }\href@noop {} {\bibfield  {journal} {\bibinfo  {journal} {Journal of modern optics}\ }\textbf {\bibinfo {volume} {61}},\ \bibinfo {pages} {1219–1224} (\bibinfo {year} {2014})}\BibitemShut {NoStop}%
\end{thebibliography}

%

\end{document}



\title{Supplemental Material\\
Low-temperature structural instabilities of the halide double perovskite \cabb{} investigated via x-ray diffraction and infrared phonons}

\author{Collin Tower}

\author{Fereidoon S. Razavi}%
\author{Jeremy Dion}%

\author{Maureen Reedyk}%

\affiliation{%
 Brock University, Department of Physics, St. Catharines, ON, Canada
}%

\author{J\"urgen Nuss}

\affiliation{
 Max Planck Institute for Solid State Research, Stuttgart, Germany
}%
\author{Reinhard K. Kremer}

\affiliation{
 Max Planck Institute for Solid State Research, Stuttgart, Germany
}%

\date{\today}

\maketitle

\begin{widetext}
This supplemental material contains information regarding crystal synthesis, details of the experimental techniques used, a discussion of water contamination of the MIR reflectance spectrum, extrapolations for Kramers-Kronig analysis, a discussion of phase transitions in double perovskites, the results of the low temperature single crystal X-ray diffraction measurements, and powder transmission measurements.

\end{widetext}

\section{Experimental Details}

\subsection{Crystal Synthesis}

The precursor chemicals CsBr (99.9\% Acros Organics), BiBr$_3$ (99\% Alfa Aesar), and AgBr (99.5\% Alfa Aesar) were dissolved in a solution of HBr acid. 2~mmol of CsBr, 1~mmol of BiBr$_3$, and 1~mmol of AgBr were dissolved in an 13~ml solution of 48 wt\% HBr (solution in water, Acros Organics) diluted to 40\% with distilled water and heated at 95 $^\circ$C. The solution was cooled to 85 $^\circ$C at a rate of 3 $^\circ$C/hr then cooled to 82 $^\circ$C at a rate of 0.5 $^\circ$C/hr. This latter temperature was held for 3 hours, and finally the solution was cooled to 20~$^\circ$C at a rate of 1~$^\circ$C/hr.

\subsection{Heat Capacity}

The heat capacity of a 9.8~mg crystal  was measured with a physical property measurement system (PPMS, Quantum Design). 
The crystal was attached to the sapphire platform secured by a minute amount of 
ApiezonN vacuum grease which was included in the addenda that was first measured. The sample was cooled at a rate of 0.1~K/min and measured upon heating from 2.5~K to 300~K. 

\subsection{Dielectric Permittivity}

The dielectric permittivity was determined with an Andeen-Hagerling AH2700A automatic capacitance bridge applying silver paint contacts to a \cabb{} crystal to form a plate capacitor. The dielectric function $\epsilon (T)$ was determined from the dielectric capacitance and loss measured at 50~Hz from 4-300~K.

\subsection{X-ray Powder Diffraction}

X-ray powder diffraction patterns were collected with a Bruker D8 Discovery X-ray diffractometer system (Bragg-Brentano scattering geometry) using monochromatized CuK$\alpha_1$ radiation). A Phenix closed cycle cryostat (Oxford Cryosystems) was used to set the sample temperature. Each pattern was collected at stabilized temperature. The powder samples were thermally anchored with ApiezonN vacuum grease to the sample platform equipped with an inlay of a specially cut silicon wafer to minimize background scattering.

\subsection{Single Crystal X-ray Diffraction}
Crystals suitable for single-crystal X-ray diffraction were selected under high viscosity oil, and mounted with Paratone-N oil on a loop made of Kapton foil (Micromounts$^{\rm TM}$, MiTeGen, Ithaca, NY). Diffraction data were collected at 32~K, 60~K, 100~K, and room temperature with a SMART APEXII CCD X-ray diffractometer (Bruker AXS, Karlsruhe, Germany), using graphite-monochromated Mo-$K\alpha_1$ radiation, and an N-Helix low-temperature device (Oxford Cryosystems, Oxford, U.K.).\cite{Cakmak2009} The reflection intensities were integrated with the SAINT subprogram available in the Bruker Suite software.\cite{Bruker2019} A multi-scan absorption correction was applied using SADABS\cite{Sheldrick2016}, and the structures were solved by direct methods and refined by full-matrix least-square fitting with the SHELXTL software package\cite{Sheldrick2008,Sheldrick2015} in case of room temperature measurement. In case of the low temperature data, the JANA2006\cite{Petricek2014} software was used for structure refinement which allows data setup and refinement for multiple twinning, Pseudo-merohedral twinning with 6 ($I{\rm 4}/m$, 100~K and 60~K) or 24 ($P\bar{1}$, 32~K) different orientations was assumed. Noticeable splitting of Bragg reflections could not be detected at 32~K.

\subsection{IR Reflectance}

IR reflectance was measured via a Bruker-IFS-66v/s Fourier transform IR spectrometer coupled to a Janis 2T400 cryostat. The FIR region was measured using a mercury arc lamp with a liquid-helium cooled bolometer as detector. A germanium coated mylar beamsplitter was used in the Michelson interferometer giving a range of 20-600 \wav. The mid-infrared (MIR) range was measured using a ceramic globar source and Mercury Cadmium Telluride (MCT) detector with a KBr beamsplitter giving a range of 600-5000 \wav. The near-infrared (NIR) range was measured using a tungsten filament source and MCT detector with a CaF$_2$ beamsplitter giving a range of 5000-11000 \wav. The ultraviolet (UV) to visible reflectivity was measured at room temperature using a Photon Technology International model A1010 spectrometer. The data compared favourably to the calculated reflectivity data obtained from ellipsometry data measured by J{\"o}bsis \etal.~\cite{UV_extrap} The data of J{\"o}bsis \etal~\cite{UV_extrap} was digitized and ultimately used to extend the high frequency range as the ellipsometry data had a much higher signal to noise ratio giving a range of 6000-33000 \wav.

\subsubsection{Water Contamination}
\begin{figure}[]
\includegraphics[width=0.5\textwidth,keepaspectratio]{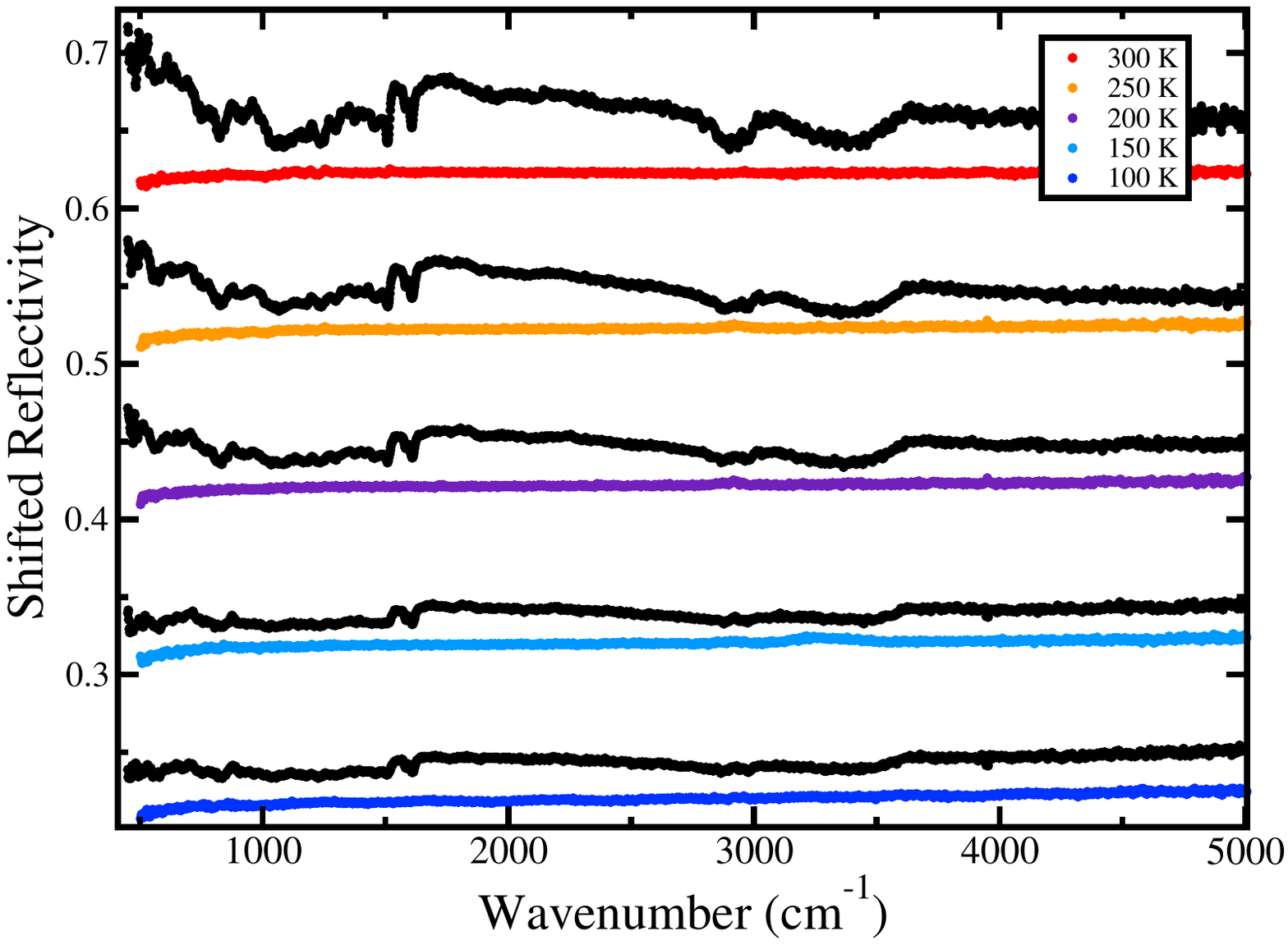}
\caption{\label{fig:MIR_anneal} The reflectivity of \cabb{} in the MIR region at selected temperatures. The spectra shown in black were collected before annealing at 100~$^{\circ}C$ for several days. The colored spectra were measured after annealing. Spectra collected above 100~K temperature have been shifted from the previous temperature by 0.1 for visual clarity.}
\end{figure}

Even though the crystals are stored in a closed dry desiccated environment they were found to contain water as could be clearly seen in reflectivity spectra in the mid-infrared (MIR) region illustrated in FIG.\ref{fig:MIR_anneal}. \cabb{} is well known to be hygroscopic and has been proposed as a humidity sensor\cite{weng2019lead} and as a photocatalyst for water splitting for hydrogen generation as it has the ability to adsorb H$_2$O and desorb H$_2$.\cite{wang2021thickness} It is unlikely that the water observed in the MIR spectra is due to water adsorbed onto the surface since the optical measurements are carried out in vacuum and a small surface layer of adsorbed water is typically easily removed. It is thus more likely that it is
the result of minute amounts of trapped water due to the synthesis in the aqueous hydrobromic acid environment. Trapped water molecules will typically form a hydrogen bond between an atom with high electronegativity in the host material and a hydrogen atom.\cite{bartalucci2023probing} Refinement of the location of the water molecules via X-ray diffraction methods is difficult due to the low scattering factor of hydrogen atoms but can be achieved with neutron scattering or nuclear magnetic resonance spectroscopy;\cite{bartalucci2023probing} however this is beyond the scope of this work.
First principles calculations for water adsorption in \cabb{} found that the most stable location for H$_2$O is to bind to the bromine atoms.\cite{liu2023situ}
The effect of exposure to humid laboratory conditions on the Raman scattering spectrum and the band gap has been investigated.\cite{jockel2024solar} It was found that exposure to 20-50\% humidity for a period of 140 hours had no significant effect on the band gap or on the Raman scattering spectrum.
In contrast, we have found that the mid-infrared reflectivity measurements are particularly sensitive to even minute amounts of water as that is the region of the spectrum where the intense absorption due to  O-H  stretching and H-O-H bending vibrations occurs.

To remove the influence of the water on the infrared reflectivity spectra, the samples were annealed in vacuum at 100 $^{\circ}C$ for several days before the measurements were performed. As shown in FIG.\ref{fig:MIR_anneal} this treatment was effective in removing spectral features due to the presence of water in the sample.
Measurements of the heat capacity, capacitance and Xray diffraction were carried out on as-grown samples stored in a closed dry desiccated environment as is the convention and were not noticeably affected by the presence of water.

\subsubsection{Extrapolations for Kramers-Kronig Analysis}

\begin{figure}[]
\includegraphics[width=0.5\textwidth,keepaspectratio]{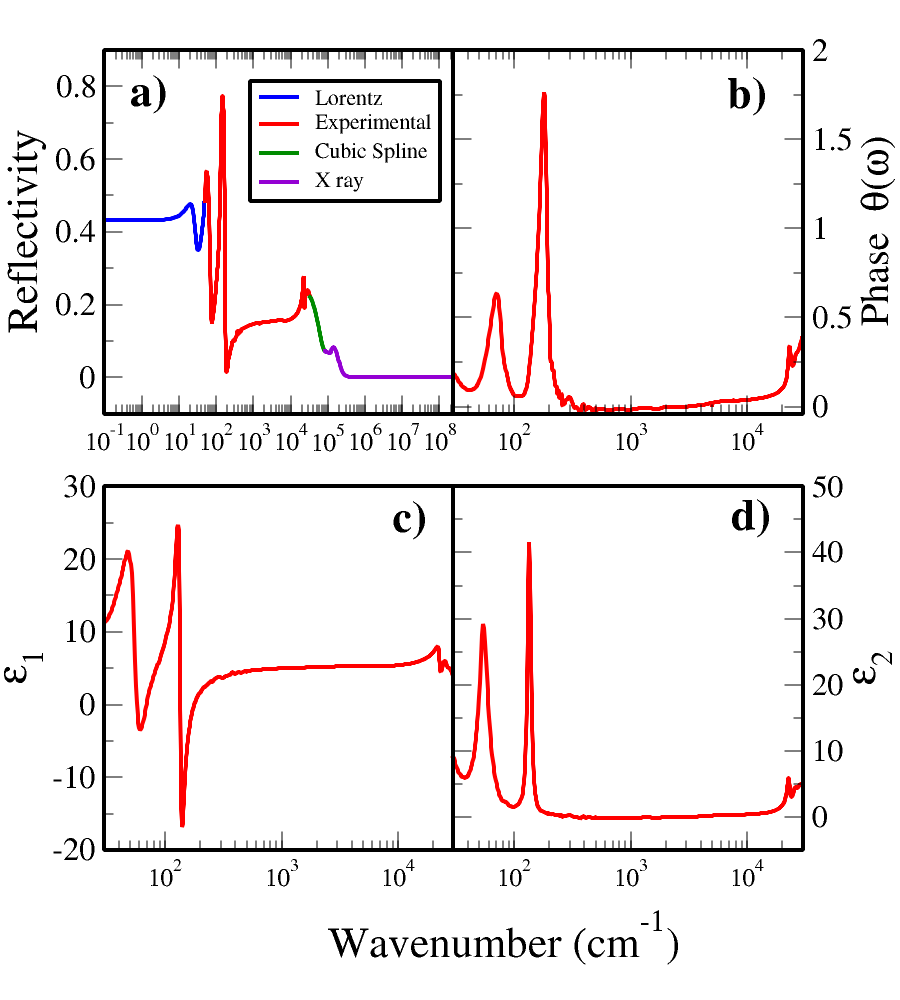}
\caption{\label{fig:KK_290K} Kramers Kronig results for \cabb{} at 290 K.  The (a) reflectance spectra  displaying the low frequency Lorentz oscillator extrapolation, the experimental data, the cubic spline and the high frequency X ray extrapolation used to obtain the (b) phase. (c) The real part of the dielectric function. (d) The imaginary part of the dielectric function.  }
\end{figure}

An example of the high and low frequency reflectance extrapolations used for the Kramers Kronig method of obtaining the dielectric function is shown in FIG. \ref{fig:KK_290K}(a). The integration was carried out at each experimentally determined wavenumber, integrating over the entire extrapolated range of wavenumbers with a phase correction accounting for the reflectance past the extra extrapolation. The Kramers-Kronig derived phase is shown in (b) and the real and imaginary parts of the dielectric function in (c) and (d) respectively. For more information on the process see the main manuscript.

\section{ Phase Transitions in Double Perovskites}

As is the case for  \cabb{}, other double perovskites such as Rb$_2$KScF$_6$, have been found to exhibit a low temperature structural phase transition from a tetragonal to a monoclinic structure.~\cite{Additional_Structural_Transition} Especially oxide double perovskites with composition A$_2$BB´O$_6$  have been widely studied with respect to their structural crystal chemistry. For a review see e.g. Ref.~\onlinecite{Vasala2015}. Depending on the Goldschmidt tolerance factor, $t = ({R_{\rm A}}~+~R_{\rm O})/\sqrt{2}~(\bar{R}_{_{\rm B}}~+~R_{\rm O})$, wherein $R_A$ and $R_O$ denotes the the ionic radii of the A cation and oxygen respectively and $\bar{R}_{\rm B}$ denotes the average of the ionic radii of the atoms B and B´, a variety of crystal structures have been observed for the oxide double perovskites. For tolerance factors $t\lesssim$~0.97 most common are compounds crystallizing with the monoclinic space group \mono{} (no. 14),  whereas for $t\gtrsim$~0.97 the cubic structure (space group \cubic, no. 225) is prevalent. Other space groups like \tetra{} (no. 87), \monotwo{} (no. 12), or $R\bar{\rm 3}$ (no. 148) are less frequent and their occurrence is centered at around tolerance factors of $t\sim$~0.97. Observation of the triclinic space group \triclinic{} (no. 2)  appears to be very rare and its existence was also questioned.\cite{Fu2010} Symmetry reduction in double perovskites, e.g. as  function of temperature starting from \cubic{}  commonly proceed via  \tetra{} and \monotwo{} to \mono\cite{Faik2012} Sometimes the intermediate step via space group $I$2/$m$ is skipped. The differentiation of the crystal structures between the different space group descriptions may be minute. A prominent example of the early period of structure investigations on double perovskites by powder diffraction is Ba$_2$LaRuO$_6$. The room temperature crystal structure initially described in the monoclinic space group \mono\cite{Battle1981} was somewhat later found to be triclinic (\triclinic) isotypic to its congener Ca$_2$LaRuO$_6$.\cite{Battle1983} Similarly, below 11~K Ba$_2$LaIrO$_6$ also shows triclinic symmetry.\cite{Zhou2009}

\section{Low Temperature Crystal Structure of 
C\MakeLowercase{s}$_2$A\MakeLowercase{g}B\MakeLowercase{i}B\MakeLowercase{r}$_6$}

\subsection{Results}
\begin{figure}[]
\includegraphics[width=0.45\textwidth,keepaspectratio]{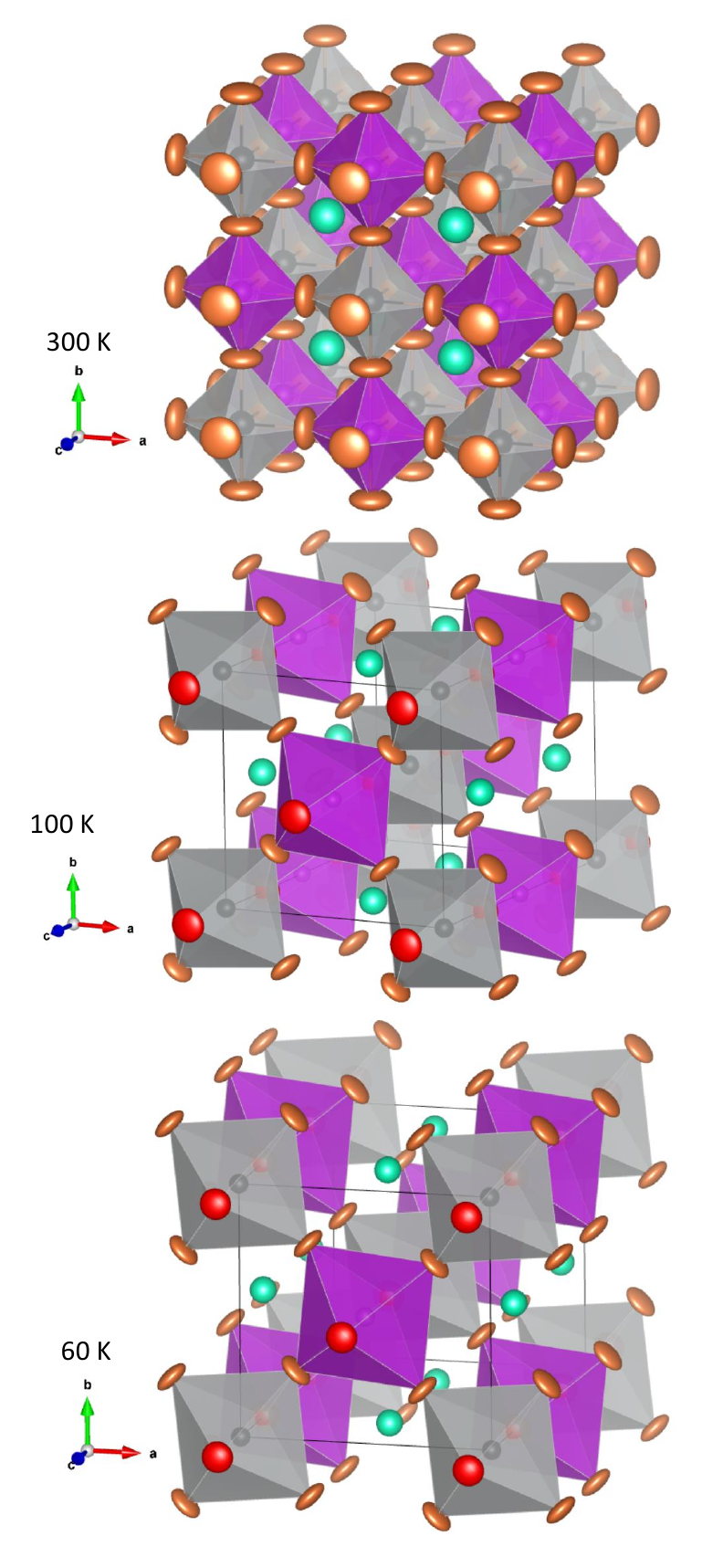}
\caption{\label{fig:ThreeStructures} Crystal structures of Cs$_2$AgBiBr$_6$ at 300~K, 100~K, and 60~K, from top to bottom, as obtained from the single crystal X-ray diffraction measurement. For 300~K the structure refinement was done in space group $Fm\bar{\rm3}m$, for 100~K, and 60~K the space group $I\rm{4}/m$ was assumed.
The two different Br atoms are shown in red and orange, the Ag and Bi atoms centering the Br octahedra are depicted in grey and violet, respectively, and the Cs atoms are represented by the light green spheres.}
\end{figure}

FIG. \ref{fig:ThreeStructures} shows the crystal structures of Cs$_2$AgBiBr$_6$ as derived from the single crystal X-ray diffraction measurements at 300~K, 100~K, and 60~K. The structure refinements were determined assuming space group $Fm\bar{\rm 3}m$ for the  300~K and $I{\rm 4}/m$ for the 100~K and the 60~K data, respectively. The structural parameters are summarized in TABLE \ref{Table2}.

\begin{table}[]
\caption{Structural parameters of Cs$_2$AgBiBr$_6$ as obtained from the refinement of the single crystal X-ray diffraction data (wavelength 0.71073~\AA, Mo $K_{\alpha}$ radiation, Bruker CCD 4-circle diffractometer). For Wyckoff site occupancy of Cs, Ag, Bi, and Br1 and Br2 atoms see TABLE III of the main manuscript. The listed isotropic displacement parameters $B_{\rm iso}$~=~8$\pi^2 U_{\rm iso}$ were calculated from the refined anisotropic displacement parameters.}
\label{Table2}
\begin{ruledtabular}
\begin{tabular}{l c  c c c }
$T$ (K)  &  32  &   60  & 100  & 300 \\

 space grp.& $I\rm{4}/m$ & $I\rm{4}/m$ & $I\rm{4}/m$ & $Fm\bar{\rm 3}m$\\

\hline
\\
$a, b$  ({\AA})   & 7.923(2)  & 7.9400(9)  & 7.9381(15)  &  11.2857(16)\\
$c$  ({\AA})   & 11.205(3) & 11.2288(13) & 11.226(3)  & 11.2857(16)\\
\\
$V_{\rm cell}$ ({\AA}$^3$) & 703.3(39) & 707.9(1) &707.4(3) & 2$\times$718.7(4)\\
\\
Cs   & &  & &  \\
$B_{\rm{iso}}$ ({\AA}$^2$)    & 0.98    &  1.25 & 1.49  & 3.96\\ 
\\
Ag  &  & & &   \\
$B_{\rm{iso}}$ ({\AA}$^2$)& 
    0.46  &  0.61       & 0.68  & 2.09\\ 
\\    
Bi   & &  & &  \\
$B_{\rm{iso}}$ ({\AA}$^2$)   & 0.49  &  0.58 & 0.57  & 1.48\\ 
\\
Br1  &  & & &    \\
$x$ & 0.21864(14)  &  0.27414(16)    & 0.26932(17)  & 0.24910(3) \\
$y$ & -0.27640(14) &  -0.22227(16) & -0.22679(15)  & 0\\
$z$ &  0   &  0        & 0 &  0 \\
$B_{\rm{iso}}$ ({\AA}$^2$)  & 
    1.24    &  1.58  & 1.82  & 4.25\\ 
\\
Br2  &  & & &    \\
$x$ &  0           &  0        & 0         & - \\
$y$ &  0           &  0        & 0         & -\\
$z$ &  0.25031(19) & -0.2490(3) & -0.2494(3)  &  - \\
$B_{\rm{iso}}$ ({\AA}$^2$)    &
         1.02  &   1.37 & 1.70 & -\\ 
\\
$wR$ (\%) & 4.35  & 3.3  & 2.9 & 1.6\\
$GoF$         & 0.72  & 0.67  & 0.79 & 1.19\\
\end{tabular}
\end{ruledtabular}
\end{table}

\begin{figure}[]
\includegraphics[width=0.45\textwidth,keepaspectratio]{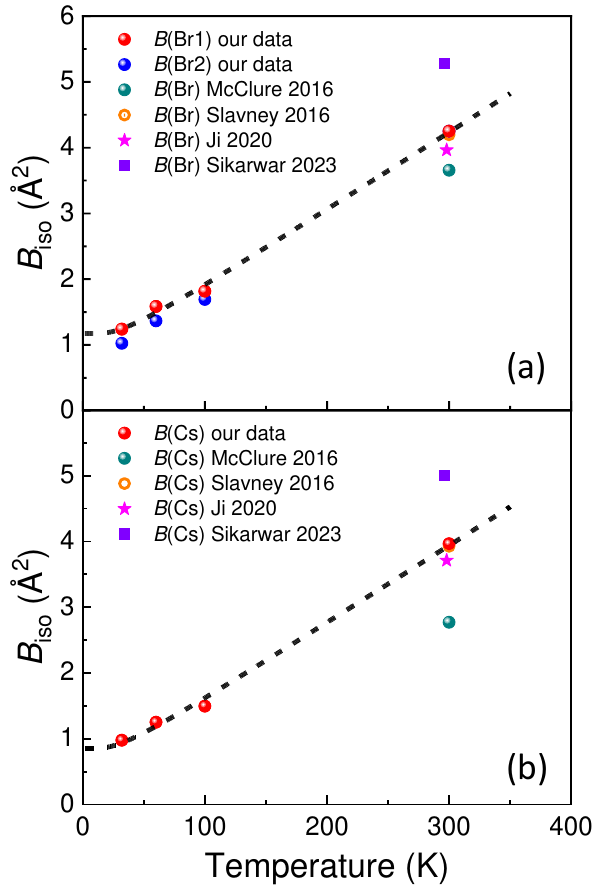}
\caption{\label{fig:Bisotropic} Isotropic displacement factors of the Br and the Cs atoms as function of temperature. In addition to our data, literature results at room temperature have also been included.\cite{McClure2016,Slavney2016,Ji2020,Sikarwar2023} The dashed lines represents fits of our data to eq.~(S1)\cite{Dunitz1988} assuming typical vibrational frequencies of $\sim$60~cm$^{-1}$ and reduced masses of $\sim$45~amu.}
\end{figure}

The atom displacement parameters exhibit a marked size difference between the metal atoms Ag, and Bi and the Cs and Br atoms, the values for the latter being by a factor of two larger than those of the former.
Whereas the displacement parameters in the tetragonal phases are almost spherical for the metal atoms Cs, Ag and Bi (they are spherical by symmetry in the cubic phase), the Br atoms show
marked oblate displacement ellipsoids, already in the cubic phase. Especially the displacement parameters of Cs and Br noticeably shrink with decreasing temperature, as displayed  in FIG. \ref{fig:Bisotropic}.  

The temperature dependence of the mean square displacement amplitude of an atom, $\langle U^2 \rangle$ is given e.g. by  Dunitz as \textit{et al.}\cite{Dunitz1988} 
\begin{equation}
\langle~U_{\rm iso}^2\rangle~=~U_0^2~+~\frac{\hbar}{8\pi^2 \omega \mu}~\coth(\frac{\hbar\omega}{2k_{\rm B}T}),
\label{S1}
\end{equation}
wherein the hyperbolic cotangent of the quotient of the vibrational energy and the thermal energy, $\hbar\omega/2k_{\rm B}T$ results from the Boltzmann distribution over the quantized energy levels of a harmonic oscillator with reduced mass $\mu$.  We have  added $U_0^2$  to account for static displacements.
For 2$k_{\rm B} T>> \hbar\omega$, $\langle U^2 \rangle$ is proportional to the temperature as observed experimentally.
Using eq.~(\ref{S2}) to calculate  the reduced mass
\begin{equation}
\frac{1}{\mu} = \sum_{i}\frac{1}{m_i},
\label{S2}
\end{equation}
where the summation is carried out over the atom masses of the vibrating atoms. 
Assuming, for example, that the vibrating entity consists of two Br atoms and either a Ag or Bi atom and taking the molar mass of Br (79.9~g/mol) and of Ag (107.9~g/mol) or Bi (209~g/mol)  results in a reduced mass of a Br~-~Ag~-~Br trimer of $\sim$29~amu and for a  Br~-~Bi~-~Br trimer 33.5~amu. Using these reduced masses to fit the temperature dependence of the displacement parameter to eq.~(\ref{S1})  vibrational frequencies between 68~cm$^{-1}$ and $\sim$73~cm$^{-1}$ result,  comparing well with  typical phonon frequencies observed for CsAgBiBr$_6$ (see main manuscript).
The static displacements are small, typically 0.3~-~0.5~ \AA$^2$ indicating that the majority of the displacements are of thermal origin.

\section{Powder Transmission}
\begin{figure}[]
\includegraphics[width=0.45\textwidth,keepaspectratio]{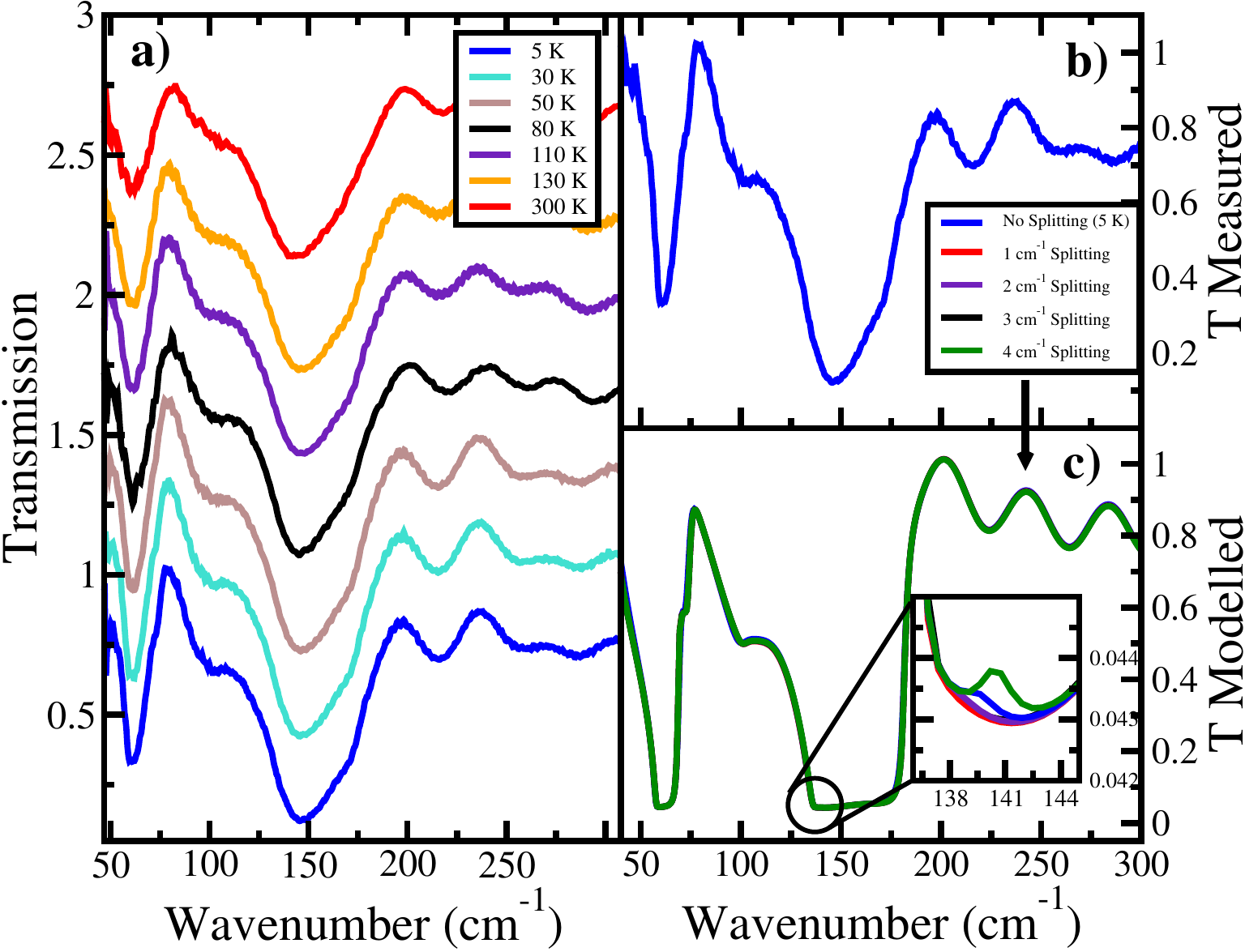}
\caption{\label{fig:PolyTrans} Powder Transmission of \cabb{}. (a) Experimental powder transmission at select temperatures. (b) Detail of the experimental powder transmission at 5 K for comparison to the model transmission shown below in panel (c). Panel (c) shows that there is negligible difference in the model spectrum for  splittings of 1-4 \wav{} (see inset for scale of differences) compared to no splitting.}
\end{figure}

Powder transmission was measured at selected temperatures on a crushed single crystal of \cabb{} in order to attempt to resolve weak additional phonons expected as a result of the phase transitions. The results are shown in FIG:\ref{fig:PolyTrans}(a). The powder sample was secured by melting it between two 30 $\mu$m thick sheets of low density polyethylene. No additional modes were, however, resolved beyond those obtained using the Kramers-Kronig extrapolated dielectric function derived from reflectance measurements, FIG.12 of the main article. The powder transmission was modeled as a rough surface bulk material with the thickness of the powder measured at $\sim$3 $\mu$m. To account for surface roughness the transmission was modeled using complex material Fresnel equations\cite{stratton} averaging over incidence angles from 0-90$^{\circ}$ in steps of 1$^{\circ}$ accounting for multiple reflections and Poynting vector cross terms arising from interference of waves at the back surface of the complex material\cite{poynting-cross}. The complex index of refraction input to the Fresnel equations was determined from fitting the 5 K Kramers-Kronig extrapolated complex part of the dielectric function to a series of Lorentzian oscillators. Oscillators were placed at 57.75, 68.93, 73.18, 97.05, and 104.46 \wav{} to model the contributions of the 55 \wav{} and 85 \wav{} modes which displayed clear splitting, as discussed in the main article. A mode was placed at 123.69 \wav{} for the low frequency branch that split off the 135 \wav{} mode as shown in FIG.11 of the main article. Finally two modes were placed centered around 138.66 \wav{} to model the unresolved splitting of the 135 \wav{} mode below the cubic-tetragonal phase transition, as shown in FIG.11(b) of the main article, and to assess whether such splitting could be resolved in powder transmission measurements. The results shown in FIG. S5(c) demonstrate that the model transmission spectrum is largely insensitive  to small wavenumber splittings. In these simulations, the modes were separated by 1, 2, 3, and 4 \wav{}, illustrating the difficulty of resolving closely spaced phonon modes because of strong absorption from nearby phonon features. The fringes with $\sim$40 \wav{} spacing are due to dividing by a  reference of two melted polyethylene sheets to account for the absorption of polyethylene.

\clearpage
\bibliography{apssamp}
